\documentstyle[12pt,epsf,a4wide]{article}
\newcommand{\xsection}[1]{\section{#1}\setcounter{equation}{0}}

\newcommand{\beq}{\begin{equation}}
\newcommand{\eeq}{\end{equation}}
\newcommand{\beqa}{\begin{eqnarray}}
\newcommand{\eeqa}{\end{eqnarray}}
\newcommand{\lto}{\longrightarrow}
\newcommand{\rto}{\rightarrow}
\newcommand{\mto}{\mapsto}
\newcommand{\vlto}{-\!\!\!-\!\!\!-\!\!\!\!\longrightarrow}
\newcommand{\kz}{I\!\!\!\!C}

\newcommand{\gz}{Z\!\!\!Z}
\newcommand{\nz}{I\!\!N}
\newcommand{\rz}{I\!\!R}
\newcommand{\qz}{I\!\!\!\!Q}
\newcommand{\unmat}{1 \;\!\!\!\! 1}
\newcommand{\lsup}{\limsup_{N\rightarrow \infty}}
\newcommand{\GaH}{\Gamma\backslash\cal H}
\renewcommand{\H}{I\!\!H}
\newcommand{\db}{\overline{d}}
\newcommand{\dcup}{\stackrel{.}{\cup}}

\def\C{\makebox[.6em]{\makebox [-.18em]{C}\rule{.1em}{1.5ex}}}
\def\Q{\makebox[.6em]{\makebox [-.25em]{Q}\rule{.08em}{1.5ex}}}
\def\3{\ss}

\def\cD{{\cal D}}

\def\cF{{\cal F}}

\def\cH{{\cal H}}

\def\cL{{\cal L}}
\def\cM{{\cal M}}
\def\cN{{\cal N}}
\def\cO{{\cal O}}

\def\cR{{\cal R}}

\newcommand{\al}{\alpha}
\newcommand{\be}{\beta}
\newcommand{\ga}{\gamma}
\newcommand{\Ga}{\Gamma}
\newcommand{\Gab}{\overline{\Gamma}}
\newcommand{\de}{\delta}
\newcommand{\De}{\Delta}
\newcommand{\io}{\iota}
\newcommand{\la}{\lambda}
\newcommand{\La}{\Lambda}
\newcommand{\om}{\omega}
\newcommand{\Om}{\Omega}
\newcommand{\si}{\sigma}
\newcommand{\Si}{\Sigma}
\newcommand{\ro}{\rho}
\newcommand{\th}{\theta}
\newcommand{\ze}{\zeta}
\newcommand{\ve}{\varepsilon}
\newcommand{\vp}{\varphi}
\newcommand{\vt}{\vartheta}
\newcommand{\lb}{\overline{\lambda}}

\begin{document}

\vspace*{2cm}
\begin{center}
{\bf{\LARGE Some Studies on Arithmetical Chaos \\
\vspace{5mm}
in Classical and Quantum Mechanics}} \\
\vspace*{4.5cm}
{\large Jens Bolte} \\
\vspace*{3.5cm}
II. Institut f\"ur Theoretische
Physik\\ Universit\"at Hamburg\\ Luruper Chaussee 149\\ 2000
Hamburg 50\\ Federal Republic of Germany
\end{center}
\newpage
\section*{Abstract}
Several aspects of classical and quantum mechanics applied to
a class of strongly chaotic systems are studied. The latter
consists of single particles moving without
external forces on  surfaces of constant negative Gaussian
curvature whose corresponding fundamental groups are supplied
with an arithmetic structure.

It is shown that the arithmetical features of the considered
systems lead to exceptional properties of the corresponding
spectra of lengths of closed geodesics (periodic orbits).
The most significant one is an exponential growth of
degeneracies in these geodesic length spectra. Furthermore,
the arithmetical systems are distinguished by a structure that
appears as a generalization of geometric symmetries.
These pseudosymmetries occur in the quantization of
the classical arithmetic systems as Hecke operators, which
form an infinite algebra of self-adjoint operators commuting
with the Hamiltonian.

The statistical properties of quantum energies in the arithmetical
systems have previously been identified as exceptional.
They do not fit into the general scheme of random matrix
theory. It is shown with the help of a simplified model for
the spectral form factor how the spectral statistics in
arithmetical quantum chaos can be understood by the properties of
the corresponding classical geodesic length spectra. A decisive
role is played by the exponentially increasing multiplicities
of lengths. The model developed for the level spacings
distribution and for the number variance is
compared to the corresponding quantities
obtained from quantum energies for a specific arithmetical
system.

Finally, the convergence properties of a representation for the
Selberg zeta function as a Dirichlet series are studied. It
turns out that the exceptional classical and quantum mechanical
properties shared by the arithmetical systems prohibit
a convergence of this important function in the physically
interesting domain.
\newpage
\tableofcontents

\vspace*{4mm}
\noindent \hspace*{4mm}
{\bf References\hfill 85}
\newpage
\xsection{Introduction}
Two classes of
theoretical activities in the field of natural sciences may
be distinguished. The first one consists of efforts
towards  unraveling the kinematics of a given system and of
finding the underlying dynamics governing the time evolution
of the system; the second one is formed by the development
of methods and techniques to obtain numerical data from the
theory describing the dynamics in order to be able to
compare them with empirical data. A sensible and
satisfying theory of natural phenomena then has to meet these
two requirements: it should be logically consistent and
ought produce testable predictions. Both of these aspects have
to be well distinguished. One consequence of this consideration
is that a theory containing deterministic dynamics need not be
predictive with respect to its time evolution. This realization lies
at the heart of what nowadays is widely known as
{\it deterministic chaos}.

To be more specific we now consider Hamiltonian dynamical
systems (with a finite number $N$ of degrees of freedom) in
classical mechanics. These may be described on N--dimensional
configuration manifolds $M$ with Riemannian metrics
$ds^2 =g_{ij}dq^i dq^j$ defined on them. ($\vec{q}=(q^1 ,
\dots ,q^N )$ are local coordinates on $M$.) The dynamics
is specified by providing $M$ with a Lagrangian function
$L(\vec{q},\dot{\vec{q}})=\frac{1}{2}(\frac{ds}{dt})^2 -
V(\vec{q})$. The equations of motion for this system may
then be obtained by Hamilton's principle as the
Euler--Lagrange equations
\beq
\label{ELG}
\frac{d}{dt}\,\frac{\partial L}{\partial \dot{q}^i}-\frac{\partial L}
{\partial q^i}=0\ ,\ \ \ \ i=1,\dots ,N\ .
\eeq
Specifying initial values $\vec{q}_0$ and $\dot{\vec{q}}_0$
at a time $t_0$ then uniquely fixes the time evolution $\vec{q}(t)$
of the system for all times $t\geq t_0$. This is the manifestation
of determinism in classical mechanics. As yet, nothing is said,
however, about predictability. Since in practice it is never
possible to prepare a system at an initial time $t_0$ to
be in a definite state $(\vec{q}_0 ,\dot{\vec{q}}_0 )$,
one has to allow the initial values to be taken from
$[\vec{q}_0 -\vec{\ve},\vec{q}_0 +\vec{\ve}] \times [\dot{\vec{q}}_0
-\vec{\de}, \dot{\vec{q}}_0 +\vec{\de}]$ for some small
uncertainties $\vec{\ve}$ and $\vec{\de}$. Predictability
would now require the uncertainties to grow only modestly
under the time evolution dictated by (\ref{ELG}). By this
we mean an increase of $|\vec{\ve}|$ and $|\vec{\de}|$ at
most like a power of $t$. It is, however, by no means
clear from the information provided so far that this will
be the case.

The class of dynamical systems possessing the most regular
kind of time evolution is given by the {\it integrable}
ones. For them there exist $N$ independent constants of motion with
pairwise vanishing Poisson brackets. If one considers their
time evolution in phase space (i.e.\ on the cotangent
bundle $T^* M$), this is found to take place on an
$N$--dimensional torus. The equations of motion can then
be integrated by quadratures. Integrable systems are predictable
in the sense just introduced.

The other extreme is given by irregular systems sharing the
property of {\it ergodicity}. Almost all of their phase space
trajectories fill the $(2N-1)$--dimensional hypersurface of
constant energy in phase space densely. The probability of finding an
arbitrary phase point in some bounded region of the
hypersurface of constant energy is proportional to the volume
of that region. This means that the trajectories of ergodic systems
are uniformly distributed in phase space. Among such systems
one can find a hierarchy of even higher irregularities:
mixing systems, Anosov--systems, K--systems, $\dots$; see e.g.\
\cite{Cornfeld} as a reference.

In between the two extremes of integrable and ergodic systems
there exists any kind of intermediate behaviour. These
systems, however, will not be dealt with in the course
of the present investigation.

It now appears to be useful to reformulate the setting a
little bit. Suppose the system under study is in a state of
energy $E$. According to the conservation of energy it
can visit only those parts of $M$ during its time
evolution, where $V(\vec{q})\leq E$ is fulfilled. On this
domain of $M$ one introduces the {\it Jacobi metric}
$dS_E^2 :=[E-V(\vec{q})]ds^2$. Maupertuis' principle of least
action, being equivalent to the equations of motion (\ref{ELG}),
is now also equivalent to the statement that the trajectories
of the system on $M$ are geodesics in the Jacobi metric
$dS_E^2$, see e.g.\ \cite{Arnold} for details. Hence
every Hamiltonian dynamics can be viewed as the geodesic
flow on some Riemannian manifold. Hopf \cite{Hopf} has shown
that a negative curvature associated with the Jacobi metric
is a sufficient condition to render the system ergodic.

A guiding principle in the present study will be to keep
things as simple as possible, without giving up the essential
structures that determine the properties of a --hopefully--
general enough class of systems. Hence we agree to restrict
our attention to the following kind of examples. They will
have two degrees of freedom, since this is the minimal dimensionality
required for a Hamiltonian system to be non--integrable; the
reason for this being that conservation of energy renders
every system of one degree of freedom integrable. Having
in mind the above remark on the Jacobi metric and its
curvature, we choose geodesic flows on two dimensional
Riemannian surfaces of constant negative Gaussian curvature
as our prime examples. We are then going to study the interplay
between the classical and quantum dynamics for these systems.

A central issue of this investigation will be to identify
fingerprints of the classical properties of a given dynamical
system in its quantum version. Since the time evolution in
(non-relativistic) quantum mechanics, which is governed
by the Schr\"odinger equation
\beq
\label{Schroedingereq}
i\hbar \frac{\partial}{\partial t}\psi (t,\vec{q})=H \psi (t,\vec{q})
\eeq
for the time dependent wave function $\psi (t,\vec{q})$,
is linear, no chaotic phenomena like an exponential sensibility
to initial conditions can occur. Thus the question arises,
how a potential irregularity of the classical system can show up
in the semiclassical limit $\hbar \rto 0$. Although this
limit is not ``smooth'' in that at the value $\hbar =0$ typical
quantum mechanical structures and quantities no longer exist,
it is expected that for small values of $\hbar$ one should be
able to detect classical properties of the system. One such
sign of the structure of the classical phase space in quantum
mechanics is the applicability of semiclassical quantization
rules such as the WKB--method (for one-dimensional systems)
or the EBK--method (for multi-dimensional systems), which
is restricted to the integrable case.
Therefore a first and obvious question would be to find a
semiclassical quantization procedure e.g.\ for strongly
chaotic classical systems.

An answer to this question is provided by Gutzwiller's
{\it periodic-orbit theory} \cite{Gutzwiller,GutzBuch},
in which the
spectrum of the quantum Hamiltonian $H$ is determined in a
semiclassical approximation by the set of actions evaluated along the
classical periodic trajectories ({\it periodic orbits}) of the system.
{\it Plane billiards} seem to be well suited for an explanation
of the general procedure. Let therefore $D\subset \rz^2$
be some connected domain on the euclidean plane. Its boundary
$\partial D$ is assumed to be piecewise smooth. A plane
billiard then consists of a particle moving freely inside $D$
and being elastically reflected on $\partial D$, which will
be chosen such that the classical dynamics is strongly
chaotic. This should mean that the system is ergodic and that
all periodic orbits are isolated in phase space and are
unstable. The action $S_{\ga}$ along a periodic orbit $\ga$
is then given by $S_{\ga}=pl_{\ga}$, where $p$ denotes the
particle's momentum and $l_{\ga}$ is the length of the orbit
$\ga$. The quantum mechanical Hilbert space for such a
billiard is given by the space of square integrable functions
$\psi (x,y)$ on $D$ that vanish on $\partial D$. In units where
$\hbar =1=2m$ the quantum Hamiltonian is $H=-\De_E =
-(\partial_x^2 +\partial_y^2 )$
and possesses a purely discrete spectrum $0<E_1 \leq E_2 \leq
E_3 \dots$, $E_n =p_n^2$. Weyl's famous law for the
number $N(E)$ of energy eigenvalues not exceeding $E$ states
that asymptotically for $E\rto\infty$
\beq
\label{Weyl}
N(E)\sim \frac{\mbox{area}(D)}{4\pi}\,E\ .
\eeq
Thus $E_n \sim \frac{4\pi}{\mbox{area}(D)}\,n$, $n\rto\infty$, and
hence the resolvent operator $(H-E)^{-1}$ is not of trace class.
Introducing, however, a suitable smearing allows to investigate
trace class operator valued functions of $H$. In \cite{SieberPLA}
one can find a regularized version of {\it Gutzwiller's trace
formula}. Asymptotically in the semiclassical limit it reads
\beq
\label{GutzwillerTF}
\sum_{n=1}^{\infty} h(p_n )\sim 2\int_0^{\infty} dp\,p\,\overline{d}(p)
h(p)+\sum_{\ga}\sum_{k=1}^{\infty} \frac{\chi_{\ga}^k \, l_{\ga}\, g(k
l_{\ga})}{e^{kl_{\ga}\la_{\ga}/2}-\si_{\ga}^k e^{-kl_{\ga}\la_{\ga}/2}}
\ .
\eeq
In this formula $h(p)$ is an even function, holomorphic in the strip
$|Im\,p|\leq \tau -\frac{\lb}{2}+\ve$, $\ve >0$, that decreases
faster than $|p|^{-2}$ for $|p|\rto\infty$. $g(x):=\int_{-\infty}^{+
\infty}\frac{dp}{2\pi}e^{ipx}h(p)$, and $\overline{d}(p)$ is a mean
spectral density expressed as a function of momentum $p=\sqrt{E}$.
Under these assumptions the integral and the sums involved in
(\ref{GutzwillerTF}) are absolutely convergent. $\chi_{\ga}=(-1)^{
j_{\ga}}$, where $j_{\ga}$ is the number of reflections on $\partial
D$ when traversing $\ga$ once. $\la_{\ga}=u_{\ga}/l_{\ga}$ is the
Lyapunov exponent of $\ga$ and $u_{\ga}$ is its stability exponent.
$\si_{\ga}$ denotes the sign of the
trace of the monodromy matrix for $\ga$.
The {\it topological entropy} $\tau$ describes the proliferation
of the number $\cN (l)$ of periodic orbits of lengths not
exceeding $l$ by
\beq
\label{PGT}
\cN (l)\sim \frac{e^{\tau l}}{\tau l}\ ,\ \ \ \ l\rto\infty\ .
\eeq
$\lb$ denotes the asymptotic average of the $\la_{\ga}$'s and is
also known as the {\it metric entropy}.

The trace formula (\ref{GutzwillerTF}) can be viewed as an identity
in the sense of distributions
for the spectral density $d(E)=\sum_{n=1}^{\infty}\de (E-E_n )$
(expressed by the variable $p$) of the form $d(p)\sim
\overline{d}(p)+d_{fl}(p)$. $d_{fl}(p)$ is an oscillatory
correction to the mean density $\overline{d}(p)$, determined by
the periodic-orbit sum in (\ref{GutzwillerTF}).
The conditions on $h(p)$ stated
after (\ref{GutzwillerTF}) then define the space of test functions.

It is possible to use Gutzwiller's trace formula in order to
derive semiclassical quantization rules that yield approximations
for the quantum energies $\{ E_n \}$ in terms of the actions
of periodic orbits in the corresponding classical system, see e.g.\
\cite{GutzBuch} and the references therein. In some sense the
so-obtained quantization rules are a substitute for the EBK-quantization
of classically integrable systems. A fundamental difference
between the semiclassical quantization schemes for classically
integrable and chaotic systems is provided by the computational
effort to be spent in order to resolve some quantum energy
$E_n$. In the integrable case this is independent of the energy,
whereas (\ref{PGT}) requires for chaotic systems
an exponentially increasing number of periodic orbits to be
taken into account when trying to compute higher and higher
energies. This observation reflects the high degree of
irregularity and complexity of chaotic systems.

Historically, the first chaotic systems under investigation were
of mathematical nature and mainly served as examples to
develop the mathematical theory of dynamical systems.
Hadamard \cite{Hadamard} studied the geodesic flow on closed
surfaces endowed with Riemannian metrics of constant negative Gaussian
curvature, which also go under the notion of {\it hyperbolic
surfaces}, and discovered an instability of the trajectories
of such flows. Since the surfaces he studied cannot be realized
as being embedded in $\rz^3$, they were considered as purely
mathematical examples. For a specific hyperbolic surface
Artin could later prove \cite{Artin} the ergodicity of the geodesic
flow on it. Moreover, at this occasion he introduced a symbolic
dynamics for this system. This was the first time that
the property of ergodicity could ever be rigorously demonstrated
for a dynamical system. One could thus view Artin's paper as
the foundation of the modern theory of dynamical systems.
Only after the development of powerful computers that allow
for a quantitative analysis of irregular dynamical systems
these subjects arose interest among a considerable number of
physicists. Starting with the series of papers \cite{Gutzwiller}
by Gutzwiller a discussion of the quantization of chaotic classical
systems was rendered possible. Again it was Gutzwiller who noticed
\cite{GutzPRL} that his trace formula turned into an exact
identity when applied to geodesic flows on surfaces of constant
negative curvature. In mathematics this identity had long before
been introduced by Selberg \cite{Selberg}, who intended to
understand the Riemann zeta function and the Riemann
hypothesis with the help of his trace formula.

Since Gutzwiller's observation hyperbolic surfaces have been
intensively studied in the context of the quantization of
classically chaotic systems. The first numerical results on
quantum energies for a hyperbolic triangle
obtained from a solution of the stationary
Schr\"odinger equation by Schmit have been presented in
\cite{Cuernavaca,BalazsVoros}. Aurich and Steiner \cite{Aurich88}
first calculated lengths of periodic orbits and also quantum
energies \cite{AuSieSt,Aurich89} on a specific
hyperbolic surface, the so-called {\it regular octagon}. The
lengths were used as an input to evaluate the Selberg trace formula
\cite{AuSieSt,Aurich89}.
By chance, the regular octagon that had been chosen for the
numerical studies turned out to be very specific: its
fundamental group is of an arithmetical origin. It was realized
\cite{Aurich88,AurichBogo}
that the arithmetic structure underlying the regular octagon
leads to exponentially increasing multiplicities of lengths of
periodic orbits. As was later noticed \cite{Aurich90,AurichProc},
this property is exceptional and has remarkable
consequences. One of these is the occurrence
of unexpected statistical properties of the related quantum energy
spectra. Although the classical system is as chaotic as it could
be, the quantum spectral statistics are more alike those of
classically integrable systems than those expected for generic
classically chaotic ones. Namely, due to a conjecture of
Bohigas, Giannoni and Schmit \cite{BoGiSch} which was mainly
based on numerical observations, generic chaotic systems excel
by quantum energy spectra that can be well described by
eigenvalues of large random matrices. The resulting spectral statistics
then differ drastically from those in the integrable case as
obtained before by Berry and Tabor \cite{BerryTabor}. Somewhat later
it was shown \cite{Aurich90}, however, that the energy fluctuations
of 30 non-arithmetic octagons are in agreement with the
predictions of random matrix theory. At the time
the unexpected statistical properties of the eigenvalues for
the regular octagon were first observed it was, however, not
clear that these follow from the arithmetic properties of the
fundamental group. This was later clearly expressed to hold for all
arithmetical systems in the two simultaneously appearing papers
\cite{BoGeGiSch} and \cite{BoStSt}.

The objective of the present investigation now is to study the
class of dynamical systems arising from geodesic flows on
hyperbolic surfaces with arithmetic fundamental groups both
from a classical as well as from a quantum mechanical point of view.
The ultimate goal then will be to understand the exceptional
statistics of the quantum energy spectra. The problem of
studying the wave functions arising on the quantum mechanical
side, however, will
not be treated here. It seems that the eigenfunctions of
the arithmetical systems are not as exceptional as the
eigenvalue spectra, see \cite{Hejhal92,Sarnak,Aurichwave}.

The organization of this paper is as follows. Chapter
2 reviews some general properties of (discrete) quantum energy
spectra and introduces the necessary means and notions for
their investigation in order to prepare for studying the spectral
statistics of arithmetical systems. Since in order to be ready
for the latter the classical aspect of the problem has to be understood
in quite some detail, chapter 3 is devoted to an investigation
of geodesic length spectra on arithmetic hyperbolic surfaces.
A central result, obtained in section 3.4, is the observation
of exponentially growing degeneracies in arithmetic length spectra.
Another peculiarity arising from the arithmeticity of a
fundamental group is the existence of infinitely many
{\it pseudosymmetries} on the corresponding surface. It will be
attempted to give a geometric picture of this structure in
section 3.5. Chapter 3 will be closed by an investigation of
fluctuations of the lengths of closed geodesics on both
arithmetic and non-arithmetic hyperbolic surfaces. These
fluctuations are closely related to the fluctuation properties
of the corresponding quantum energies. Using the latter in order
to learn about the former is also known as {\it inverse quantum
chaology}.

Chapter 4 then deals with quantum energy spectra
of arithmetical systems. At first {\it Hecke operators} are
discussed as representations of pseudosymmetries on the
wave functions and it is found that this structure results
in constraints on the eigenvalue spectra. This realization
is taken as a first hint towards exceptional spectral statistics.
Thereafter the spectral form factor is reviewed and its
properties in the arithmetical case, derived from the exponentially
increasing multiplicities of lengths of periodic orbits, are
employed to introduce a simplified model form factor.
The latter can then be used
to derive a model for the level spacings distribution as well as
for the number variance in arithmetical chaos. The model will
also be compared to existing numerical data in order to
test its quality. Finally, convergence properties of the
Selberg zeta function \cite{Selberg}, which arises from the
Selberg trace formula and plays an important role for obtaining
quantization rules, are studied. It is observed that the spectral
statistics influence the convergence properties of this function
considerably, and that the arithmetic systems suffer from a lack
of convergence in the physically interesting domain.

Chapter 5
finally summarizes the findings of the preceding investigations.
Furthermore, three appendices are included. The first one
reviews the theory of the Riemann zeta function, which is in
many respects similar to the Selberg zeta function. The analogy
of these two functions is particularly useful for inverse
quantum chaology. Appendix B treats the desymmetrization of
a class of hyperbolic surfaces explicitly in order to illustrate
the general procedure of getting rid of unwanted symmetries
on hyperbolic surfaces. The final appendix collects the definitions
of $O$--, $o$--, and $\Om$--estimates for remainder terms that
occur at several instants in the main body of the text.

\xsection{Some General Remarks on Discrete Quantum Energy Spectra}
Let a classical dynamical system be given with a finite number of
degrees of freedom. After quantization its
Hamiltonian $H$ should have a discrete spectrum $0<E_1
\leq E_2 \leq E_3 \dots$. A {\it quantization rule}
then is the specification of a function $f:\,\nz \rto \rz$
such that $E_n =f(n)$. The explicit knowledge of such a
quantization function completely solves the problem
of determining the energy spectrum of the system. If one
is merely interested in energy eigenvalues and not in
wavefunctions, a quantization rule is equivalent to solving
the stationary Schr\"odinger equation $H\psi_n =E_n \psi_n $.
One possibility to obtain a quantization function is to study
the {\it spectral staircase}
\beq
\label{staircase}
N(E):= \# \{E_n\ ;\ E_n \leq E\,\} = \sum_{E_n \leq E} 1\ .
\eeq
Defining then $N_0 (E):=\frac{1}{2}\lim_{\ve\rto 0}\left[ N(E+\ve)
+N(E-\ve)\right]$ yields the quantization rule
\beq
\label{quantrule}
N_0 (E_n )=n-\frac{1}{2}\ ,
\ \ \ \ \ \ n\in\nz\ ,
\eeq
which can be converted to the condition $\cos(\pi N_0 (E))=0$
for $E$ to be an energy eigenvalue, see \cite{Aurich92.1,Aurich92.2}.
This consideration stresses the importance of studying the
spectral staircase thoroughly if one is interested in the
quantization of a dynamical system.

There is, however, only a very restricted number of
examples where an exact quantization function is explicitly
known, among which are the typical textbook examples
of a particle in a box, the harmonic oscillator, or the
Coulomb potential. Semiclassically, quantization rules are
provided by the WKB-- or EBK--methods for all classically
integrable systems. For these there exist canonical
transformations of the classical phase space variables
$(\vec{p}, \vec{q})$ to action-angle variables $(\vec{I},
\vec{\om})$, such that the Hamiltonian function $H(\vec{p},
\vec{q})$ transforms to one being only dependent on the
actions $\vec{I}$, $H=H(\vec{I})$. Then
\beq
\label{EBK}
E_{\vec{m}}=H({\vec{I}}_{\vec{m}})\ ,\ \ \ {\vec{I}}_{\vec{m}}=
\vec{m}+\frac{1}{4}\vec{\al} \ ,
\eeq
is the desired (semiclassical) quantization rule. In
(\ref{EBK}) $\vec{m}$ runs through $\gz^N$ and $\vec{\al}$ is
the vector of Maslov indices.

It is one of the major goals of {\it quantum chaology} to
obtain an analogue of (\ref{EBK}) for strongly chaotic systems.
Since there do not exist action-angle variables and even
no remnants thereof for ergodic systems, the quantization problem
has to be tackled in a totally different manner. A starting
point one could think of would be to study the spectral
staircase in as much detail as possible, thereby keeping in
mind the relation (\ref{quantrule}). For such an analysis it
proves useful to split $N(E)$ into a smooth part $\overline{
N}(E)$, which is an approximation to $N(E)$ in that the steps
have been smeared out, and a remaining contribution $N_{fl}
(E)$. This then describes the fluctuations of $N(E)$ about
its mean $\overline{N}(E)$. A priori there is no preferred
way to define the smoothing $\overline{N}(E)$. The splitting
\beq
\label{split}
N(E)=\overline{N}(E)+N_{fl}(E)
\eeq
should only be subject to some general requirements.
$\overline{N}(E)$ should be smooth without ``too many''
oscillations (ideally one would require it to be
monotonically increasing), and it should be asymptotically
identical to the true staircase, i.e.\ $\overline{N}(E)\sim
N(E)$, for $E\rto\infty$. $N_{fl}(E)$ then should describe
fluctuations about $\overline{N}(E)$, i.e.\ $N_{fl}(E)=
N(E)-\overline{N}(E)$ should fluctuate about zero in the
limit $E\rto \infty$. This requirement can be expressed
in a variety of different ways, one of which is
\beq
\label{arithmean}
\lim_{L\rto\infty}\ \frac{1}{N(L)} \sum_{E_n \leq L} N_{fl}(E_n )=0\ ,
\eeq
i.e.\ an asymptotic vanishing of an arithmetic mean of $N_{fl}(E)$.
Another, although similar, requirement uses an integral
instead of a sum, namely
\beq
\label{intmean}
\lim_{L\rto\infty}\ \frac{1}{L}\int_0^L dE\ N_{fl}(E) =0\ .
\eeq

Provided there is a Gutzwiller trace formula available, like
(\ref{GutzwillerTF}) for chaotic plane billiards, a natural
splitting (\ref{split}) is yielded by the two contributions to
the r.h.s.\ of the trace formula. In order to obtain this one
introduces the complex variable $s=\frac{\lb}{2}-ip$, i.e.\
$E(s)=p^2=-(s-\frac{\lb}{2})^2$. A regularized trace of the
resolvent operator may be introduced by choosing the function
$h(p)=\frac{1}{E(s)-p^2}-\frac{1}{E(\si)-p^2}$, for $Re\,s$,
$Re\,\si\,>\tau$, which fulfills the requirements for
a test function to be used in (\ref{GutzwillerTF}). Inserting
this into the trace formula yields
\beqa
\label{resolvent}
\sum_{n=1}^{\infty}\left[\frac{1}{E(s)-E_n}-\frac{1}{E(\si)-E_n}
\right] &\sim& 2\int_0^{\infty}dp\,p\,\overline{d}(p)\left[
               \frac{1}{E(s)-p^2}-\frac{1}{E(\si)-p^2}\right]
               \nonumber \\
        &    & -\frac{1}{2s-\lb}\sum_{\ga}\sum_{k=1}^{\infty}
               \frac{\chi_{\ga}^k l_{\ga}\ e^{-(s-\frac{\lb}{2})kl_{
               \ga}}}{e^{kl_{\ga}\la_{\ga}/2}-\si_{\ga}^k\,e^{-kl_{\ga}
               \la_{\ga}/2}} \\
        &    & +\frac{1}{2\si-\lb}\sum_{\ga}\sum_{k=1}^{\infty}
               \frac{\chi_{\ga}^k l_{\ga}\ e^{-(\si-\frac{\lb}{2})kl_{
               \ga}}}{e^{kl_{\ga}\la_{\ga}/2}-\si_{\ga}^k\,e^{-kl_{\ga}
               \la_{\ga}/2}} \ . \nonumber
\eeqa
Choosing now $E(\si)\in \rz$, i.e.\ $Re\,\si \rto \frac{\lb}{2}+$,
and keeping $Re\,s >\frac{\lb}{2}$, i.e.\ $Im\,E(s)>0$,
taking the imaginary part on both sides of (\ref{resolvent}),
and multiplying the result with $-\frac{1}{\pi}$ yields
($E=p^2$, $p>0$)
\beq
\label{dvonE}
\sum_{n=1}^{\infty} \de (E-E_n )\sim \overline{d}(E)-
Im\,\left\{ \frac{1}{2ip\pi}\frac{Z'(\frac{\lb}{2}-ip)}
{Z(\frac{\lb}{2}-ip)}\right\} \ ,
\eeq
where the {\it dynamical zeta function} $Z(s)$ has been introduced.
It is for $Re\,s>\tau$ defined by the {\it Euler product}
\beq
\label{zetafct}
Z(s):=\prod_{\ga}\prod_{n=0}^{\infty}\left( 1-\chi_{\ga} \si_{\ga}^n
e^{-(s+n\la_{\ga}+\frac{1}{2}(\la_{\ga}-\lb))l_{\ga}}\right) \ .
\eeq
Integrating both sides of (\ref{dvonE}) from zero to $E$ yields
(assuming $\mbox{arg}\,Z(\frac{\lb}{2})=0$)
\beq
\label{NvonE}
N(E)\sim \overline{N}(E)+\frac{1}{\pi}\, \mbox{arg}\, Z\left(\frac{\lb}
{2}+i\sqrt{E}\right) \ ,
\eeq
where $\overline{N}(E)=\int_0^E dE'\,\overline{d}(E')$.
One has now obtained, in the semiclassical limit $E\rto\infty$,
that
\beq
\label{Nfluc}
N_{fl}(E)\sim\frac{1}{\pi}\,\mbox{arg}\, Z\left(\frac{\lb}{2}+i
\sqrt{E}\right)\ .
\eeq
This relation stresses the importance of dynamical zeta
functions for the study of quantum spectral properties of
chaotic dynamical systems. In addition, dynamical zeta functions
turn out to be powerful tools to obtain quantization rules
in periodic-orbit theory directly \cite{SieberPRL,Matthies,%
Tanner,Bogomolny,Keating,BerryKeating2,AurichProc}.
Their zeroes $s_n$ on the {\it critical line} $Re\,s =\frac{\lb}{2}$
yield semiclassical approximations to the quantum energies via
$E_n =E(s_n )$. This result may be obtained by integrating both
sides of (\ref{resolvent}).

In \cite{Aurich89,Aurich92.1,Aurich92.2}
a periodic-orbit expression for the r.h.s.\ of (\ref{Nfluc})
has been obtained that enabled the authors to compute
$N_{fl}(E)$ numerically in some approximation from the lengths
of primitive periodic orbits for several chaotic systems.
To do so one inserts the Gaussian $h(p)=\frac{1}{\ve \sqrt{\pi}}
(e^{-(p-q)^2 /\ve^2}+e^{-(p+q)^2/\ve^2})$,
with $g(x)=\frac{1}{\pi}\cos (qx)\,e^{-\ve^2 x^2 /4}$,
into the trace formula (\ref{GutzwillerTF}). Then one integrates
the resulting formula in $q$ from zero to $p$, $p^2 =E$.
Since $\lim_{\ve \rto 0} h(p)=\de (p-q)+\de (p+q)$, the l.h.s.\
of (\ref{GutzwillerTF}) yields for $\ve >0$
a smoothing $N_{\ve}(E)$ of
the spectral staircase $N(E)$ with $\lim_{\ve \rto 0} N_{\ve}(E)
=N(E)$,
\beq
\label{dreg}
\lim_{\ve \rto 0}\ \sum_{n=1}^{\infty}\frac{1}{\ve\sqrt{\pi}}
\int_0^p dq\ \left( e^{-\frac{(p_n -q)^2}{\ve^2}}+e^{-\frac{(p_n
+q)^2}{\ve^2}} \right) =\sum_{n=1}^{\infty}\Theta (E-E_n )
=N(E)\ .
\eeq
On the r.h.s.\ of (\ref{GutzwillerTF}) one obtains $\overline{N}_{\ve}
(E)+N_{\ve ,fl}(E)$, with
\beq
\label{Nflucreg}
N_{\ve ,fl}(E)=\frac{1}{\pi}\sum_{\ga}\sum_{k=1}^{\infty}\frac{1}{k}
\frac{\chi_{\ga}^k \ e^{-\ve^2 k^2 l_{\ga}^2 /4}}{e^{kl_{\ga}
\la_{\ga}/2}-\si_{\ga}^k e^{-kl_{\ga}\la_{\ga}/2}} \sin (\sqrt{E}
kl_{\ga}) \ .
\eeq
As long as $\ve >0$, the double sum on the r.h.s.\ of (\ref{Nflucreg})
converges absolutely, since then the Gaussian damping factor
$e^{-\ve^2 k^2 l_{\ga}^2 /4}$ overcompensates the exponential
proliferation (\ref{PGT}) of periodic orbits.

It will now be shown, using the smoothing $N_{\ve ,fl}(E)$
for the fluctuating part $N_{fl}(E)$ of the spectral staircase,
that the two conditions (\ref{arithmean}) and (\ref{intmean})
are satisfied by the splitting (\ref{NvonE}). Thus, in the
semiclassical limit, $N_{fl}(E)$ indeed describes the fluctuations
of $N(E)-\overline{N}(E)$ about zero, which is a major justification
to identify $\overline{N}(E)$ as a mean spectral staircase.
The essential part of demonstrating that the requirement
(\ref{arithmean}) is met is to study, after inserting
(\ref{Nflucreg}) into (\ref{arithmean}),
\beq
\label{mean}
\frac{1}{N(L)}\sum_{p_n \leq \sqrt{L}}
\sin(kl_{\ga}p_n)
\eeq
for every term in the summation over periodic orbits $\ga $.
For the following the reasonable assumption has to be made that the
momenta $p_1 ,\dots,p_N $ are linearly independent over $\gz$ for
any $N\in \nz$. It is then known \cite{Kac}
(see also Bohigas' contribution in \cite{LesHouches})
that the functions $\cos p_1 t,\dots,\cos p_N t$ are statistically
independent. If $A$ is any Lebesgue-measurable set in $\rz$
and $l(A)$ is its Lebesgue-measure, define the relative measure
of $A$ as $l_r (A):=\lim_{T\rto\infty}\,\frac{1}{2T}l\{A\cap (-T,+T)\}$.
If now
\beq
x_N (t):=\sqrt{\frac{2}{N}}\,\sum_{n=1}^N \cos p_n t \ ,
\eeq
and $A_N (a,b):= \{t;\ x_N (t)\in (a,b)\}$ denotes the set
of the $t\in\rz$ such that $x_N (t)$ takes its values in the interval
$(a,b)$, then \cite{Kac}
\beq
\lim_{N\rto\infty}l_r \{ A_N (a,b)\} =\frac{1}{\sqrt{2\pi}}
\int_a^b dx\ e^{-\frac{1}{2}x^2}\ .
\eeq
In other words, the values of the function $x_N (t)$ approach
a normal distribution with zero mean and unit variance in
the limit $N\rto\infty$. If one then replaces $\cos p_n t$
by $\sin p_n t$ (for $t=kl_{\ga}$), which does not change
the above argument, the values of the function
\beq
\label{fE}
f_L (kl_{\ga}):=\frac{1}{\sqrt{N(L)}}
\sum_{p_n \leq \sqrt{L}}\sin (kl_{\ga}p_n)
\eeq
become, in the limit $L\rto\infty$, Gaussian distributed with
zero mean and variance $\si_f^2 =\frac{1}{2}$.
Inserting (\ref{Nflucreg}) into (\ref{arithmean}) leads to a
triple summation. The one over the energies $E_n$ yields
(\ref{mean}). The remaining ones over periodic orbits $\ga$
and their repetitions $k$ sum $f_L (t)$ taken at $t=kl_\ga$
times a prefactor,
\beq
\label{sumfL}
\frac{1}{N(L)}\sum_{E_n \leq L}N_{\ve ,fl}(E_n )
=\frac{1}{\pi\sqrt{N(L)}}\sum_{\ga}\sum_{k=1}^{\infty}\frac{1}{k}
\frac{\chi_{\ga}^k \ e^{-\ve^2 k^2 l_{\ga}^2 /4}}{e^{kl_{\ga}
\la_{\ga}/2}-\si_{\ga}^k e^{-kl_{\ga}\la_{\ga}/2}}f_L (kl_\ga )\ .
\eeq
Since $f_L (kl_\ga )$ to be summed over is Gaussian distributed
about zero, this is for $kl_\ga \rto\infty$ an effectively bounded
function. Therefore the sum on the r.h.s.\ of (\ref{sumfL})
converges due to the regularizing Gaussian damping as long as
$\ve >0$. In the limit $L\rto\infty$ then the complete
expression (\ref{sumfL}) vanishes like $\frac{1}{\sqrt{N(L)}}$.
Weyl's law (\ref{Weyl}) thus determines the rate of vanishing
to be of the order of $L^{-\frac{1}{2}}$.

In order to proof the validity of (\ref{intmean}), when $N_{\ve,
fl}(E)$ from (\ref{Nflucreg}) is used instead of $N_{fl}(E)$,
one has to calculate
\beqa
\label{integral}
\frac{1}{L}\int_0^L dE\ \sin(kl_{\ga}\sqrt{E})
   &=& \frac{2}{L}\int_0^{\sqrt{L}}
       dp\ p\,\sin(pkl_{\ga}) \nonumber \\
   &=& -\frac{2}{kl_{\ga}\sqrt{L}}\cos(kl_{\ga}\sqrt{L})
       +\frac{2}{(kl_{\ga})^2 L} \sin(kl_{\ga}\sqrt{L})\ ,
\eeqa
which vanishes for $L\rto\infty$. The double sum over periodic orbits
and their repetitions one is in analogy to (\ref{sumfL}) left
with therefore also vanishes. It is interesting to notice that both
mean values (\ref{arithmean}) and (\ref{intmean}) extended over a finite
interval of length $L$ vanish as $L^{-\frac{1}{2}}$ for $L\rto\infty$.
For every $\ve >0$ the conditions
(\ref{arithmean}) and (\ref{intmean}) are thus fulfilled, and
hence also in the limit $\ve\rto 0$. It seems that any other
reasonable requirement expressing the same statement on the
fluctuations of $N_{\ve ,fl}(E)$ will also be satisfied, so that the
splitting (\ref{NvonE}) seems to be a natural one for a semiclassical
analysis of the spectral staircase $N(E)$. This point may
stress the importance and usefulness of employing Gutzwiller's
trace formula and dynamical zeta functions.

It should be remarked that completely analogous results can be
obtained for the spectral density $d(E)=\frac{d}{dE}\,N(E)$.
Splitting $d(E)=\overline{d}(E)+d_{fl}(E)$ according to
(\ref{dvonE}) and using the regularization (\ref{Nflucreg}) yields
\beq
\label{dflucreg}
d_{\ve ,fl}(E)=\frac{1}{2\pi\sqrt{E}}\sum_{\ga}\sum_{k=1}^{\infty}
\frac{\chi_{\ga}^k \ l_\ga\ e^{-\ve^2 k^2 l_{\ga}^2 /4}}{e^{kl_{\ga}
\la_{\ga}/2}-\si_{\ga}^k e^{-kl_{\ga}\la_{\ga}/2}} \cos (\sqrt{E}
kl_{\ga}) \ .
\eeq
Again, the two averaging procedures (\ref{arithmean}) and
(\ref{intmean}) can be employed, leading to the result that
$d_{\ve ,fl}(E)$ vanishes with respect to those mean values
if one extends the interval
$(0,L)$ to be averaged over to infinity, i.e.\ for $L\rto\infty$.
Compared to $N_{\ve ,fl}(E)$ these means of $d_{\ve ,fl}(E)$
vanish faster, namely like $L^{-1}$.

As described above a major objective in quantum chaology
is to explain and classify the statistical properties of
quantum energy spectra for systems with a chaotic classical
limit, and one way to achieve this is to describe the
spectral staircase as explicitly as possible. Since $\overline{N}
(E)$ is smooth and rather ``harmless'', the more interesting
object is $N_{fl}(E)$. To be able to compare systems with
different mean behaviours of their respective spectral
staircases one conventionally introduces an {\it unfolding}
of spectra. Define $x:=\overline{N}(E)$, $x_n =\overline{N}
(E_n)$. The unfolded spectrum $\{ x_n \}$ has a spectral
staircase that shall be denoted by $N(x)$. By its very
definition, $\overline{N}(x)=x$ and thus
\beq
\label{unfoldstair}
N(x)=x+N_{fl}(x) \ .
\eeq
After being unfolded different spectra can be compared quite
easily. They then only differ in the fluctuating parts of
their respective spectral staircases.

A conventional measure to investigate spectral statistics
is the {\it spectral rigidity} $\De_3 (L;x)$, originally
introduced by Dyson and Mehta \cite{Dyson} to analyze
spectra of complicated atomic nuclei. It is defined as the integrated
quadratic deviation of the spectral staircase $N(x)$ from
the best fitting straight line over an interval $[x-L,x+L]$,
\beq
\label{rigidity}
\De_3 (L;x):= < \min\limits_{(A,B)}\ \frac{1}{2L}\int_{
-L}^{+L}dy\,\left[ N(x+y)-A-By \right]^2 > \ .
\eeq
$<\dots >$ denotes an averaging in $x$ over an interval
of length $\De x$, such that $\De x\ll x$ and $\De x \gg 1$.
The latter condition means that the averaging should take place over
many eigenvalues. (The mean distance of neighbouring eigenvalues
is one due to $\overline{N}(x)=x$.) At the same time
the interval  should be much smaller than $x$ itself. This
procedure presupposes that $x\gg 1$, i.e.\ it is only
possible in the semiclassical regime, and is thus also
referred to as a {\it semiclassical averaging} \cite{BerryA400}.
Notice that (\ref{rigidity}) differs from the definition in
\cite{BerryA400} by the replacement $L\rto 2L$.

Introducing the splitting (\ref{unfoldstair}) in (\ref{rigidity})
one obtains the minimum on the r.h.s.\ for
\beqa
\label{ABmin}
A_{min} &=& x+\frac{1}{2L}\int_{-L}^{+L}dy\ N_{fl}(x+y)
            \nonumber \\
B_{min} &=& 1+\frac{3}{2L^3}\int_{-L}^{+L}dy\ y\,N_{fl}(x+y)\ .
\eeqa
We will soon see that $A_{min}\rto x$ and $B_{min}\rto 1$ in
the limit $L\rto \infty$. Thus one finds that
\beq
\label{deltainf}
\De_{\infty}(x):=\De_3 (\infty ;x)= \lim_{L\rto \infty}
< \frac{1}{2L}\int_{-L}^{+L}dy\,[N_{fl}(x+y)]^2 >\ ,
\eeq
which means that $\De_{\infty}(x)$ is the quadratic mean
of $N_{fl}(x)$. This expression will also sometimes be
denoted as $<N_{fl}(x)^2 >$.

When considering the limit $L\rto\infty$ one first
encounters the problem to define $N_{fl}(x)$ for $x<0$,
a situation occurring in (\ref{rigidity}) for $L>x$. The
only reasonable definition seems to be $N_{fl}(x)=0$ for
$x<0$. Since $N(E)=0$ for $E<0$, an obvious choice for
$\overline{N}(E)$ seems to be $\overline{N}(E)=0$ for
$E<0$. Then also $N_{fl}(E)=0$ for $E<0$. This is consistent with
the fact that $s=\frac{\lb}{2}-i\sqrt{E}\in \rz$ for $E<0$, and thus
$\mbox{arg}\, Z(\frac{\lb}{2}+i\sqrt{E})=0$. We therefore cut off,
in the limit $L\rto\infty$, the integrations on the r.h.s.\
of (\ref{ABmin}) at $-x$.

As our whole discussion deals with two dimensional
problems, Weyl's law (see e.g.\ (\ref{Weyl})) yields
$\overline{N}(E)\sim const.\ E$, $E\rto\infty$. Thus
$\overline{d}(E)\sim \overline{d}=const.\,$, $E\rto\infty$.
We are interested in the semiclassical limit and hence use
$x=\overline{N}(E)\sim \overline{d}\,E$ from now on. Thus
\beqa
\label{rigidityint}
\frac{1}{2L}\int_{-x}^L dy\, N_{fl}(x+y) &\sim& \frac{\overline{d}}
{2L}\int_0^{E+\frac{L}{\overline{d}}}dE'\ N_{fl}(E')\ ,\nonumber \\
\frac{3}{2L^3}\int_{-x}^L dy\,y\,N_{fl}(x+y) &\sim& \frac{3
\overline{d}^2}{2L^3}\int_0^{E+\frac{L}{\overline{d}}}dE'\ (E'-E)\,
N_{fl}(E')\ .
\eeqa
A possible way to treat the integrals on the r.h.s.\ of
(\ref{rigidityint}) is to use the smoothing (\ref{dreg}).
This means that $N_{\ve ,fl}(E)$ for $\ve >0$ from
(\ref{Nflucreg}) has to be inserted into (\ref{rigidityint}),
and the limit $\ve \rto 0$ has to be taken after the
evaluation of the resulting integrals. The vanishing of the r.h.s.\
of the first line of (\ref{rigidityint}) for $L\rto\infty$
has already been demonstrated, see (\ref{integral}). The second line
can be treated similarly, leading to
\beqa
\label{integrals}
\frac{3\overline{d}^2}{2L^3}\int_0^{E+\frac{L}{\overline{d}}}dE'
\ E'\,\sin(\sqrt{E'}kl_{\ga})
   &=& \frac{3\overline{d}^2}{L^3}\left[ \frac{6\sqrt{E+\frac{L}
       {\overline{d}}}}{(kl_{\ga})^3}-\frac{\left( E+\frac{L}
       {\overline{d}}\right)^{3/2}}{kl_{\ga}}\right] \cos\left( kl_{\ga}
       \sqrt{E+\frac{L}{\overline{d}}}\right) \nonumber \\
   & & +\frac{3\overline{d}^2}{L^3}\left[ \frac{3(E+\frac{L}
       {\overline{d}})}{(kl_{\ga})^2}-\frac{6}
       {(kl_{\ga})^4 }\right] \sin\left(kl_{\ga}
       \sqrt{E+\frac{L}{\overline{d}}}\right) \ ,
\eeqa
which also vanishes for $L\rto\infty$. Thus for any $\ve >0$
the smoothings of (\ref{rigidityint}) vanish in the limit
$L\rto\infty$. From this we conclude that also after
performing $\ve \rto 0$ the integrals on the r.h.s.\ of
(\ref{ABmin}) vanish in the limit $L\rto\infty$, leading
to $A_{min}\rto x$ and $B_{min}\rto 1$, as it has been
claimed above.

For finite values of $L$ the spectral rigidity measures
deviations of the spectral staircase from the best fitting
straight line over an interval of length $2L$. Therefore
$\De_3 (L;x)$ indicates correlations in a quantum energy
spectrum on a scale $L$. For $L\rto 0$ the fact that $N(x)$
is a step function causes the rigidity to be $\frac{2}{15}L$.
For slightly larger values of $L$ $\De_3 (L;x)$ begins to
measure spectral correlations. A completely uncorrelated
spectrum excels by a rigidity of $\De_3 (L;x)=\frac{2}{15}L$
throughout the whole range of $L$--values, $0\leq L<\infty$.
Such a spectrum is obtained by a Poisson process through
$x_{i+1}=x_i +s_i$, where the $s_i$'s are independent
outcomes of measurements of the random variable $s$, which is
distributed according to the probability density $P(s)
=e^{-s}$. The $s_i =x_{i+1}-x_i$ are called the {\it
nearest-neighbour level spacings}, or briefly {\it level
spacings}, of the unfolded spectrum $\{ x_i \}$. Due to its
universal behaviour for $L\rto 0$ the rigidity is, however,
not the appropriate tool to investigate short-range correlations
in a spectrum. It becomes useful only for medium- and long-range
correlations corresponding to $L\gg 1$, i.e.\ on scales
taking several levels into account. The level spacings are then
suitable quantities to measure correlations on small scales
$L\approx\frac{1}{2}$.

The systems to be considered from now on shall be completely
desymmetrized in order to avoid degeneracies in their spectra
that superimpose the effects one is looking for. For such an
arbitrary unfolded quantum energy spectrum one can
then study the distribution of the level spacings of the
$x_n \leq x$ and ask the question, whether there will exist a
limiting distribution $P(s)$,
\beq
\label{levelspace}
\lim_{x\rto\infty}\ \frac{ \# \{n;\ x_n \leq x,\ s_n \in
(a,b)\} }{N(x)}=\int_a^b ds\ P(s)\ ,
\eeq
and what it will look like if it exists. The question for the
existence of $P(s)$ is a very subtle one and no mathematically
rigorous answer is known in general. For some specific
examples, where one does have a rigorous theory, see
\cite{Sinai,Bleher}. Extensive numerical calculations in
numerous examples, however, indicate that the existence
of a limiting distribution cannot be doubted. Therefore this
question will not be pursued any further.

It turns out that answering
the question for  the precise form of $P(s)$ allows for a
classification of quantum systems according to the properties
of their respective classical limits. Going further on one
can also study medium- and long-range correlations in the
spectra and try to extend this classification to the spectral
rigidity $\De_3 (L;x)$. Berry developed in \cite{BerryA400}
a semiclassical theory of the spectral rigidity based on
periodic-orbit theory
(see also Berry's contribution in \cite{LesHouches}).
He found a universal behaviour of
$\De_3 (L;x)$ for $1\ll L\ll L_{max}$; $L_{max}$ is
proportional to $\frac{1}{T_{min}}$, where $T_{min}$
is the period of the shortest periodic orbit
of the classical system. (For billiard systems $T_{min}=\frac{l_{min}}
{2p}$, so that $L_{max}\propto\sqrt{x}$.)
In the range $1\ll L\ll L_{max}$ classically integrable
systems share a rigidity of $\De_3 (L;x)=\frac{2}{15}L$,
whereas for classically chaotic systems possessing a
time-reversal invariance Berry found that $\De_3 (L;x)=\frac{1}{\pi^2}
\log L +const.$.
For very large values of $L$, $L\gg L_{max}$,
he obtained for both cases a saturation of the rigidity at a
value that is determined by the contributions of short
periodic orbits to $d_{fl}(x)$. The energy dependence of
the saturation value $\De_{\infty}(x)$ now is again
characteristic for the classical limit of the quantum system.
If this is integrable, Berry can show that $\De_{\infty}(x)\sim
const.\,\sqrt{x}$, $x\rto\infty$. Chaotic time-reversal
invariant systems, however, are demonstrated to yield $\De_{\infty}
(x)\sim \frac{1}{2\pi^2}\log x$, $x\rto\infty$. According to
(\ref{deltainf}) $\De_{\infty}(x)$ is just the mean square
$<N_{fl}(x)^2 >$ of the fluctuation part of the spectral
staircase. Hence Berry found, in the semiclassical limit,
a universal behaviour for this quantity that only depends
on whether the classical limit of the considered system is integrable
or chaotic. It should be mentioned that the characterization
of systems as being {\it time-reversal invariant} means that
generically the lengths of periodic orbits are twofold
degenerate, i.e.\  the multiplicities $g(l_{\ga})$ of lengths
should asymptotically, for $l_{\ga}\rto\infty$, approach two.
There are, however, classically completely chaotic time-reversal
invariant systems with exponentially increasing multiplicities
of lengths to which Berry's results do not apply. These systems
are subsumed under the notion of {\it arithmetical chaos}
and are the main objects of the present investigation.

A question that might immediately arise is whether or not
this universality of the spectral properties of quantum
systems carries over to short-range correlations, i.e.\
to the level spacings distributions $P(s)$. Berry and Tabor
demonstrated \cite{BerryTabor} that generic classically integrable
systems show a Poissonian level spacing, $P(s)=e^{-s}$. Thus their
quantum energy spectra are close to totally uncorrelated
ones, since the
short-, medium-, and long-range correlations are Poissonian.
The only difference occurs with the saturation of $\De_3 (L;x)$
for very large $L$, $L\gg L_{max}$. Concerning chaotic
systems, however, there do not exist theoretical results
for the level spacings. But it is generally believed that
for these systems $P(s)$
may be well described by the distribution of spacings of
eigenvalues in an ensemble of large random matrices.
In {\it random matrix theory} (RMT) the statistical
properties of the eigenvalues of random matrices from
several ensembles have been thoroughly studied, see e.g.\
\cite{Mehta,Brody,Haake} and Bohigas' contribution
in \cite{LesHouches}.
For time-reversal invariant chaotic systems the appropriate ensemble
is the Gaussian orthogonal ensemble (GOE) of real
symmetric matrices, whose level
spacings distribution may be well approximated by Wigner's
surmise $P(s)=\frac{\pi}{2}s\,e^{-\pi s^2 /4}$. Dropping the
requirement of time-reversal invariance one has to consider the
Gaussian unitary ensemble (GUE) of complex hermitian matrices.
Historically, the conjecture that quantum spectra may be described
by RMT goes back to Wigner and Landau and Smorodinsky,
who applied this to the resonance levels of complicated
atomic nuclei, see \cite{Porter} for a collection of the
original contributions. Bohigas, Giannoni and Schmit extended
the conjecture of an RMT-behaviour of quantum energy spectra to
systems with only a few degrees of freedom possessing chaotic
classical limits \cite{BoGiSch}, supporting this by detailed
numerical studies for several systems.
In addition, the spectral rigidity
$\De_3 (L;x)$ for $1\ll L\ll L_{max}$ found by Berry agrees
with the RMT--prediction. Again, as in the integrable
case, the difference lies in the saturation of the rigidity
of the actual spectra for $L\rto\infty$, as described by
Berry's periodic-orbit theory.

In summary, the quantum spectral properties
of classically integrable and classically chaotic systems
differ in that for small level spacings, $s\rto 0$, the former
ones show $P(s)\sim 1-s$, whereas the latter ones behave as
$P(s)\sim \frac{\pi}{2}s$. For classically integrable systems
hence the phenomenon of {\it level attraction} occurs, whereas
classically chaotic systems excel by a {\it level repulsion}.
On larger scales $L$ rather strong correlations, measured
by the spectral rigidity $\De_3 (L;x)$, are found in the chaotic
case leading to an only logarithmically increasing rigidity.
In contrast, integrable systems possess nearly uncorrelated
spectra; only on very large scales correlations do occur.

Up to now several measures to study the statistical properties
of quantum energy spectra have been introduced, most of
them involving the fluctuating part $N_{fl}(E)$ of the
spectral staircase. In sections 3.6 and 4.5 then the strengths
of spectral fluctuations as measured by $N_{fl}(E)$ or $\De_\infty
(E)$ play an important role and decide on the applicability of
certain methods of inverse quantum chaology.
It hence seems to be worthwhile to
study $N_{fl}(E)$ more thoroughly to get hands on
quantum energy spectra of classically chaotic systems.
Thus it seems to be justified to devote some fraction
of the following investigations of quantum energy spectra to a
study of their fluctuations, and to the question
how these are being influenced by the properties of the
classical limits of the considered systems.

\xsection{Classical Aspects of Arithmetical Chaos}
In this chapter the classical dynamics of the systems
subsumed as showing {\it arithmetical chaos} shall be
introduced. In order to make the whole presentation
self-contained, the first section of this chapter
is devoted to a review of important definitions and
results in two dimensional hyperbolic geometry. The second
section then contains a discussion of geodesic length
spectra of hyperbolic surfaces and introduces some useful
relations among quantities referring to length spectra. Then,
in the following two sections, arithmetic Fuchsian groups
are defined and their geodesic length spectra are studied,
leading to the final result on the exponential growth of the
mean multiplicities of lengths. The results of sections
3.2 to 3.4 have previously been published in \cite{Bolte}.
The next section then is devoted
to a discussion of pseudosymmetries on arithmetic surfaces.
Finally, in section 3.6, statistical properties of length
spectra are studied.

\subsection{A Brief Review of Hyperbolic Geometry}
The classical dynamical systems to be considered from now
on are geodesic flows on hyperbolic surfaces of finite
area. In physics terms they are given by single
point-particles of mass $m$ moving on surfaces of
constant negative Gaussian curvature without being
subject to any external force. The absence of a potential
causes the Jacobi metric to be proportional to the
hyperbolic metric the surface has been endowed with.
Thus the classical trajectories of a particle on such a surface
are the geodesics of the hyperbolic metric. In the
following the geometric setting shall be reviewed in
a rather sketchy manner in order to recall the necessary
notions and to introduce the notations used further on.
A rather extensive review of the physical and mathematical
aspects of the problem may be found in \cite{BalazsVoros}.

A convenient model for hyperbolic geometry in two dimensions
is the upper complex half-plane $\cH =\{z=x+iy;\ y>0\}$
endowed with the Poincar\'e metric $ds^2 =y^{-2}(dx^2 +dy^2 )$.
In this setting the Gaussian curvature is $K=-1$
everywhere on $\cH$. This is dimensionless since the internal
length scale $R$ has been normalized to $R=1$. Otherwise
the metric would have to be replaced by $ds^2 \rto R^2 \,ds^2 $.
The hyperbolic distance $d(z,w)$ of two points $z,w\in\cH$
is the infimum of the set of lengths of curves connecting
$z$ and $w$, measured with $ds^2$. As $\cH$ endowed with
$ds^2$ is geodesically complete the infimum is attained
for the geodesic segment connecting $z$ and $w$. Its length
is given by
\beq
\label{distance}
\cosh d(z,w)=1+\frac{|z-w|^2}{2\,Im\,z\ Im\,w}\ .
\eeq
There exists an operation of the group $PSL(2,\rz)=SL(2,\rz)/
\{\pm \unmat \}$ on $\cH$ by fractional linear transformations.
If $\ga = \left( a\ b \atop c\ d \right) \in SL(2,\rz)$,
and $z\in\cH$, then
\beq
\label{linfrac}
\ga z=\frac{az+b}{cz+d}\ .
\eeq
This operation is transitive and compatible with the group
structure of $SL(2,\rz)$. The stabilizing subgroup of $z_0 =i
\in\cH$ is the maximal compact subgroup $K=SO(2,\rz)\subset
SL(2,\rz)$. Thus the hyperbolic plane may also be
realized as the symmetric space $\cH\cong SL(2,\rz)/SO(2,\rz)$.
Since both matrices $\ga$ and $-\ga$ from $SL(2,\rz)$ map
a $z\in\cH$ to the same image it is the projective group
$PSL(2,\rz)$, where the centre $\{\pm\unmat\}$ of $SL(2,\rz)$
has been factored out, that effectively operates on $\cH$.
{}From now on the distinction between matrices $\ga\in
SL(2,\rz)$ and classes $[\ga ]\in PSL(2,\rz)$ will be
dropped and an identification of $\ga$ and $-\ga$ will
be understood automatically. Matrices $\ga\in SL(2,\rz)$
will be chosen such that $tr\,\ga \geq 0$. This somewhat
sloppy notation seems to be convenient and should not
cause confusion.

$PSL(2,\rz)$ operating on $\cH$ turns out to be the group
of orientation preserving isometries of the Riemannian
space $(\cH ,ds^2 )$, i.e.\ $\ga :\ \cH\rto\cH$ preserves the
lengths of curves on $\cH$. Three classes of elements
of $PSL(2,\rz)$ have to be distinguished according to
their traces $(\ga\not= \unmat)$:
\begin{enumerate}
\item {\it Elliptic elements} $\ga$ have $0\leq tr\,\ga <2$.
Such a $\ga$ has one fixed point in the interior of $\cH$. It
is conjugate within $SL(2,\rz)$ to an element $\left( \ \cos\vt
\ \sin\vt \atop -\sin\vt\ \cos\vt \right)$, $\vt\in [0,2\pi)$.
\item {\it Parabolic elements} $\ga$ have $tr\,\ga =2$. Such a
$\ga$ has one fixed point on the boundary $\partial\cH$
of $\cH$ ($\partial\cH =\rz \cup \{i\infty\}$). It is
conjugate within $SL(2,\rz)$ to an element $\left( 1\ x \atop
0\ 1 \right)$, $x\in\rz$.
\item {\it Hyperbolic elements} $\ga$ have $tr\,\ga >2$. Such a
$\ga$ has two fixed points on $\partial\cH$. It is conjugate
within $SL(2,\rz)$ to an element $\left( {N^{\frac{1}{2}}\atop 0}
{0\atop N^{-\frac{1}{2}}}\right)$. $N>1$ is called the
{\it norm} of $\ga$.
\end{enumerate}
The geodesics of the Poincar\'e metric on $\cH$ are the
half-circles and the straight half-lines perpendicular to
the real axis. The two end points of a geodesic are the
fixed points of a (unique) hyperbolic $\ga\in PSL(2,\rz)$.
This $\ga$ maps the geodesic onto itself, which is therefore
also called the {\it invariant geodesic} of $\ga$.

A discrete subgroup $\Ga\subset PSL(2,\rz)$ gives rise to
the orbit space ${\GaH}=\{ {\Ga}z ; \ z \in \cH \}$. This means
that an equivalence relation on $\cH$ is introduced
by an identification of points that are related by a
$\Ga$-transformation. As the Poincar\'e metric is
$\Ga$-invariant it induces a metric on the orbit space $\GaH$,
which will also be called Poincar\'e (or hyperbolic) metric.
If $\Ga$ contains no elliptic elements, i.e.\ no $\ga\in\Ga$
has a fixed point on $\cH$, $\GaH$ is a regular surface.
If, however, elliptic elements are present in $\Ga$, $\GaH$
will be a regular surface outside the respective fixed
points. Including these turns $\GaH$ into an {\it orbifold}.
Despite this slight complication any $\GaH$, irrespective
of a possible existence of orbifold-points, will be called
a {\it hyperbolic surface}. If $\GaH$ has finite area
(measured with the Poincar\'e metric), $\Ga$ is called a
{\it Fuchsian group of the first kind}, otherwise a
{\it Fuchsian group of the second kind}. The latter ones
will be excluded from the further investigations. If a
Fuchsian group of the first kind contains no parabolic
elements, $\GaH$ is compact. Therefore such a $\Ga$ is
called {\it cocompact}.

The abstract orbit space $\GaH$ may be realized explicitly
as a {\it fundamental domain} $\cF$ of $\Ga$ on $\cH$.
This is a simply connected region $\cF\subset\cH$ such that
the interior of $\cF$ contains no two $\Ga$-equivalent points
and the union of all $\Ga$-translates of $\cF$ covers all of $\cH$.
The boundary $\partial\cF$ of $\cF$ consists of segments
that have to be identified under the operation of $\Ga$
appropriately in order to yield a closed surface. If $\Ga$
is {\it strictly hyperbolic}, i.e.\ it contains besides
the identity only hyperbolic elements, $\GaH$ is a compact
surface of genus $g\geq 2$.  Thus it is topologically
equivalent to a sphere with $g$ handles. It is then always
possible to find a fundamental domain $\cF$ that is bordered
by $4g$ geodesic segments (a $4g$-gon). Since the Gaussian
curvature is $K=-1$ the Gau\3-Bonnet theorem yields for the
area of $\GaH$ $\mbox{area}({\GaH})=\mbox{area}({\cF}) = 4 \pi (g-1)$.
If $\Ga$ is not cocompact the $\Ga$-conjugacy classes of
parabolic elements are called {\it cusps}. A fundamental domain
then extends to infinity (i.e.\ to $\partial\cH$). For
each cusp $\partial\cF$ contains a point of $\partial\cH =
\rz \cup \{i\infty \}$.

A convenient presentation of a Fuchsian group $\Ga$ may
be given in terms of generators related to a given
fundamental domain $\cF$, and relations among them. As generators
one can take those transformations that identify pairs of
geodesic segments of $\partial\cF$, together with their inverses.
If $\Ga$ is strictly hyperbolic, $\partial\cF$ has $4g$, $g\geq 2$,
components. Thus $\Ga$ has $4g$ generators which are
conventionally denoted as $a_1 ,b_1 ,\dots,a_g ,b_g ,a_1^{-1},
\dots,b_g^{-1}$. One frequently chooses the order of
the generators such that the one relation they obey is
\beq
\label{relation}
a_1 b_1 a_1^{-1}b_1^{-1}\dots a_g b_g a_g^{-1}b_g^{-1}=\unmat \ .
\eeq
This presentation seems to be well suited for explicit
calculations since the group may be constructed explicitly
in terms of $SL(2,\rz)$-matrices once a fundamental domain
$\cF$ is given. (See e.g.\ \cite{Aurich90} and appendix B for
examples.) The procedure will then be as follows: draw an arbitrary
geodesic $4g$-gon ($g\geq 2$) with $\mbox{area}(\cF)=4\pi (g-1)$. It
only has to share the further property that the geodesic
segments have to come in pairs of identical lengths. Construct
the $2g$ matrices (from $SL(2,\rz)$) that identify the
pairs of edges. Together with the $2g$ inverse matrices
$4g$ generators for the desired Fuchsian group $\Ga$ have
been found. $\Ga$ itself can then be obtained by forming all
possible products of the generator matrices. (These are
also called {\it words} in the generators.) In general
the $4g$ generators will be ``arbitrary'' matrices, which
means that it does not seem possible to obtain an explicit
law that determines their matrix entries. Forming words in
the generators will also yield matrices with seemingly
arbitrary entries. In general there will thus not exist
some (algebraic) structure for the matrix entries of the
group elements. A consequence of this fact for practical
purposes hence is that there seems to be no explicit
enumeration scheme for the group elements other than
the generator method just described.

This observation is indeed true for ``generic'' Fuchsian
groups. An exception to this rule, however, does exit. This
is formed by the class of {\it arithmetic Fuchsian groups}.
For these the arithmetic structure appearing in their
definition determines their matrix entries. The basic and
most prominent example for an arithmetic Fuchsian group is
the {\it modular group} $SL(2,\gz)$. It consists of all
$2\times 2$ matrices with rational integers as entries
and being of unit determinant. Leaving aside the determinantal
condition the determination of the matrix entries could not
be easier. For general arithmetic groups the definition
is a bit more involved and will therefore be postponed
to section 3.3. But still, the characterization of the set
of traces of group matrices is rather simple compared to
the generic case.

For applications in the context of quantum chaos the
knowledge of the set of traces of elements of a Fuchsian
group is of major importance, since it is tightly connected
to the knowledge of the geodesic length spectrum. The latter in turn
is the input on the classical side of periodic-orbit sum
rules. Since the set out of which the traces of an arithmetic
Fuchsian group have to be taken is known (see section 3.4),
it is possible to develop an algorithm that allows to calculate
the geodesic length spectrum (including multiplicities) for
an arithmetic group up to a certain length completely, see
\cite{AurichBogo,Schleicher,Matthies,Ninnemann}.
For non-arithmetic groups there seems to be no better
way to calculate the length spectrum numerically other than
forming words in the group generators up to a given number
of generators per word. This method, however, does not yield
the length spectrum completely up to some value of the length.

Because of their importance for periodic-orbit theory the following
section will be devoted to a discussion of geodesic length spectra
connected with Fuchsian groups.

\subsection{Geodesic Length Spectra of Hyperbolic Surfaces}
By the very construction of hyperbolic surfaces $\GaH$,
where $\Ga$ is a Fuchsian group of the first kind, the
hyperbolic plane $\cH$ is the universal covering space of
$\GaH$. A closed geodesic $c$ on $\GaH$ lifts to a geodesic
$\hat c$ on $\cH$, which is invariant under some hyperbolic
$\ga_c \in \Ga$. For any $\ga\in\Ga$ the transformation
$\ga\ga_c \ga^{-1}$ has some invariant geodesic $\hat c_{\ga}$
on $\cH$ that projects down to the same closed geodesic
$c$ on $\GaH$. There exists in fact a one-to-one correspondence
between $\Ga$-conjugacy classes of hyperbolic elements
of $\Ga$ and closed geodesics on $\GaH$. The hyperbolic length
$l$ of $c$ can be related to $\ga_c$ as follows: the lift
$\check{c}$ of $c$ (being a segment of $\hat c$) on $\cH$
connects two points $z$ and $z'$ on $\cH$ which are $\Ga$-equivalent.
Thus there exists a $\ga\in\Ga$ such that $z'=\ga z$. Then
$l=d(z,z')=d(z,\ga z)$. Since $\check{c}$ is a geodesic segment
$d(z,\ga z)$ minimizes the lengths of all curves on $\cH$
connecting $z$ and $\ga z$. Varying $z$ continuously along
$\hat c$ cannot change the choice of $\ga$. Since all these
$z$'s and $\ga z$'s lie on $\hat c$, this has to be the invariant
geodesic of $\ga$, and hence $\ga =\ga_c$. To compute
$l=d(z,\ga_c z)$ one can conjugate $\ga_c$ in $PSL(2,\rz)$
to obtain the diagonal matrix $\ga' =\left( {N^{\frac{1}{2}}\atop 0}
{0\atop N^{-\frac{1}{2}}} \right)=g\ga_c g^{-1}$, for some
$g\in SL(2,\rz)$ and $N>1$. Then $l=d(z,\ga_c z)=d(z,g^{-1}\ga'
gz)=d(gz,\ga' gz)$. The invariant geodesic of $\ga'$ is
obviously the imaginary axis, and choosing $gz$ to be on it
one obtains from (\ref{distance}) that $\cosh l=1+\frac{1}{2N}
(N-1)^2 $. This finally leads to
\beq
\label{lengths}
2\,\cosh \frac{l}{2}=tr\,\ga' =tr\,\ga \ .
\eeq
Because of the one-to-one correspondence between conjugacy
classes of hyperbolic elements $\ga\in\Ga$ and closed
geodesics $c$ on $\GaH$ we denote the length $l$ of $c$
by $l=l(\ga)$, where $\ga$ is some representative of the
conjugacy class $\{ \ga \}_{\Ga}=\{\hat \ga \ga \hat \ga^{-1};
\ \hat \ga \in \Ga \}$ related to $c$.

The set of lengths of closed geodesics on $\GaH$ is called the
{\it geodesic length spectrum} $\cL (\Ga)$ of $\GaH$, or
briefly, of $\Ga$,
\beq
\label{lengthspec}
\cL (\Ga)=\{ l(\ga);\ \ga\in\Ga \ \mbox{hyperbolic}\}\ .
\eeq
Introducing the notation $\cL (\Ga)=\{ l_1 <l_2 <l_3 \dots \}$
the counting function $\hat \cN (l)$ for the geodesic length
spectrum is
\beq
\label{nhat}
\hat \cN (l):= \#\,\{n;\ l_n \leq l \}\ .
\eeq
In general a length spectrum of a hyperbolic surface will be
degenerate, i.e.\ there exist several closed geodesics on
the surface of the same length. If e.g.\ $\Ga$ is strictly
hyperbolic, no $\ga\in\Ga$ different from the identity is
conjugate to its inverse, so that every length at least
occurs twice. Interpreted in physics terms this means that
the dynamical system possesses a time-reversal symmetry,
since the geodesic corresponding to the inverse of a
hyperbolic transformation is the original classical
trajectory traversed backwards in time. In general the
multiplicity of any $l\in\cL (\Ga)$ will be denoted by
$g(l)\in \nz$. The counting function including
multiplicities is the {\it classical staircase function}
\beq
\label{Nvonl}
\cN (l):= \#\,\{\,\{\ga \}_{\Ga};\ \ga\in\Ga\ \mbox{hyperbolic and }
l(\ga)\leq l\}\ .
\eeq
Asymptotically, for $l\rto\infty$, the magnitude of $\cN (l)$
is universally determined by {\it Huber's law} \cite{Huber,Hejhal},
also known as the {\it prime geodesic theorem} (PGT),
\beq
\label{Huber}
\cN (l)\sim Ei(l)\sim \frac{e^l}{l}\ ,\ \ \ l\rto\infty\ .
\eeq
The remainder to this asymptotic relation will in more detail
be dealt with in section 3.6.
$Ei(l)$ is related to the logarithmic integral $li(x)=
P\int_0^x \frac{dt}{\log t}$ by $Ei(\log x)=li(x)$, see
e.g.\ \cite{Bateman}. ($P\int dt\dots$ thereby denotes the principal
value of the integral.) Comparing (\ref{Huber}) with (\ref{PGT})
yields that the topological entropy for geodesic flows
on hyperbolic surfaces is $\tau =1$.

A closed geodesic is called {\it primitive}, if it is not
a multiple traversal $(\geq 2)$ of some other closed geodesic
on the same surface. The corresponding conjugacy class
$\{ \ga \}_{\Ga}$ is then also primitive, i.e.\ a
$\ga' \in \{ \ga \}_{\Ga}$ is not a power $(\geq 2)$ of
some other (hyperbolic) element of $\Ga$. The primitive
closed geodesics give rise to the {\it primitive geodesic
length spectrum} $\cL_p (\Ga ) =\{ l_{p,1}<l_{p,2}<l_{p,3}<\dots
\}$. All quantities defined for the full length spectrum
can be introduced analogously for the primitive one as well.
To distinguish the respective quantities one introduces
an index $p$ in all notations referring to the primitive
length spectrum. The PGT (\ref{Huber}) remains true also
for $\cN_p (l)$. Since
\beq
\label{Nhatprim}
\hat\cN (l)=\sum_{l_n \leq l}1 = \sum_{r=1}^{[l/l_1 ]}\sum_{r
l_{p,n}\leq l}1=\sum_{r=1}^{[l/l_1 ]}\hat\cN_p (l/r) \ ,
\eeq
and $\hat\cN_p (l)$ is positive and monotonically increasing,
the counting functions for $\cL (\Ga)$ and $\cL_p (\Ga)$,
respectively, show the same asymptotical behaviour,
\beq
\label{Nhatasymptotic}
\hat\cN (l)\sim \hat\cN_p (l)\ ,\ \ \ l\rto\infty\ .
\eeq

In the following sections the multiplicities $g_p (l)$
of lengths of primitive closed geodesics will play a major
role. As it seems to be impossible to determine the
multiplicities explicitly, the only quantity one can lay
hands upon appears to be an average $<g_p (l)>$, and its
asymptotic behaviour for $l\rto\infty$. In number theory
\cite{Kraetzel} two functions $f(n)$ and $h(n)$, $n\in\nz$, are
said to be of the same {\it average order}, if
\beq
\label{aseq}
\sum_{n=1}^N f(n) \sim \sum_{n=1}^N h(n)\ ,\ \ \ N\rto\infty \ .
\eeq
Led by this definition a mean multiplicity $<g_p (l)>$ is
introduced as a continuous function of $l$ that is of the average
order of $g_p (l)$,
\beq
\label{asmult}
\cN_p (l)=\sum_{l_{p,n}\leq l}g_p (l_{p,n}) \stackrel{!}{\sim}
\sum_{l_{p,n}\leq l}<g_p (l_{p,n})> = \int_0^l d\hat\cN_p (l')\,
<g_p (l')>\ ,
\eeq
for $l\rto\infty$. Substituting the multiplicities by their mean
hence does not violate the PGT.
Since $\cN_p (l) \sim Ei(l)$, $l\rto\infty$,
one observes the asymptotic relation
\beq
\label{multdef}
<g_p (l)>\sim \frac{e^l}{l}\cdot\left[ \frac{d\hat\cN_p}{dl}
\right]^{-1}\ ,\ \ \ l\rto\infty \ ,
\eeq
by differentiating (\ref{asmult}). This result may be
interpreted in more visual terms as follows. $\cN_p (l)$
is a step function with step-width $\De l_n :=l_{p,n} -l_{p,n-1}$
and step-height $g_p (l_{p,n})$ at $l=l_{p,n}$. A mean of
the staircase $\cN_p (l)$ should then have a slope of
$<g_p (l)>/<\De l>$. As $<\De l>^{-1}$ is the mean density of
primitive lengths it is asymptotically given by $\frac{d
\hat\cN_p}{dl}$. Thus $\frac{d\cN_p}{dl}\sim <g_p (l)>\frac{
d\hat\cN_p}{dl}$, which is equivalent to (\ref{multdef}).
{}From this relation one concludes that one has to gain
information on the asymptotics of the counting function
$\hat\cN_p (l)$ of different lengths of primitive closed
geodesics in order to be able to determine the asymptotic
behaviour of the mean multiplicity $<g_p (l)>$. Notice that
the same procedure may also be carried through for the full
length spectrum instead of the primitive one. Since both
staircase functions $\cN (l)$ and $\hat\cN (l)$ share the same
asymptotic behaviour as $\cN_p (l)$ and $\hat\cN_p (l)$,
respectively, the mean multiplicities agree asymptotically.

Next we are interested in relating the counting functions
$\hat\cN_p^{(1)}(l)$ and $\hat\cN_p^{(2)}(l)$ for $l\rto\infty$.
Thereby $\hat\cN_p^{(i)}(l)$ shall correspond to the
length spectra $\cL_p (\Ga_i )$, where $\Ga_1$ is a subgroup
of index $d>1$ in the Fuchsian group of the first kind
$\Ga_2$. This means that $\Ga_2$ decomposes into $d$ cosets
of $\Ga_1$ according to
\beq
\label{cosets}
\Ga_2 =\Ga_1 \dcup \Ga_1 \ga_1 \dcup \dots \dcup \Ga_1 \ga_{d-1}\ ,
\eeq
where $\ga_1 ,\dots,\ga_{d-1}$ are elements of $\Ga_2$, but not
of $\Ga_1$. Let $\cL_p (\Ga_2 )$ be given as
\beq
\label{largespectrum}
\cL_p (\Ga_2 )=\{ l_{p,1}<l_{p,2}<l_{p,3}<\dots \}\ ;
\eeq
$\cL_p (\Ga_1 )$ shall now be determined in terms of the
$l_{p,n}$'s. To this end one observes that if $\ga\in\Ga_2$,
then there exists a $k\in\nz$, such that $\ga^k \in\Ga_1$.
To see this take an arbitrary $\ga\in\Ga_2$ and form
$\cup_{m\in\gz}\Ga_1 \ga^m \subset\Ga_2 $. The union on the
l.h.s.\ cannot be disjoint, since according to (\ref{cosets})
$\Ga_1$ is of finite index in $\Ga_2$. Therefore there exists
a $\ga_0 \in\Ga_1 \ga^r \cap\Ga_1 \ga^s $ for some pair
$r\not= s$. Thus $\ga^{r-s}$ and $\ga^{s-r}$ lie in $\Ga_1$.
Choosing $k=|r-s|$ proves the assertion.

Let now $\ga_p$ be a primitive hyperbolic element of $\Ga_2$,
hence $l(\ga_p )\in\cL_p (\Ga_2 )$. Then either i) $\ga_p
\in\Ga_1$: $\ga_p$ is also primitive hyperbolic in $\Ga_1$
and thus $l(\ga_p )\in\cL_p (\Ga_1 )$, or ii) $\ga_p \not\in
\Ga_1$. Then, by the above remark, there exists a $k\in\nz$
with $\ga_p^k \in\Ga_1$. The minimal such $k$ takes care
for $l(\ga_p^k )=kl(\ga_p )$ to be an element of $\cL_p (\Ga_1)$.
Altogether, the primitive length spectrum of $\Ga_1$ can be
characterized as
\beq
\label{smallspectrum}
\cL_p (\Ga_1 )=\{ k_1 \,l_{p,1},\ k_2 \,l_{p,2},\ k_3 \,l_{p,3},
\dots \}
\eeq
with positive integers $k_j$. Since the $k_j$'s may take
arbitrary values in $\nz$, this enumeration of elements of
$\cL_p (\Ga_1 )$ will in general not be an ordered one.

The determination of the asymptotic behaviour of $\hat\cN_p^{(1)}
(l)$ in terms of that of $\hat\cN_p^{(2)}(l)$ requires to
know how often (asymptotically) a certain value of the $k_j$'s
in (\ref{smallspectrum}) occurs. This question may be
answered with the help of the decomposition (\ref{cosets})
of $\Ga_2$ into cosets of $\Ga_1$. On average, a fraction
of $\frac{1}{d}$ of the elements of $\Ga_2$ are also elements of
$\Ga_1$. Thus going through the conjugacy classes of hyperbolic
elements of $\Ga_2$ and picking one representative from
each class yields a fraction of $\frac{1}{d}$ of the latter
ones to lie in $\Ga_1$. This statement is meant in the
following sense: one chooses an ordering of the conjugacy
classes according to the corresponding lengths $l$ of
closed geodesics (equivalently, in ascending order of their
traces). Then, for $l\rto\infty$, the respective fraction
of elements of $\Ga_1$ approaches $\frac{1}{d}$. Hence a
fraction of $\frac{1}{d}$ of the $k_j$'s equals one.
Proceeding further in the same manner yields a fraction
of $(1-\frac{1}{d})\frac{1}{d}$ to be two, and so on.

By (\ref{largespectrum}) and (\ref{smallspectrum})
obviously $\hat\cN_p^{(1)}(l)\leq\hat\cN_p^{(2)}(l)$.
The above remark on the fraction of $k_j =1$ yields
$\hat\cN_p^{(1)}(l)\geq\frac{1}{d}\hat\cN_p^{(2)}(l)$ in
the limit $l\rto\infty$. This relation has to be an
inequality rather than an equality because there might occur
some $k_j \geq 2$ with $k_j l_{p,j}\leq l$. The number of
$k_j =k\geq 2$ such that $kl_{p,j}\leq l$ is, however,
bounded by $\hat\cN_p^{(2)}(\frac{l}{k})$. But this is
asymptotically dominated by the contribution coming from $k=1$.
One therefore observes the asymptotic relation
\beq
\label{Nhatsubgroup}
\hat\cN_p^{(1)}(l)\sim\frac{1}{d}\,\hat\cN_p^{(2)}(l)\ ,
\ \ \ l\rto\infty\ .
\eeq
Again, a corresponding relation also holds for the
counting functions $\hat\cN^{(1)}(l)$ and $\hat\cN^{(2)}(l)$
of the full length spectra.

Finally, the asymptotic relation of the counting functions
for length spectra of two commensurable Fuchsian groups
shall be dealt with, since this will be needed later when
treating arithmetic Fuchsian groups. One recalls
that two subgroups $H_1$ and $H_2$ of a group $G$ are
called {\it commensurable}, if the intersection $H_1 \cap
H_2$ is a subgroup of finite index in both $H_1$ and $H_2$.
Let therefore $\Ga_a$ and $\Ga_b$ be two commensurable Fuchsian
groups of the first kind. $\Ga_0 :=\Ga_a \cap \Ga_b$ shall be
of index $d_a <\infty$ in $\Ga_a$ and of index $d_b <\infty$
in $\Ga_b$. By (\ref{Nhatsubgroup}) the counting functions
are related through
\beqa
\label{transitive}
\hat\cN_p^{(0)}(l) &\sim& \frac{1}{d_a}\,\hat\cN_p^{(a)}(l)\ ,
                          \nonumber \\
\hat\cN_p^{(0)}(l) &\sim& \frac{1}{d_b}\,\hat\cN_p^{(b)}(l)\ ,
\eeqa
for $l\rto\infty$. From this the asymptotic relation
\beq
\label{Nhatcommensurable}
\hat\cN_p^{(a)}(l) \sim \frac{d_a}{d_b}\,\hat\cN_p^{(b)}(l)\ ,
\ \ \ l\rto\infty\ ,
\eeq
easily follows. Recalling the asymptotic behaviour
(\ref{multdef}) of the mean multiplicities  of primitive
lengths yields
\beq
\label{multcommensurable}
<g_p^{(a)}(l)> \sim \frac{d_b}{d_a}\, <g_p^{(b)}(l)>\ ,
\ \ \ l\rto\infty\ .
\eeq
Therefore, given two commensurable Fuchsian groups, the respective
mean multiplicities of primitive lengths of closed geodesics
are asymptotically, for $l\rto\infty$, proportional to one
another. The factor of proportionality is given by the
ratio of the indices with which the two groups contain
their intersection as a subgroup.

So far, the discussion of length spectra has been completely
general in that it has been valid for all Fuchsian groups
of the first kind. A major objective of the present
investigation, however, is to stress the importance of
distinguishing arithmetic Fuchsian groups from
non-arithmetic ones. Concerning length spectra, the difference
occurs with the behaviour of the mean multiplicities
$<g_p (l)>$ as $l\rto\infty$. For arithmetic groups $\Ga$ it shall
be shown in section 3.4 that $<g_p (l)>\sim c_{\Ga}\,\frac{e^{l/2}}
{l}$, $l\rto\infty$, where $c_{\Ga}$ is some constant depending
on the specific group $\Ga$. This behaviour is exceptional
and constitutes a major property by which the arithmetical
systems excel among general chaotic dynamical systems. It
is indeed known that for general hyperbolic surfaces $g_p (l)$
is always unbounded \cite{Horowitz,Randol,Buser}. The proof of
this proceeds algebraically; Buser \cite{Buser}, however,
geometrically constructs explicit examples of degenerate closed
geodesics, where the multiplicities of the respective lengths
cannot be accounted for by symmetries of the surface. He can estimate
the multiplicity of a length $l$ to which the construction is
applied to be of the order $l^a$ for $a=\frac{\log 2}{\log 5}$ and
$l\rto\infty$. Aurich \cite{Aurichpc} computed the lower parts of
length spectra for several arbitrarily chosen compact surfaces of
genus two  and never observed a multiplicity that could not be
explained by symmetries in the computed $l$--range. Thus it appears
that high multiplicities (having values beyond the expectation based
on symmetries) show up rather scarcely. All this information
indicates that for non-arithmetic Fuchsian groups the multiplicities
of lengths of closed geodesics do not grow exponentially. The latter
seems to be characteristic of surfaces with arithmetic Fuchsian groups.

Without giving the precise definition of arithmetic
Fuchsian groups at this stage we would now like to try to
illustrate the difference between arithmetic and
non-arithmetic groups in order to give an intuitive understanding
of the different properties of their respective length spectra.
As described in section 3.1 a presentation of a Fuchsian group
$\Ga$ may be yielded in terms of generator matrices
$g_1 ,\dots ,g_n$, where the $g_i$'s define fractional
linear transformations on $\cH$ that identify pairs of
edges of the fundamental domain $\cF$ of $\Ga$. If $\cF$
(and thus also $\Ga$) is arbitrary, then $g_1 ,\dots ,g_n$
will be matrices with ``arbitrary'' real entries. Consequently
the set of matrix entries of all elements of $\Ga$
has no obvious algebraic structure other than that the entries
are composed of those of the generator matrices.
Since concerning length spectra the traces of group
elements are the objects of interest, a possible
algebraic structure of the set of traces $tr\,\Ga =\{ tr\,\ga ;
\ \ga \in \Ga \}$ of elements of $\Ga$ would influence $\cL (\Ga)$.
As an example of an arithmetic group the modular group
$\Ga_{mod}=SL(2,\gz)$ should serve. From the theorem proved in
\cite{Latimer} one can conclude that $tr\,\Ga_{mod}=\gz$
(see also \cite{Schleicher}). Therefore $tr\,\Ga_{mod}$
has a nice algebraic
structure in that it is just the ring of rational integers.
According to (\ref{lengths}) one can hence enumerate the lengths of
closed geodesics on $\Ga_{mod}\backslash\cH$ by $n=2\cosh(l_n /2)$,
$n=3,4,5,\dots$. Since $\hat\cN (l_n )=n-2$, one finds
the asymptotic law $\hat\cN (l_n )\sim e^{l_n /2}$, $n\rto\infty$,
for the modular group. By (\ref{multdef}) the multiplicities
of lengths therefore grow asymptotically as
\beq
\label{multmodular}
<g(l)>\sim 2\,\frac{e^{l/2}}{l}\ ,\ \ \ l\rto\infty \ .
\eeq
It was now observed in \cite{BoStSt,BoGeGiSch}
that such an exponential increase of the
multiplicities is a common feature of arithmetic groups
and may serve as a characterization of them. The example
of the modular group shows most clearly that this law derives
from the fact that the condition on the set of traces $tr\,\Ga_{mod}$
to be the ``rigid'' set $\gz$ forces the number $\hat\cN (l)$
of different lengths to grow only like $const.\,e^{l/2}$.
One can transform this condition also into one which
applies directly to $tr\,\Ga$. If $\hat\cN(l) \sim c\,e^{l/2}$,
$l\rto\infty$, then the number of lengths in an interval
$[l,l+2e^{-l/2}]$ is asymptotically given by $c$. If
one returns to $tr\,\Ga$, then the above interval corresponds
asymptotically to the interval $[t,t+1]$, where $t=2\cosh(l/2)
\sim e^{l/2}$. Thus the number of traces of elements of $\Ga$
in the interval $[t,t+1]$ is asymptotically given by
\beq
\label{bcp}
\# \{tr\,\Ga \cap [t,t+1]\} \sim c\ ,\ \ \ t\rto\infty \ .
\eeq
This condition shows more explicitly that the condition on
$tr\,\Ga$ enforces the traces to form a ``rigid'' discrete
subset of the real line. The condition $\#\{tr\,\Ga \cap [t,t+1]\}
\leq c(\Ga)$ for some constant $c(\Ga)$ has most recently
been proven by Sarnak and Luo \cite{SarnakLuo} to be
fulfilled for all arithmetic Fuchsian groups $\Ga$ and
has been called the {\it bounded clustering property}
by them.

Our plan now is to derive (\ref{multmodular}) for general
arithmetic groups. Since this needs a lot of techniques from
number theory, at first a brief review of relevant notions and
facts concerning algebraic number fields and quaternion algebras
shall be given. This then allows to state the precise
definition of arithmetic Fuchsian groups. Only after this
has been done it will be possible to study the traces occurring
for arithmetic groups in more detail.

\subsection{Arithmetic Fuchsian Groups}
This section contains a collection of notions and facts
from number theory that are needed to state the definition
of arithmetic Fuchsian groups. More background on the
algebraic foundations and on number theory may be found in
\cite{Hilbert,Waerden,Borewicz}. As references to
find additional information on quaternion algebras and arithmetic
Fuchsian groups \cite{Shimura,Vigneras,Miyake} may be
consulted.

An extension $K$ of finite degree $n$ of the field of rational
numbers $\Q$ is a field that contains $\Q$ as a subfield and,
viewed as a vector space over $\Q$ (in the obvious manner),
is of finite dimension $n$. Let $\Q [x]$ denote the ring of
polynomials in a variable $x$ with rational coefficients. $\al
\in K$ will be called {\it algebraic}, if it is a zero of some
polynomial from $\Q [x]$. The {\it minimal polynomial} of $\al$
is the (unique) element in $\Q [x]$ of lowest degree, and with
leading coefficient one, that has $\al$ as a root. The field $K$
is called an {\it algebraic number field}, if every $\al \in K$
is algebraic. Every extension $K$ of $\Q$ of finite degree is
known to be algebraic.

If $M$ is some arbitrary subset of $K$, $\Q (M)$ is defined to be
the smallest subfield of $K$ that contains both $M$ and $\Q$.
It is given by all values of all polynomials in the elements of
$M$ with rational coefficients, and all possible quotients thereof.
$\Q (M)$ is called the {\it adjunction} of $M$ to $\Q$. One
can now show that every algebraic number field $K$ of finite degree
over $\Q$ can be realized as an adjunction of a single algebraic
number $\al \in K$ to $\Q$; therefore $K=\Q (\al )$.

Since $K$ is a vector space of dimension $n$ over $\Q$, the $n+1$
algebraic numbers $1,\al,\dots ,\al^n $ have to be linearly
dependent and thus have to obey a relation
\beq
a_n \al^n +\cdots +a_1 \al +a_0 =0
\eeq
with rational coefficients $a_i$ and $a_n \neq 0$. Normalizing
the leading coefficient to one leaves $\al$ as a root of
an irreducible polynomial $f_{\al}(x)\in \Q [x]$ of degree $n$.
$f_{\al}(x)$ is the minimal polynomial of $\al$. Since $\{ 1,\al ,
\dots ,\al^{n-1} \}$ may serve as a basis for $K$ over $\Q$,
any $x\in K$ may be expanded as a linear combination of powers
of $\al$ up to the order $n-1$,
\beq
x=b_{n-1}\al^{n-1}+\cdots +b_1 \al +b_0 \ ,
\eeq
with rational coefficients $b_i$.

The polynomial $f_{\al}(x)$ has $n$ different complex roots $\al_1  ,
\dots ,\al_n$ $(\al_1 =\al )$. One can thus define $n$ different
homomorphisms
$\vp_i :\ K\lto \C$, $i=1,\dots ,n$, that leave $\Q$ invariant, by
\beq
\vp_i (x):= b_{n-1}\al_i^{n-1}+\cdots +b_1 \al_i +b_0 \ ,
\eeq
$\vp_1 (x)=x$. The $\vp_i$'s are called the {\it conjugations} of
$K$. If all images of $K$ under these homomorphisms
are contained in the real numbers, in other words if $f_{\al}(x)$
has only real roots, then $K$ is said to be
{\it totally real}.

On $\Q$ the usual absolute value $\nu_1 (x)=|x|$, $x\in \Q$,
introduces a  topology, which is, however, not complete.
The $n$ conjugations $\vp_i $ offer $n$ distinct ways to
embed $K$ into $\rz$. Thus $n$ different absolute values
$\nu_i$ are given on $K$ by $\nu_i (x):= |\vp_i (x)|$,
and these can be used to complete $K$
to $K_{\nu_i}\cong \rz$. The $\nu_i$'s are also called the
(archimedean) {\it infinite primes} of $K$.

All algebraic numbers in $K$ whose minimal polynomials have
coefficients in the rational integers $\gz$ form a
ring $\cR_K $, which is called the {\it ring of integers of} $K$.
An element $x\in \cR_K $ is also called an {\it algebraic integer}.
In $K$ at most $n$ algebraic numbers can be linearly independent
over $\Q$. (This is equivalent to saying that these numbers are
linearly independent over $\gz$.) Suppose now that $\la_1 ,\dots ,
\la_m$, $m\leq n$, are linearly independent numbers from $K$.
Then all linear combinations of the $\la_i$'s with integer
coefficients form an additive abelian group. This is called
a $\gz$--{\it module of rank m}. A $\gz$--module
$o\subset K$ of the maximal possible rank $n$ that at
the same time is a subring of $K$ is called an {\it order} of $K$.
Since we understand a ring to contain a unity, every order $o
\subset K$ contains the rational integers $\gz$. Further it is
known that there exists a {\it maximal order} in $K$ that
contains all other orders, and that this maximal order is just the
ring of algebraic integers $\cR_K$. An order $o$ possesses a
module-basis of $n$ algebraic numbers $\om_1 ,\dots ,\om_n $
that are linearly independent over $\gz$ and hence also,
equivalently, over $\Q$,
\beq
o= \gz \om_1 \oplus \cdots \oplus \gz \om_n \ .
\eeq
The {\it discriminant} of $o$ is defined to be $D_{K/ \qz}(o):=
[det (\vp_j (\om_i ))]^2 \neq 0$. In complete analogy one can
also define a discriminant for any $\gz$--module of rank $n$ in $K$.

Another important notion, to be introduced now, is that of a
{\it quaternion algebra}. In doing so, we will mainly follow
\cite{Vigneras,Shimura}.
An algebra $A$ over a field $K$ is
called {\it central}, if $K$ is its centre; it is said to be
{\it simple}, if it contains no two-sided ideals besides
$\{ 0\}$ and $A$ itself. A quaternion algebra then is defined
to be a central simple algebra $A$ of dimension four over $K$.
In more explicit terms $A$ may be visualized as follows:
the elements of a basis $\{ 1, \al ,\be ,\ga \}$ of $A$ over
$K$ have to obey the relations $\ga =\al \be =-\be \al$,
$\al^2 =a$, $\be^2 =b$; $a,b \in K\backslash \{0\}$.
Any $X\in A$ may then be expanded as
\beq
\label{basis}
X=x_0 +x_1 \al +x_2 \be +x_3 \al\be \ ,
\eeq
with $x_0 ,\dots , x_3 \in K$. On $A$ there exists an involutory
anti-automorphism, called the {\it conjugation} of $A$, that
maps $X$ to $\bar X:= x_0 -x_1 \al -x_2 \be -x_3 \al\be$. Thus
$\overline{\bar{X}}=X$ and $\overline{X\cdot Y}=\bar{Y}\cdot \bar{X}$.
The conjugation enables one to define the {\it reduced trace}
and the {\it reduced norm} of $A$,
\beqa
tr_A (X) &:=& X+\bar{X}=2x_0 \ , \nonumber \\
n_A (X)  &:=& X\cdot \bar{X}= x_0^2 -x_1^2 a-x_2^2 b +x_3^2 ab \ .
\eeqa
If $A$ is a division algebra, i.e.\ if every $X\neq 0$ in $A$
possesses an inverse, $n_A (X) =0$ implies $X=0$. The
inverse is then given by $X^{-1}=\frac{1}{n_A (X)} \bar{X}$.

A $\gz$--module $\cO \subset A$ of (the maximal possible) rank $4n$
that also is a subalgebra in $A$ is called an {\it order} of $A$.
The introduction of a module-basis $\{ \tau_1 ,\dots ,
\tau_{4n} \}$ turns the order into
\beq
\label{order}
\cO =\gz \tau_1 \oplus \cdots \oplus \gz \tau_{4n} \ .
\eeq
We further introduce the {\it group of units of norm one}
$\cO^1 := \{ \ve \in \cO ;\ \ve^{-1}\in \cO ,\ n_A (\ve)=1\,\}$.

A well-known example of a (division) quaternion algebra
is given by {\it Hamilton's quaternions}
\beq
\H := \left\{ \left( \begin{array}{cc} z & w \\ -\bar w & \bar z
\end{array} \right) \, ;\ z,w \in \C \right\}\ .
\eeq
$\H $ is a four dimensional $\rz$--subalgebra of $M(2,\C )$,
the algebra of complex $2\times 2$-- matrices, characterized
by the parameters $a=b=-1$. The subgroup of elements of
reduced norm one is just $SU(2,\C )$. An even simpler example
of a (non-division) quaternion algebra over $\rz$ is
$M(2,\rz )$. In fact, $\H$ and $M(2,\rz )$ are the only
quaternion algebras over $\rz$.

A classification of quaternion algebras over $K$ can now be
achieved by looking at the corresponding algebras over $\rz$
with the help of the $n$ completions $K_{\nu_i}\cong \rz$.
Define $A_i := A \otimes_{\qz} K_{\nu_i} \cong A \otimes_{\qz}
\rz$, which is a quaternion algebra over $\rz$. Hence it is either
isomorphic to $\H$ (if it is a division algebra), or to
$M(2,\rz )$ (if it is a non-division algebra). For the definition
of arithmetic Fuchsian groups (see \cite{Vigneras,Takeuchi})
we consider the case $A_1 \cong
M(2,\rz )$ and $A_i \cong \H$ for $i=2,\dots ,n$.
Therefore there exists an isomorphism
\beq
\label{Iso}
\rho :\ \ A\otimes_{\qz} \rz \lto M(2,\rz )\oplus \H \oplus \cdots
\oplus \H \ ,
\eeq
where there occur $n-1$ summands of $\H$. $\rho_j$ will denote
the restriction of $\rho$ to $A$ followed by a projection onto
the $j$-th summand in (\ref{Iso}).
The several reduced traces and norms for $X\in A$ in (\ref{Iso})
are related by
\beqa
\label{det}
tr\,\rho_1 (X)    &=& tr_A (X)\ , \nonumber \\
det\,\rho_1 (X)   &=& n_A (X) \ , \nonumber \\
tr_{\H}\rho_j (X) &=& \vp_j (tr_A (X))=\vp_j (tr\,\rho_1 (X))\ , \\
n_{\H}(\rho_j (X))&=& \vp_j (n_A (X))=\vp_j (det\,\rho_1 (X))\ ,
                      \ \ \ \ j=2, \dots ,n\ .\nonumber
\eeqa
The image of $A$ under $\rho_1$ in $M(2,\rz )$ may also be
expressed in more explicit terms by using the basis $\{1,\al ,
\be ,\al \be \}$ for $A$, see (\ref{basis}): $\rho_1 (1)$ is the
$2\times 2$ unit matrix; $\rho_1 (\al )$ and $\rho_1 (\be )$
may be represented, by using the parameters $a,b >0$, as
\beq
\rho_1 (\al )=\left( \begin{array}{cc} \sqrt{a} & 0 \\ 0 &
-\sqrt{a} \end{array} \right)\ , \ \ \ \ \
\rho_1 (\be )=\left( \begin{array}{cc} 0 & \sqrt{b} \\
\sqrt{b} & 0 \end{array} \right)\ .
\eeq
For $X=x_0 +x_1 \al + x_2 \be + x_3 \al \be \in A$ the matrix
$\rho_1 (X)$ in this representation takes the form
\beq
\label{matrix}
\rho_1 (X)=\left( \begin{array}{cc} x_0 +x_1 \sqrt{a} & x_2 \sqrt{b}
+x_3 \sqrt{ab} \\ x_2 \sqrt{b}-x_3 \sqrt{ab} & x_0 -x_1 \sqrt{a}
\end{array} \right) \ .
\eeq
We are now seeking for a subset in $A$ whose image under $\rho_1$
in $M(2,\rz)$ gives a Fuchsian group $\Ga$. Therefore
$\rho_1^{-1} (\Ga)$ must be a discrete multiplicative subgroup
of $A$. Furthermore, for $\rho_1 (X)=\ga\in \Ga$ the
condition $det\,\ga =1$ must be fulfilled. Thus by (\ref{det})
$n_A (X)=1$ has to be required. Hence we are led to look
at groups of units of norm one $\cO^1$ of orders $\cO \subset
A$. Regarding their images under $\rho_1$ one finds in
\cite{Shimura,Takeuchi} the following \\

\noindent {\sc Proposition:} Let $A$ be a quaternion algebra
over the totally real algebraic number field $K$ of degree
$n$. Let $\cO \subset A$ be an order and $\cO^1$ be its
group of units of norm one. Then $\Ga (A,\cO ):=\rho_1 (\cO^1 )$
is a Fuchsian group of the first kind. Moreover, $\Ga (A,\cO )
\backslash \cH$ is compact if $A$ is a division algebra. A change
of the isomorphism $\rho$ in (\ref{Iso}) amounts to a conjugation of
$\Ga (A,\cO )$ in $SL(2,\rz )$. \\

\noindent {\sc Remark:} This proposition can be traced back to a
more general theorem of A.\ Weil that deals with an adelic setting,
see e.g.\ \cite{Weil,Shimura}. For the proposition to be true
it is essential that the quaternion algebra is such that on
the r.h.s.\ of (\ref{Iso}) there appears exactly one factor of
$M(2,\rz)$ and $n-1$ factors of $\H$. (\ref{Iso}) therefore is
an integral component of the definition of arithmetic
Fuchsian groups. \\

The proposition now tells us that we have found what we were
looking for: a class of arithmetically defined Fuchsian groups.

We are aiming at counting the numbers of distinct primitive
lengths to gain information on the mean multiplicities in the
length spectra derived from the Fuchsian groups under
consideration. For this purpose (\ref{multcommensurable}) allows
to enlarge the class of groups appearing in the proposition
a little. \\

\noindent {\sc Definition:} A Fuchsian group $\Ga$ that is a
subgroup of finite index in some $\Ga (A,\cO )$ will be
called a {\it Fuchsian group derived from the quaternion algebra A}.
(For $\Ga (A,\cO )$ the shorthand phrase {\it quaternion group}
will also be sometimes used.) A Fuchsian group $\Ga$ that is
commensurable with some $\Ga (A,\cO )$ will be called an
{\it arithmetic Fuchsian group}. \\

In the preceding sections the modular group $\Ga_{mod}$
always served as an example of an arithmetic Fuchsian group.
It will now be seen that $\Ga_{mod}$ fits into the scheme
just introduced. The number field to be considered is $K=\Q$,
and hence $n=1$. In $\Q$ the ring of integers is of course
$\cR_K =\gz$, which is also the only order in $K$. The relevant
quaternion algebra is the simplest one can think of, namely
the one characterized by the two parameters $a=1=b$. Thus $A$
is simply the matrix algebra $M(2,\Q )$, which clearly is
not a division algebra. This is in accordance with the
non-compactness of the surface $\Ga_{mod}\backslash \cH$.
The order $\cO \subset A$ (see (\ref{order})) that determines
$\Ga_{mod}$ is characterized by the $\gz$--basis
$\{ \tau_1 ,\dots ,\tau_4 \}$,
\beq
\label{taubasis1}
\tau_1 =\left( \begin{array}{cc} 1&0\\0&0 \end{array}\right) \ ,\ \ \
\tau_2 =\left( \begin{array}{cc} 0&1\\0&0 \end{array}\right) \ ,\ \ \
\tau_3 =\left( \begin{array}{cc} 0&0\\1&0 \end{array}\right) \ ,\ \ \
\tau_4 =\left( \begin{array}{cc} 0&0\\0&1 \end{array}\right)\ .
\eeq
Therefore, $\cO =M(2,\gz)$ and $\cO^1 =SL(2,\gz)$. The modular
group is of course well-studied. A lot of information about this
group and the spectral theory on the surface $\Ga_{mod}\backslash\cH$
can be found e.g.\ in \cite{Terras}.

As a second example let us introduce the regular octagon
group $\Ga_{reg}$. This is a strictly hyperbolic Fuchsian
group that leads to the most symmetric compact surface $\Ga_{reg}
\backslash\cH$ of genus two (see
\cite{BalazsVoros,Aurich88,AurichBogo}). $\Ga_{reg}$ is a
subgroup of index two of the quaternion group $\Ga (A,\cO)$
that is defined over the number field $K=\Q (\sqrt{2})$ of
degree $n=2$. The ring of integers of this field is $\cR_K =
\gz[\sqrt{2}] =\{ m+n\sqrt{2};\ m,n\in\gz \}$. The module-basis
$\{ \om_1 ,\om_2 \}=\{ 1,\sqrt{2} \}$ for $\cR_K$ may also
serve as a basis for $K$ over $\Q$. The only non-trivial
conjugation of $K$ is given by
\beq
\label{conjugation}
\vp_2 (p+q\sqrt{2}) = p-q\sqrt{2}\ ,\ \ \ p,q\in\Q\ .
\eeq
The quaternion algebra $A$ necessary to define $\Ga (A,\cO )$
is determined by the two parameters $a=1+\sqrt{2}$ and $b=1$.
The order $\cO \subset A$ can be characterized by giving the
$\gz$--basis $\{ \tau_1 \dots ,\tau_8 \}$ (see (\ref{order})).
In the present case this is $\{\om_1 \cdot 1,\dots ,\om_2 \cdot
\al\be \}$, so that an element $\ga =\rho_1 (X)$ for $X\in\cO$
looks like
\beq
\label{groupmatrix}
\ga =\left( \begin{array}{cc} x_0 + x_1 \sqrt{1+\sqrt{2}} &
x_2 + x_3 \sqrt{1+\sqrt{2}} \\ x_2 - x_3 \sqrt{1+\sqrt{2}} &
x_0 - x_1 \sqrt{1+\sqrt{2}} \end{array} \right) \ ,
\eeq
with $x_i =m_i +n_i \sqrt{2}$, $m_i ,n_i \in \gz$. The quaternion
group $\Ga (A,\cO)$ now consists of all matrices $\ga$ of the
form (\ref{groupmatrix}) with $det\,\ga =1$. The regular octagon
group $\Ga_{reg}$ is characterized by the fact that $m_0$ has
to be an odd integer. In \cite{Pignataro} it is shown that
by adjoining the additional (elliptic) matrix
$S=\left( {0\atop -1}{1\atop 0} \right) $ to $\Ga_{reg}$ one
obtains this quaternion group via $\Ga (A,\cO )=\Ga_{reg}
\dcup \Ga_{reg} S$.

\subsection{Multiplicities in Length Spectra for
Arithmetic Fuchsian Groups}
The problem of this section is to characterize the set of
traces $tr\,\Ga$ of elements of an arithmetic Fuchsian group $\Ga$,
and then to determine the asymptotics of the number of distinct
lengths corresponding to it. It, however, suffices to concentrate
on quaternion groups, since every arithmetic group is by definition
commensurable to a quaternion group. Then (\ref{multcommensurable})
can be used to obtain the desired asymptotics of the mean
multiplicities for the given group from that for the related
quaternion group.

Let therefore $\Ga=\Ga(A,\cO)$ be a quaternion group over the algebraic
number field $K$ of degree $n$ as described in the preceding section.
For $X= x_0 +
x_1 \al +x_2 \be +x_3 \al\be \in \cO$ denote $\rho_1 (X)=
\ga \in \Ga$. By (\ref{matrix}) one sees that $\frac{1}{2}tr\,\ga
=x_0 \in \cO |_K $. Since the coefficients $x_0$ of elements
$X\in\cO $ will play a major role in the following, the set of these
will be given a name. Let therefore be
\beq
\label{MM}
\cM :=\{x_0 ;\ X=x_0 +x_1 \al +x_2 \be +x_3 \al\be \in \cO \}\ .
\eeq
Then
\beq
tr\,\Ga =tr_A \cO^1 \subset tr_A \cO = 2\cM \ .
\eeq
The inclusion $ tr_A \cO^1 \subset tr_A \cO $ will in general
be a proper one and we will return to this problem later.
The aim now is to determine the number $\hat\cN_p (l)$
of distinct primitive lengths on $\Ga \backslash \cH$ for
$l\rto \infty$. By (\ref{lengths}) one hence has to count
the number of distinct traces in $\Ga$ with $2<tr\,\ga \leq
2R$, $R:= \cosh (l/2)\rto \infty$.

First we want to describe the set $\cM$ a little bit
further. To this end we have to introduce some more notation. Let
$\{ \om_1 ,\dots ,\om_n \}$ be a basis for $K$ (as a vector
space) over $\Q$. With the help of the basis $\{ 1,\al ,\be ,\al\be \}$
of $A$ over $K$ then $\{ \chi_1 ,\dots ,\chi_{4n} \} := \{ \om_1
\cdot 1,\dots ,\om_n \cdot \al\be \}$ is a basis of $A$ over
$\Q$. On the other hand the module-basis $\{ \tau_1 ,\dots ,
\tau_{4n} \}$ of $\cO$ (see (\ref{order})) consists of $4n$
linearly independent (over $\gz$ as over $\Q$) elements of
$A$ and thus may also serve as a basis for $A$ over $\Q$.
The two $\Q$--bases of $A$ are therefore related by
\beq
\label{basechange}
\tau_i =\sum_{j=1}^{4n} M_{i,j}\, \chi_j \ ,
\eeq
where $(M_{i,j})\in GL(4n,\Q )$.
The order $\cO \subset A$ then takes the form
\beq
\label{bigo}
\cO =\gz \, \sum_{j=1}^{4n} M_{1,j}\, \chi_j \oplus \cdots
\oplus \gz \, \sum_{j=1}^{4n} M_{4n,j}\, \chi_j
\eeq
after inserting (\ref{basechange}) into (\ref{order}). As the centre
$K$ of $A$ is spanned by $\{ \chi_1 ,\dots ,\chi_n \} \cong
\{ \om_1 ,\dots ,\om_n \}$, it turns out that
\beq
\label{littleo}
\cM= \gz \, \sum_{j=1}^n M_{1,j}\, \om_j + \cdots
+ \gz \, \sum_{j=1}^n M_{4n,j}\, \om_j \ .
\eeq
Obviously, $\cM$ is a $\gz$--module in $K$. Since $( M_{i,j})\in
GL(4n,\Q )$, out of the $4n$ algebraic numbers $\sum_{j=1}^n M_{i,j}
\om_j $, $i=1,\dots ,4n$, $n$ are linearly independent. One can
therefore choose the module-basis $\{ \mu_1 ,\dots ,\mu_n \}$
among them,
\beq
\label{M}
\cM= \gz \mu_1 \oplus \cdots \oplus \gz \mu_n \ .
\eeq
In general, however, $\cM$ is not a subring and hence no order
in $K$, because the multiplication in it need not close. But
in \cite{Takeuchi} one finds that $tr\,\Ga =2 \cM$ is contained
in the ring $\cR_K$ of integers of $K$. Defining $\hat \mu_i :=
2\mu_i $ for $i=1,\dots ,n$ then yields
\beq
2\cM=\gz \hat \mu_1 \oplus\cdots \oplus \gz \hat \mu_n \subset \cR_K \ .
\eeq
By (\ref{lengths}) this means for the geodesic length spectrum
$\cL (\Ga)$ that $2\cosh (l/2)\in 2\cM \subset \cR_K$ is an
algebraic integer for every $l\in \cL (\Ga)$.

So far, $tr_A \cO =2\cM$ has been described in an algebraic
way. The object of interest, however, is $tr\,\Ga =tr_A \cO^1 $.
The problem thus is the following. Let an $x_0 \in\cM$ be given,
which means that there is at least one $X\in\cO$ with
$X=x_0 +x_1 \al +x_2 \be +x_3 \al\be$. There might be,
however, several $X_k \in\cO$ all sharing the same first
coefficient $x_0$. In a more formal way this can be described
using the bases (\ref{bigo}) and (\ref{littleo}). According
to (\ref{littleo}) $x_0$ can be represented as $x_0 =
\sum_{i=1}^{4n}\sum_{j=1}^n r_i M_{i,j}\om_j $, where all
$r_i \in\gz$. The choice of $(r_1 ,\dots ,r_{4n})\in \gz^{4n}$
is not unique, since $\cM$ has only rank $n$. Thus one can
vary $(r_1 ,\dots ,r_{4n})$ over a certain subset of $\gz^{4n}$
without changing $x_0$. $X=\sum_{i,j=1}^{4n} r_i M_{i,j}\om_j$,
however, does change under this variation. The so resulting
(discrete) set of $X\in \cO$ then is the set of $X_k$'s
mentioned above. The question now is whether there exists
some $X\in\cO$ amongst the ones all sharing the same first
coefficient $x_0$ that has a reduced norm $n_A (X)=1$.
If the answer is in the affirmative, then also $2x_0 \in
tr_A \cO^1 =tr\,\Ga$. But in order to decide on this
question one has to know whether there is a solution to
\beq
\label{quadraticeq}
-x_1^2 a-x_2^2 b+x_3^2 ab=1-x_0^2
\eeq
for a given $x_0 \in \cM$ in the three variables $(x_1 ,x_2 ,
x_3 )$ so that $X=x_0 +x_1 \al+x_2 \be +x_3 \al\be \in\cO$.
The indefinite quadratic equation (\ref{quadraticeq}) defined over the
given non-trivial domain of variables appears to be difficult to deal
with in full generality. In the following this subtle problem
shall be avoided, and the determinantal condition $n_A (X)=1$
for $X\in\cO$ shall be replaced by a weaker auxiliary
requirement. The price to pay for this will be that a
hypothesis will have to be introduced below without which
no result could be obtained.

Any $X\in\cO^1$ is characterized within $\cO$ by the condition
$n_A (X)=1$. By (\ref{Iso}) and (\ref{det}) this implies
that $n_{\H} (\rho_j (X))=\vp_j (n_A (X))=1$ for $j=2,\dots ,n$.
Therefore $\rho_j (X) \in SU(2,\C)$ and hence $tr_{\H} \rho_j
(X)=\vp_j (tr\,\ga )\in [-2,+2]$ for $j=2,\dots ,n$.
We will now call
\beq
tr_I \Ga := \{ 2x_0 ;\ x_0 \in \cM ,\ |\vp_j
(x_0 )|\leq 1,\ j=2,\dots ,n\,\}
\eeq
the {\it idealized set of traces} of $\Ga$. Instead of the inclusion
$tr_A \cO^1 \subseteq tr_A \cO$ we are thus now considering
$tr_I \Ga \subseteq tr_A \cO$. By the very construction of the
idealized traces it is clear that $tr_A \cO^1 \subseteq tr_I \Ga$.
{}From this one obtains the chain of inclusions
\beq
\label{inculsionchain}
tr\,\Ga =tr_A \cO^1 \subseteq tr_I \Ga\subseteq tr_A \cO \ .
\eeq
To avoid the problems of finding solutions to
the quadratic equation (\ref{quadraticeq}), we are now going
to count the number of idealized traces instead of the number
of actual traces. Any $2x_0 \in tr_I \Ga$ that does not occur in
$tr\,\Ga$ is referred to as a {\it gap} in the length
spectrum corresponding to $\Ga$. It is the number of these
gaps we have to make a hypothesis on.

In order to count the number of idealized traces up to a
certain value, the counting function
\beq
\label{NR}
\cN_I (R):= \frac{1}{2}\,\cdot\,\#\,\{x_0 \in\cM ;\ |x_0 |\leq R,\ |
\vp_j (x_0 )|\leq 1,\ j=2,\dots ,n\,\}
\eeq
will be introduced with $R = \cosh (l/2)$.
The factor of $\frac{1}{2}$ takes care of
the overcounting by admitting both signs for $x_0$.

The determination of the asymptotics of $\cN_I (R)$ for $R\rto
\infty$ will now be achieved by investigating the number of
certain lattice points in some parallelotope. The procedure we are
going to follow uses some standard receipt from algebraic number
theory, see \cite{Borewicz} and \cite{Lang}. At first $K$ is being
mapped to $K_{\nu_1}\times \dots \times K_{\nu_n}\cong \rz^n $
by: $x\in K$, $x\mto {\bf x}=(x_1 ,\dots ,x_n ):=(\vp_1 (x),\dots
,\vp_n (x))$. In $\rz^n$ we consider, given $n$ linearly
independent vectors ${\bf e}_1 ,\dots ,{\bf e}_n $,
a lattice
\beq
\label{lattice}
L:=\gz {\bf e}_1 \oplus \cdots \oplus \gz {\bf e}_n
\eeq
with fundamental cell
\beq
F:=I {\bf e}_1 \oplus \cdots \oplus I {\bf e}_n \ ,
\eeq
$I:= [0,1)$. In $\rz^n $ we shall consider usual euclidean
volumes. $F$ then has a volume of $vol(F)=det(e_{ij})$,
where $(e_{ij})$ denotes the $n\times n$ matrix formed
by the $n$ row vectors ${\bf e}_j \in \rz^n$. We
further introduce the parallelotope
\beq
P_R :=\{ {\bf x}\in \rz^n ;\ |x_1 |\leq R,\ |x_j |\leq 1,\ j=2,
\dots ,n\, \}
\eeq
of volume $vol(P_R )=2^n \,R$. In a first obvious approximation,
the number $n_L (R)$ of lattice points in $P_R$ is given by
$vol(P_R )/vol(F)$. Corrections to this result are caused
by contributions of the surface of $P_R$;
this is of dimension $n-1$ (in $\rz^n$), whereas $P_R$
itself is of dimension $n$. One therefore expects the
corrections to be of the order of $vol(P_R )^{(n-1)/n}$.
Indeed, in \cite{Lang} it is shown that
\beq
n_L (R) =\frac{vol(P_R )}{vol(F)} + c\cdot vol(P_R )^{1-1/n} \ ,
\eeq
with some constant $c$. The surface correction is therefore
subdominant in the limit $R\rto \infty$ and one finds that
\beq
\label{nLR}
n_L (R) = \frac{2^n}{det(e_{ij})}\cdot R +O(R^{1-1/n})\ .
\eeq
One can now construct an appropriate lattice $L$ that allows to
represent $\cN_I (R)$ as the corresponding $\frac{1}{2}n_L (R)$.
To this end one notices that the module-basis $\{ \mu_1 ,\dots ,
\mu_n \}$ of $\cM\subset K$ is being mapped to a set of $n$
linearly independent vectors $\{ {\bf e}_1 ,\dots ,{\bf e}_n \}$,
with ${\bf e}_j :=(\vp_1 (\mu_j ),\dots ,\vp_n (\mu_j ))$. The
independence may be seen by $[det(e_{ij})]^2 = [det(\vp_i (
\mu_j ))]^2 = D_{K/\qz} (\cM)$ $\neq 0$. One can thus use these
${\bf e}_j $'s to define a lattice $L$ as in (\ref{lattice}).
Its fundamental cell $F$ has volume $vol(F)= \sqrt{ D_{K/\qz}(\cM)}$.
An element $x_0 =k_1 \mu_1 +\cdots +k_n \mu_n \in \cM$,
$k_j \in \gz$, is being mapped to ${\bf x_0}= k_1 {\bf e}_1 +\cdots +
k_n {\bf e}_n \in L$, and this relation is clearly bijective. One
can thus embed $\cM$ as $L$ in $\rz^n$, and hence (see
(\ref{NR})) $\cN_I (R) =\frac{1}{2} n_L (R)$. Using $R=\cosh
(l/2)\sim \frac{1}{2}e^{l/2}$, $l\rto \infty$, and (\ref{nLR}),
one concludes that
\beq
\label{35}
\cN_I (\cosh (l/2))\sim 2^{n-2}\ [ D_{K/\qz} (\cM)]^{-1/2}\ e^{l/2}
\ ,\ \ \ l\rto \infty\ .
\eeq
As already mentioned, we are not able to pin down the
exact number of gaps that might occur for a general Fuchsian
group of the type $\Ga =\Ga (A, \cO )$. We expect, however,
that the following hypothesis holds true: \\

\noindent {\sc Hypothesis:} Asymptotically, for $l\rto \infty$,
\beq
\label{Hyp}
\hat\cN (l) \sim \cN_I (\cosh (l/2)) \ .
\eeq

\noindent
In other words, it is assumed that the number of gaps grows at
most like $O(e^{(\frac{1}{2} -\de)l})$, $\de >0$, $l\rto \infty$.

We are now in a position to state our result as the \\

\noindent {\sc Theorem:} Let $\Ga$ be an arithmetic Fuchsian
group, commensurable with the group $\Ga (A, \cO )$ derived
from the quaternion algebra $A$ over the totally real
algebraic number field $K$ of degree $n$.
Denote by $d_1$ the index of the subgroup $\Ga_0 := \Ga \cap
\Ga (A,\cO )$ in $\Ga$, and by $d_2$ the respective index of
$\Ga_0$ in $\Ga (A,\cO )$. Let $D_{K/\qz}(\cM)$ be the
discriminant of the module $\cM \subset K$ that contains
$\frac{1}{2}tr\,\Ga (A,\cO )$.
Then, under the hypothesis (\ref{Hyp}), the number $\hat\cN_p (l)$
of distinct primitive lengths on $\Ga \backslash \cH$ up
to $l$ grows asymptotically like
\beq
\label{Theorem}
\hat\cN_p (l)\sim 2^{n-2}\ \frac{d_1}{d_2}\ [D_{K/\qz}
(\cM)]^{-1/2} \,\cdot\ e^{l/2}\ ,\ \ \ l\rto \infty\ .
\eeq

\noindent {\sc Proof:} Assume the validity of the hypothesis
(\ref{Hyp}) and recall the asymptotic relation $\hat\cN_p (l)
\sim \hat\cN(l)$, $l\rto \infty$, from section 3.2. Therefore
also $\hat\cN_p (l) \sim \cN_I (\cosh (l/2))$. Using
(\ref{Nhatcommensurable}) and (\ref{35}) then leads
to the assertion.\\

The main objective of this study was not the counting function
$\hat\cN_p (l)$ but rather the mean multiplicity $<g_p (l)>$.
As these two quantities are asymptotically related by
(\ref{multdef}) one can easily obtain from the theorem the
following \\

\noindent {\sc Corollary:} The local average of the primitive
multiplicities in the cases described in the theorem
behaves asymptotically like
\beq
\label{Corollary}
<g_p (l)>\sim 2^{3-n}\ \frac{d_2}{d_1}\ \sqrt{D_{K/\qz} (\cM)}\ \cdot\,
\frac{e^{l/2}}{l}\ ,\ \ \ l\rto \infty\ .
\eeq

To confirm the results of the theorem and the corollary the
two examples of arithmetic Fuchsian groups introduced at the end
of the preceding section shall be investigated now.

Again, at first the modular group will be treated. For this
$\cO =M(2,\gz)$ and $\cO^1 =SL(2,\gz)$.
Expand $X\in \cO$ into the basis (\ref{taubasis1}), $X=k_1 \tau_1
+\cdots + k_4 \tau_4 $, $k_i \in \gz$, from which one observes that
$\frac{1}{2}tr_A X =x_0 =\frac{1}{2}(k_1 +k_4 )$. This yields
$\cM =\frac{1}{2}\gz$ and $\mu_1 =\frac{1}{2}$, see (\ref{M}). It is
known \cite{Latimer,Schleicher} that for the modular group $tr\,\Ga
=\gz =2\cM$ and therefore no gaps in the set of traces occur.
The discriminant of $\cM$ now is trivially obtained, and
$\sqrt{D_{\qz}(\cM)}=\mu_1 =\frac{1}{2}$. As the modular group
$\Ga_{mod}$ is the quaternion group $\Ga (A,\cO )$ itself, one
concludes using $d_1 =d_2 =1$,
\beqa
\hat\cN_p (l)&\sim& e^{l/2} \ , \nonumber \\
<g_p (l)>   &\sim& 2\,\cdot\ \frac{e^{l/2}}{l}\ ,\ \ \ l\rto \infty \ ,
\eeqa
which agrees with the previously known result (\ref{multmodular}).

The second example, the regular octagon group $\Ga_{reg}$,
can almost as easily been dealt with. For this one has to go
back to the quaternion group $\Ga (A,\cO)$ described by
(\ref{groupmatrix}). One observes that $x_0 \in\cM =\gz[\sqrt{2}]$,
for which one can use the basis $\{ \mu_1 ,\mu_2 \}=\{ 1,
\sqrt{2} \}$. With the help of the conjugation $\vp_2$
(see (\ref{conjugation})) one obtains the basis $\{ {\bf e}_1 ,
{\bf e}_2 \}$ for the lattice $L\subset\rz^2 $ as $\{ (1,1),
(\sqrt{2},-\sqrt{2})\}$. This allows to determine the discriminant
of $\cM$, leading to $\sqrt{D_{K/\qz}(\cM)}=|det(e_{ij})|=
2\sqrt{2}$. Since $\Ga_{reg}$ is a subgroup of index two
in $\Ga (A,\cO)$, one has to choose $d_1 =1$ and $d_2 =2$
in (\ref{Theorem}) and (\ref{Corollary}). Thus
\beqa
\hat\cN_p (l) &\sim& \frac{1}{4\sqrt{2}}\,\cdot \ e^{l/2} \ ,
                    \nonumber \\
< g_p (l)>   &\sim& 8\sqrt{2}\,\cdot \ \frac{e^{l/2}}{l}\ ,
                    \ \ \ l\rto \infty \ .
\eeqa
This is exactly the result obtained in \cite{Aurich88,AurichBogo}.
In \cite{Aurich88} the fact that $\frac{1}{2}tr\,\ga =m+n\sqrt{2}$,
$m,n\in\gz$, for $\ga\in\Ga_{reg}$ was found for the first
time. The condition that $|\vp_2 (\frac{1}{2}tr\,\ga)|=|m-n\sqrt{2}|
\leq 1$ was then observed empirically. In \cite{AurichBogo}
the numbers $x_0 =m+n\sqrt{2}$ fulfilling $|m-n\sqrt{2}|\leq 1$
were called {\it minimal numbers} and the necessity of this
condition for the regular octagon group was shown. Also, by
a numerical computation of the primitive length spectrum
up to $l=18$ it was demonstrated that gaps do exist for
the regular octagon, but that their existence does not influence
the numerically calculated mean multiplicity $<g_p (l)>$.

A final remark on a more constructive approach to determine
the asymptotics of $<g_p (l)>$ for a given arithmetic
group $\Ga$ will be added. First, one has to know the quaternion
group $\Ga (A,\cO)$ which is commensurable with $\Ga$,
and the indices $d_1$ and $d_2$ describing $\Ga\cap\Ga (A,\cO)$
as a subgroup in $\Ga$ and $\Ga (A,\cO)$. In \cite{Takeuchi}
it is shown that one can get the relevant number field $K$
by adjoining $tr\,\Ga (A,\cO )$ to $\Q$. Furthermore, the algebra
$A$ can be obtained as the linear span of $\Ga (A,\cO )$
over $K$.
Analogously, the linear span of $\Ga (A,\cO )$ over $\cR_K$
yields the order $\cO \subset A$. One then has
to find the module-basis $\{ \tau_1 ,\dots ,\tau_{4n} \}$ of
$\cO$. This can be used to obtain the matrix $(M_{i,j})$
appearing in (\ref{basechange}). Given this one has to
identify the module $\cM$ containing $\frac{1}{2}tr\,\Ga (A,\cO )$
(see (\ref{littleo}) and (\ref{M})) in order to determine
its discriminant $D_{K/\qz}(\cM)$. One can now plug all this
information into (\ref{Corollary}) to get the answer
to the problem. This procedure may, however, be quite
formal and in special cases it may be more convenient to
try a direct approach to determine $<g_p (l)>$.
Nevertheless, the above result is general and sometimes it
will only be necessary to know that $<g_p (l)>\sim const.\,
\frac{e^{l/2}}{l}$, without specifying the constant.
The latter, however, is included in the expressions
(\ref{spacemodel}) and (\ref{Sisaturation}) for the
model describing the statistical properties of the related
quantum energy spectra. For a quantitative description a knowledge
of the constants appears to be necessary.

\subsection{Pseudosymmetries}
Up to now those classical aspects of arithmetical chaos
have been investigated that are connected with geodesic
length spectra. There is a further peculiarity of the
arithmetical systems, which is a property of the classical system
also appearing in its quantum version. The phenomenon
to be discussed in this section is the occurrence of infinitely
many {\it pseudosymmetries} for hyperbolic surfaces $\GaH$ with
arithmetic Fuchsian groups $\Ga$. As these are closely related
to the {\it Hecke ring} for $\Ga$, which will be represented
on the quantum mechanical wave functions by {\it Hecke operators},
the pseudosymmetries somehow mediate between the classical
and quantum aspects of arithmetical chaos. Before going
into the details, the description of symmetries of hyperbolic
surfaces will be briefly reviewed in order to get an
intuitive understanding of pseudosymmetries as some
generalizations of symmetries.

A {\it symmetry} $g$ of a hyperbolic surface $\GaH$ is an
isometry of this surface, and therefore necessarily also
an isometry of the hyperbolic plane $\cH$. Thus $g$ is
a fractional linear transformation on $\cH$. The
corresponding matrix from $SL(2,\rz)$ will also be denoted
by $g$. The symmetry group $\Si =\{\unmat ,g_1 ,\dots ,g_{N-1} \}$
of the surface $\GaH$ is hence a subgroup of $SL(2,\rz)$.
In order that a $g\in SL(2,\rz)$ is a symmetry it has to
commute with the Fuchsian group $\Ga$. This may be seen
as follows: $z\in\cH$ is being identified with $\ga z$ for all
$\ga\in\Ga$, thus also $gz$ with $\ga gz$. If $g$ is a
symmetry, then the identification of $z$ with $g^{-1}\ga gz$
has to be the same as of $z$ with $\ga' z$ for all $\ga ,\ga'
\in\Ga$. This will be the case, if (and only if) $g^{-1}\Ga g
=\Ga$. Defining the group $\Ga' :=\Ga \cup g_1 \Ga\cup \dots
\cup g_{N-1}\Ga$ this condition means that $\Ga$ is a
normal subgroup of $\Ga'$, and $\Si \cong \Ga'/\Ga$.
The fact that the surface $\GaH$ possesses symmetries can
thus be formulated in an algebraic way. The Fuchsian group
$\Ga$ is a normal subgroup of some other Fuchsian group
$\Ga'$ and the symmetry group $\Si$ is the factor group
$\Ga'/\Ga$. This algebraic setting then allows to deal
with symmetries in the context of the Selberg trace formula,
see e.g.\ \cite{Venkov,VenkovZograf} and appendix B.
Since $\Ga'$ is also a Fuchsian group one can construct
the surface $\Ga'\backslash\cH$. This can be viewed as the
result of a desymmetrization procedure. It is, loosely
speaking, a fundamental domain for the operation of the
symmetry group $\Si$ on the surface $\GaH$.

A certain generalization of this concept of symmetries
is provided by the pseudosymmetries of arithmetic surfaces.
The starting point of their discussion will be the algebraic
properties satisfied by arithmetic Fuchsian groups. From this
the geometric properties of the related surfaces will be studied.
The algebraic side of the problem lies at the heart of the
construction of arithmetic groups and may be found e.g.\ in
\cite{Shimura,Venkov,Miyake}. It resulted from a generalization
of Hecke's investigation \cite{Hecke} of automorphic forms
for the modular group. This is the reason why in number theory
the notion of {\it modular correspondences} has been
introduced for the algebraic setting. To the author's knowledge,
it was Sarnak who recently introduced \cite{Sarnak} the name
{\it pseudosymmetries} for the geometric setting related to the
modular correspondences. Since the geometric construction
may be easier to understand the problem intuitively,
henceforth the notion of pseudosymmetries will be used throughout.
It is the purpose of this section to explain the algebraic
as well as the geometric structures accompanying the
pseudosymmetries as explicitly as it seems possible. This,
however, inevitably requires some algebraic notions that shall be
introduced first.

A major role will be played in the following by the
{\it commensurator} $\Gab$ of a Fuchsian group $\Ga$. It is
a subgroup of $G:=GL^+ (2,\rz)=\{g\in GL(2,\rz);\ det\,g >0\}$.
Then $\Gab :=\{g\in G;\ g^{-1}\Ga g\sim\Ga \}$, where ``$\sim$''
denotes the commensurability of two subgroups of a group.
A $g\in\Gab$ hence transforms the Fuchsian group $\Ga$ by conjugation
into a group $g^{-1}\Ga g$ that is still commensurable
with the original group $\Ga$. Thus $\Ga' (g):=g^{-1}\Ga g\cap
\Ga$ is a subgroup of finite index in $\Ga$ and in $g^{-1}\Ga g$.
$\Gab$ clearly is a group, and it obviously contains $\Ga$
as a subgroup. If $g$ is a symmetry of the surface $\GaH$,
then $g^{-1}\Ga g =\Ga$, leading to $\Ga' (g)=\Ga$, and
thus $g\in\Gab$. Therefore, $\Gab$ contains $\Ga'$ (see
above) as a subgroup. We will speak of a non-trivial
commensurator, if $\Gab$ contains $\Ga'$ as a proper
subgroup. The objects of $\Gab$ of interest are then the
cosets of $\Ga' \backslash\Gab$. We will henceforth tacitly
assume to mean representatives of these cosets when speaking of the
commensurator. These will give rise to pseudosymmetries.

A commensurator $\Gab$ is defined for any Fuchsian group
$\Ga$ and no use has so far been made of the arithmeticity
of the groups of interest. The difference between
commensurators of arithmetic and non-arithmetic groups is
clarified by a theorem of Margulis \cite{Margulis,Swinnerton}:\\

\noindent {\sc Theorem:} If $\Ga$ is an arithmetic Fuchsian
group, then its commensurator $\Gab$ is dense in $G$. If
$\Ga$ is non-arithmetic, then $\Gab$ is commensurable with $\Ga$.\\

\noindent {\sc Remark:} If $\Gab$ is commensurable with $\Ga$,
then $\Gab\cap\Ga$ is of finite index in $\Gab$. Now, $\Gab\cap\Ga
=\Ga$, since $\Ga\subseteq\Gab$. Thus $\Ga$ is a subgroup of
finite index $d$ in $\Gab$, $\Gab =\Ga\cup\Ga\ga_1 \cup\dots\cup
\Ga\ga_{d-1}$. The set of non-trivial elements of the
commensurator hence is $\{\unmat ,\ga_1 ,\dots ,\ga_{d-1}\}$,
which is finite, and therefore at most finitely many
pseudosymmetries exist for non-arithmetic groups. Since in the
arithmetic case $\Gab$ is dense in $G$ infinitely many
pseudosymmetries are then present. This criterion thus may serve
as a characterization of arithmeticity of Fuchsian groups.
Although in the non-arithmetic case finitely many
pseudosymmetries might exist, to the author's knowledge
no explicit example of a non-trivial pseudosymmetry is known
for this case. \\

By the definition of the commensurator, for every $g\in\Gab$
the group $\Ga' (g)$ is a subgroup of finite index $n$ in
$\Ga$, thus
\beq
\label{decompGa}
\Ga =\Ga'(g)\ga_1 \cup\dots\cup\Ga'(g)\ga_n \ ,\ \ \ \ga_1
=\unmat\ .
\eeq
The pseudosymmetry related to this $g$ is then said to be
of {\it order n}. A symmetry is in this notation a pseudosymmetry
of order $n=1$, since if $g\in\Si$, then $\Ga =\Ga'(g)$. For
the following it appears to be useful not to deal with $g\in
\Gab$, but rather with the double cosets $\Ga g\Ga$. By
(\ref{decompGa}) these decompose as
\beq
\label{doublecoset}
\Ga g\Ga =\Ga g\ga_1 \cup\dots\cup\Ga g\ga_n =\Ga\al_1
\cup\dots\cup\Ga\al_n \ ,
\eeq
where the definition $\al_i :=g\ga_i$ has been used. The double
cosets $\Ga g\Ga$ for $g\in \Gab$ are the quantities to
construct the Hecke ring for $\Ga$ from. Its definition
and discussion will be postponed to section 4.1 where the
quantum aspects of arithmetical chaos will be investigated,
since it directly leads to the definition of Hecke
operators that are relevant for the quantum mechanical problem.

The decomposition (\ref{decompGa}) of $\Ga$ into cosets of
$\Ga'(g)$ now enables one to interpret the pseudosymmetry
related to $g$ geometrically. We require $g$ to be a
pseudosymmetry of order $n\geq 2$ in order to deal with
a proper generalization of a symmetry. Then  $\Ga'(g)$
is a proper subgroup of $\Ga$, and the surface $\Ga'(g)
\backslash\cH$ is an $n$--sheeted covering of the original
surface $\GaH$. To illustrate the situation one can draw
the following commutative diagram (see e.g.\ \cite{Farkas,Shimura}):

$$\begin{array}{rcl} \cH & \stackrel{id}{\vlto} & \cH \\
        \vp_g \downarrow &    &\downarrow  \vp  \\
\Ga'(g)\backslash\cH & \stackrel{f}{\vlto} & \GaH \end{array} $$ \\

\noindent
$\vp$, $\vp_g$ and $f$ are the natural projections. For $z\in\cH$
denote by $\Ga_z :=\{\ga\in\Ga; \ga z=z\}$ the subgroup of
$\Ga$ that stabilizes $z$; the corresponding subgroup of
$\Ga'(g)$ then is $\Ga'_z :=\Ga_z \cap\Ga'(g)$. $\Ga_z$ and
$\Ga'_z$ consist of elliptic elements (or parabolic elements
if one admits cusps as fixed points). For $z\in\cH$ denote
its image under the projection on the surface $\GaH$
by $p=\vp (z)$. The preimage of $p$ under $f$ on $\Ga'(g)
\backslash\cH$ then consists of the $h\leq n$ points
$f^{-1}(p)=\{q_1 ,\dots,q_h \}$. If $e_j$ denotes the
ramification number of $f$ over $q_j$, then $\sum_{j=1}^h
e_j =n$. On $\cH$ one then chooses points $w_1 ,\dots,w_h$
such that $q_j =\vp_g (w_j)$, i.e.\ $p=f(\vp_g (w_j ))$,
$j=1,\dots,h$. Ramification numbers $e_j$ different from one
can only occur at elliptic points of $\GaH$, since $e_j$
is the index of $\Ga'_{w_j}$ as a subgroup of $\Ga_{w_j}$.
These two groups are non-trivial only at elliptic points.
The fractional linear transformations $\{\si_1 ,\dots ,\si_h \}$,
defined by $\si_j z =w_j$, mediate the mappings that
interchange the sheets of the covering $\Ga'(g)\backslash\cH
\rto\GaH$, when projected down onto the surfaces by $\vp_g$
and $\vp$. Since the above diagram is commutative one finds
for all $\ga\in\Ga$ that $f^{-1}(p)=f^{-1}(\vp (z))=f^{-1}(\vp (
\ga z))=\vp_g (\ga z)$. Thus there exists a unique index $j$ such
that $\vp_g (\ga z)=q_j$; hence $\vp_g (\ga z)=q_j=\vp_g (w_j )
=\vp_g (\si_j z)$. Since $\ga z$ and $\si_j z$ project to
the same point on $\Ga'(g)\backslash\cH$, there is some
$\de\in\Ga'(g)$ with $\ga z =\de \si_j z$. Therefore
$\ga^{-1}\de\si_j \in \Ga_z$, implying by inverting the l.h.s.,
that $\ga\in\Ga'(g)\si_j \Ga_z$. One hence obtains that
\beq
\label{coverdecomp}
\Ga =\Ga'(g)\si_1 \Ga_z \dcup\dots\dcup\Ga'(g)\si_h \Ga_z \ .
\eeq
It is not difficult to show that this decomposition of $\Ga$
is disjoint. Choosing $z$ not to be an elliptic point,
hence $\Ga_z =\{\unmat\}$ and $h=n$, and comparing
(\ref{coverdecomp}) with (\ref{decompGa}) leads to the conclusion
that $\si_j =\ga_j' \ga_j$, for some $\ga_j' \in \Ga'(g)$,
$j=1,\dots ,n$. The transformations $\{\ga_1
,\dots ,\ga_n \}$ appearing in (\ref{decompGa}) can therefore
be given an interpretation as mediating the interchanging of
the sheets of the covering outside branch points. If $\Ga$
is strictly hyperbolic, i.e.\ if $\GaH$ is a compact surface
of genus $g\geq 2$ without elliptic points, then $\Ga_z =\{
\unmat\}$ for all $z\in\cH$ and hence all ramification numbers
are one. In this case $\Ga'(g)\backslash\cH \rto\GaH$ is an
unramified $n$--sheeted covering.

In conclusion one can give the following geometric picture
of a non-trivial pseudosymmetry of order $n\geq 2$ (see
also \cite{Sarnak}): it leads to an $n$--sheeted covering
of the surface $\GaH$ that is unramified, if $\Ga$ contains
no elliptic elements. Otherwise it is ramified over the
elliptic fixed points of $\Ga$. The stabilizing group
$\Ga_z$ of an elliptic fixed point $z$ then determines
the ramification number at this point. In case the operation
under consideration is a symmetry, it is a pseudosymmetry
of order $n=1$, and thus no non-trivial covering occurs.
The generalization of symmetries to non-trivial pseudosymmetries
in this picture consists in the spreading out of the
coverings over the base surface. But still,
by definition, these always have finite numbers of sheets.

Another interpretation of a pseudosymmetry related to a non-trivial
$g\in\Gab$ can be given in terms of the effect of $g$ on the
closed geodesics on $\GaH$. The operation of $g$ on $\cH$ is
equivalent to its operation on $\Ga$ by conjugation. Thus
its effect on a closed geodesic can be described by the
mapping $\ga \mapsto g^{-1}\ga g$ for the hyperbolic $\ga\in \Ga$
related to the geodesic. Since $g^{-1}\Ga g$ is
commensurable with $\Ga$, the fraction of $\ga\in\Ga$ that are
being mapped onto $\Ga$ is given by the index of $\Ga'(g)=g^{-1}
\Ga g\cap \Ga$ in $\Ga$. By (\ref{decompGa}) this is just
the order $n$ of the pseudosymmetry. Thus a finite fraction of
$\frac{1}{n}$ of the closed geodesics on $\GaH$ are mapped
by $g$ again to closed geodesics on the same surface. Notice that this
reasoning is similar to the one leading to (\ref{Nhatsubgroup}).

We are now going to consider the modular group $\Ga_{mod}=
SL(2,\gz)$ as an example to illustrate the above construction
explicitly. As already mentioned, this was historically also
the first case to be studied, and where modular correspondences
have been introduced.
A nice presentation of several facts about the modular group
can be found in \cite{Terras}; also \cite{Shimura,Miyake} are
useful to be consulted. The first task in this context is to obtain
the commensurator $\Gab$ of the modular group. This can be found to
be $\Gab =GL^+ (2,\Q)$, a fact that is relatively easy to prove.
(See \cite{Shimura,Miyake} for details.) Notice that $GL^+ (2,\Q)$
is dense in $GL^+ (2,\rz)$, as it is predicted by Margulis' theorem.
Let now be $M_n (\gz):= \{ g\in M(2,\gz);\ det\,g=n \}$,
$n\in \nz$. This set is not a group, as the multiplication
does not close in it. However, it may be decomposed disjointly
as (see \cite{Terras})
\beq
\label{mnz}
M_n (\gz)=\bigcup\limits_{ad=n \atop 0\leq b<d}\Ga_{mod}
\left( \begin{array}{cc} a & b \\ 0 & d \end{array} \right) \ .
\eeq
Now define the semi-group
\beq
\label{Delta}
\Delta := \{g\in M(2,\gz);\ det\,g >0 \}=
\bigcup\limits_{n\in\nz} M_n (\gz)\ ,
\eeq
and take a $g\in\Gab$. Then there exists a $q\in\Q$ such that
$g' :=qg\in\Delta$. Since $g^{-1}\Ga g=g'^{-1}\Ga g'$, ones
attention may be restricted from $\Gab$ to $\Delta$. In
\cite{Miyake} one can now find the following result: let $g\in
\Delta$, then there exist $l,m \in\nz$, $l|m$ (this notation
means that $l$ is a divisor of $m$), such that $\Ga_{mod}g\Ga_{mod}
=\Ga_{mod}\left( l\ 0 \atop 0\ m \right) \Ga_{mod}$. It is then
further shown in \cite{Miyake} that $\Ga_{mod}\left( l\ 0 \atop
0\ m \right) \Ga_{mod}=\cup\ \Ga_{mod}\left( a\ b \atop 0\ d \right)$,
$ ad=lm$, $0\leq b<d$, $(a,b,d)=l$. (The latter notation
means that the largest common divisor of $a$, $b$ and $d$ is $l$.)
Thus the $\{ \al_i \}$ in (\ref{doublecoset}) are given by
$\{ \left( a\ b \atop 0\ d \right);\ ad=lm,\ 0\leq b<d,\ (a,b,d)=l \}$.
Conventionally one looks at all $g\in\Delta$ with $det\,g=n$,
$n\in\nz$, simultaneously and finds that
\beq
\label{mnzdecomp}
\bigcup\limits_{g\in\Delta \atop det\,g =n}\Ga_{mod}g\Ga_{mod}
=M_n (\gz)=\bigcup\limits_{ad=n \atop 0\leq b
<d}\Ga_{mod}\left( \begin{array}{cc} a & b \\ 0 & d \end{array}
\right)\ .
\eeq
The simplest non-trivial case is $n=2$, where one has to choose
$g=\left( l\ 0 \atop 0\ m \right)$ with $l=1$, $m=2$. Then
\beq
\label{decompn2}
\Ga_{mod}\left( \begin{array}{cc} 1 & 0 \\ 0 & 2 \end{array}\right)
\Ga_{mod}=
\Ga_{mod}\left( \begin{array}{cc} 1 & 0 \\ 0 & 2 \end{array}\right)\cup
\Ga_{mod}\left( \begin{array}{cc} 1 & 1 \\ 0 & 2 \end{array}\right)\cup
\Ga_{mod}\left( \begin{array}{cc} 2 & 0 \\ 0 & 1 \end{array}\right)\ .
\eeq
Since the r.h.s.\ of (\ref{decompn2}) consists of three cosets,
the pseudosymmetry related to this $g\in\Delta$ is of order
three. The corresponding covering $\Ga'(g)\backslash\cH\rto\Ga_{mod}
\backslash\cH$ will be found by studying the effect of $g$ on
an arbitrary $\ga\in\Ga_{mod}$.
Let therefore be $\left( a\ b \atop c\ d \right)\in\Ga_{mod}$. Then
$g^{-1} \left( a\ b \atop c\ d \right)g=\left( {a\atop c/2}{2b\atop
d} \right)$. Thus
\beq
\label{Gagmod}
\Ga'(g)=g^{-1}\Ga_{mod}g\cap\Ga_{mod}=\left\{ \left( \begin{array}{cc}
a & b \\ c & d \end{array} \right) \in\Ga_{mod};\ b,c\mbox{ even}
\right\}
\eeq
is the group that defines the desired covering over the modular
surface $\Ga_{mod}\backslash\cH$.

The modular group certainly provides the simplest example to consider.
In \cite{Shimura,Miyake} also congruence subgroups of $\Ga_{mod}$
are treated explicitly. Miyake even goes one step further in that he
deals with arithmetic groups that are unit groups of quaternion
algebras over $\Q$. Already then he has to argue adelicly,
which complicates the discussion considerably. To the
author's knowledge there does not exist any explicit treatment
of arithmetic groups derived from quaternion algebras over
number fields $K$ of degree $n\geq 2$. In such cases a first major
obstacle is to identify the commensurator group $\Gab$. An
explicit knowledge of the relations (\ref{decompGa}) and
(\ref{doublecoset}) for a given arithmetic group is mandatory
to construct the Hecke ring and the Hecke operators for that group
explicitly. From the case of the modular group one learned that
this knowledge helped a lot for the numerical determination
of the energy eigenvalues and the eigenfunctions, see
\cite{Steil}. To apply the
methods used for the modular group also to e.g. the regular
octagon group, however, seems at the moment too hard a problem.

\subsection{Statistical Properties of Geodesic Length Spectra}
This final section of the present chapter contains a
topic that might serve as a link to the following chapter
dealing with the quantum mechanics of arithmetical chaos.
The reason for this is that the statistical properties
of length spectra on hyperbolic surfaces are investigated
using the Selberg trace formula and the Selberg zeta function,
which provide a duality relation between classical and quantum
aspects. It is then possible to use results on quantum energy
spectra to gain information on classical length spectra. Also,
some analytic methods employed here using properties of the
Selberg zeta function are quite similar to those that will be
applied in chapter 4.

In the same way
as the spectral staircase $N(E)$ plays a major role in the
investigation of quantum energy spectra, the classical
counting function $\cN_p (l)$ is an important tool to study
geodesic length spectra. Since the PGT (\ref{Huber}) states
that the respective counting
functions $\cN_p (l)$ share the same asymptotic behaviour
for all hyperbolic surfaces,
it is the remainder to the asymptotic value that is the
quantity of interest. It plays a similar role as $N_{fl}(E)$
does for quantum energy spectra. The aim of this section
now is to gain information on this remainder; and by what
has been said so far about geodesic length spectra of hyperbolic
surfaces it seems to be clear that arithmetic Fuchsian
groups have to be distinguished from non-arithmetic ones.
For the analytic part of the study an analogy to the Riemann zeta
function $\ze (s)$ appears to be constructive, since an
important tool to obtain information on the remainder term
to the PGT will be the {\it Selberg zeta function} $Z(s)$
\cite{Selberg,Hejhal,Venkov}, which is in many respects
similar to the Riemann zeta function. The theory of the latter
is briefly reviewed in appendix A. A detailed
analysis of $\ze (s)$ then leads to the {\it prime
number theorem} (PNT) \cite{Titchmarsh,Ingham},
which states that the number of primes $\pi (x)$ not exceeding the
value $x$ is asymptotically given by $\pi (x)\sim li(x)\sim\frac{x}
{\log x}$, $x\rto\infty$. Identifying the $n$--th prime $p_n$ with
$e^{l_n}$ then clearly shows a close resemblance of the PNT
with the PGT (\ref{Huber}). It shall now be explained how far
the analogy between primes and primitive closed geodesics
on a hyperbolic surface can go.

As already mentioned, the Selberg zeta function $Z(s)$ is needed
in order to proof the PGT and to estimate the remainder term.
The notation already indicates that the Selberg zeta function
is the dynamical zeta function (\ref{zetafct}) for geodesic
flows on hyperbolic surfaces. In the same way as the dynamical
zeta function is derived from Gutzwiller's trace formula with
a special test function (\ref{resolvent}), the Selberg zeta
function is obtained from the {\it Selberg trace formula}
\cite{Selberg,Hejhal,Venkov}. The latter is an exact analogue
of the smeared version (\ref{GutzwillerTF}) of Gutzwiller's
trace formula, i.e.\ it is an exact identity and not only
a semiclassical approximation.

For ease of notation from now on only strictly hyperbolic
Fuchsian groups will be considered in this section.
The general case of
Fuchsian groups of the first kind yields nothing new regarding
the present problem, since the contributions of the
hyperbolic elements are the relevant ones concerning length spectra.
But of course, elliptic and parabolic elements can also be
treated, see \cite{Selberg,Hejhal,Venkov} for details.

The quantization of the geodesic flow on a hyperbolic surface
$\GaH$ is determined by the stationary Schr\"odinger equation
\beq
\label{hypSchrodingereq}
-\De \psi (z)=E\psi (z)\ .
\eeq
The {\it hyperbolic Laplacian} is given in terms of the
coordinates of $\cH$ by
\beq
\label{hypLaplace}
\De =y^2 (\partial_x^2 +\partial_y^2 )\ ,
\eeq
and the wave functions are required to be invariant under the
operation of $\Ga$ on $\cH$, $\psi (\ga z)=\psi (z)$ for all
$\ga\in\Ga$, in order to yield functions on the orbit space
$\GaH$. $-\De$ can then be defined as a self-adjoint operator
on $L^2 (\GaH)$ with a purely discrete spectrum $0=E_0 <E_1
\leq E_2 \leq \dots$, $E_n =p_n^2 +\frac{1}{4}$.
The scalar product on $L^2 (\GaH)$ is derived from the hyperbolic
metric as
\beq
\label{scalarprod}
<\psi , \vp>=\int_{\GaH}\frac{dx\,dy}{y^2}\ \overline{\psi (z)}
\,\vp (z)\ ,
\eeq
for $\psi ,\vp\in L^2 (\GaH)$. Quantum energies
$E_k$ that are related to complex momenta $p_k$, $0<E_k <
\frac{1}{4}$, are called {\it small eigenvalues}. Their existence
depends on the geometry of $\GaH$, and it is known that only
finitely many can exist on a single surface, see
\cite{Buser} for a review. On compact surfaces
of genus $g=2$ at most one might occur \cite{Schmutz}. Small
eigenvalues play a special role in connection with the PGT and
have to be treated separately.

The Selberg trace formula now reads \cite{Selberg,Hejhal,Venkov}
\beq
\label{SelbergTF}
\sum_{n=0}^\infty h(p_n )=
\frac{\mbox{area}(\cF)}{4\pi}\int_{-\infty}^{+\infty}
dp\ p\,h(p)\tanh (\pi p) + \sum_{\{\ga \}_p }\sum_{k=0}^\infty
\frac{l(\ga)\ g(kl(\ga))}{2\sinh (kl(\ga)/2)}\ .
\eeq
The outer sum on the r.h.s.\ of (\ref{SelbergTF}) runs over
all $\Ga$--conjugacy classes of primitive hyperbolic $\ga\in\Ga$,
thus equivalently, over all primitive closed geodesics on $\GaH$.
Comparing (\ref{SelbergTF}) with Gutzwiller's trace formula
(\ref{GutzwillerTF}) shows that all Lyapunov exponents are
identical, $\la_{\ga}=1$ for all $\ga$, and hence are also identical
to the metric entropy $\lb =1$. An admissible test function
$h(p)$ has to be even, and to be holomorphic in the strip $|Im\,p|\leq
\frac{1}{2}+\ve$, $\ve >0$. Also, $h(p)=O(|p|^{-2-\ve})$ for
$|p|\rto\infty$. $g(x)=\int_{-\infty}^{+\infty}\frac{dp}{2\pi}
e^{ipx}h(p)$ denotes the Fourier-transform of the test function.
The Selberg zeta function $Z(s)$ arises from (\ref{SelbergTF}) as
the dynamical zeta function has been obtained from (\ref{resolvent}),
see (\ref{resolvent})--(\ref{zetafct}). Its Euler product
converges for $Re\,s>\tau =1$,
\beq
\label{Eulerprod}
Z(s)=\prod_{\{ \ga \}_p }\prod_{n=0}^\infty \left( 1-e^{-(s+n)
l(\ga)} \right) \ ,
\eeq
where the variable $s=\frac{1}{2}-ip$ is related to the energy
variable through $E=s(1-s)$. Choosing the test function $h(p)=
\frac{1}{p^2+(s-\frac{1}{2})^2}-\frac{1}{p^2 +(\si -\frac{1}{2})^2}$
for $Re\,s$, $Re\,\si\,>1$ yields a regularized trace
of the resolvent operator for $-\De$, which is the appropriate
analogue of (\ref{resolvent}). From this relation one can
obtain the analytic properties of $Z(s)$. It is an entire
holomorphic function with {\it trivial zeros} at s=0 of
multiplicity $\frac{\mbox{area}(\cF)}{2\pi}+1$, at
$s=1$ of multiplicity one, and at $s=-k$, $k\in\nz$, of multiplicities
$\frac{\mbox{area}(\cF)}{2\pi}(k+1)$. Its {\it non-trivial zeros}
are related to the eigenvalues of $-\De$ through $s_n =\frac{1}{2}\pm
ip_n$, $E_n =s_n (1-s_n )$; their multiplicities are given
by the respective multiplicities of the eigenvalues. Small eigenvalues
therefore correspond to zeros of $Z(s)$ in the interval $(0,1)$.
Leaving aside the latter ones $Z(s)$ thus fulfills an analogue
of the Riemann hypothesis (RH) for $\ze (s)$ in that $Re\,s_n
=\frac{1}{2}$.

In the case of the Riemann zeta function the magnitude of the
remainder term in the PNT is determined by the non-trivial
zero with largest real part $\si_0$, see appendix A. Since for
the Selberg zeta function the RH is known to be true, and hence
$\si_0 =\frac{1}{2}$, once the contributions of small
eigenvalues have been extracted explicitly, one would
expect the remainder term $Q_R (l)$
in the PGT (see also (\ref{generalPGT})) to grow like
\beq
\label{QRexpect}
|Q_R (l)|=e^{\frac{1}{2}l}\cdot\om (l)\ ,
\eeq
where $\om (l)$ is a combination of powers and logarithms of $l$.
Up to now it was, however, not possible to prove this. The
analogy with $\ze (s)$ will in the following be pushed as far
as possible, adopting the strategy employed in appendix A for
$\ze (s)$ to construct a Dirichlet series whose abscissa of
conditional convergence should act as a ``detector'' for $\si_0$.
It will then become clear where the obstacle comes from
that prevents one from proving the assertion (\ref{QRexpect})
about the magnitude of the remainder term.

The Selberg zeta function has a simple zero at $s=1$ due to the
eigenvalue $E_0 =0$. Hence in the vicinity of $s=1$ it behaves
like $Z(s)=Z'(1)\,(s-1)+O((s-1)^2 )$. One could now think of
using a Dirichlet series for $Z(s)$ itself, since it has no pole
at $s=1$, as $\ze (s)$ has, and thus there seems to be no
obvious obstruction
in pushing the domain of convergence of the Dirichlet series
further to the left in the $s$--plane. However, $Z(s)$ is entire
holomorphic so that even on the critical line $Re\,s=\frac{1}{2}$
there are no poles serving as such obstructions. Therefore
the use of $Z(s)$ would not yield
the desired effect. To circumvent this problem
one would like to use $Z(s)^{-1}$ instead, which has the desired
poles at the non-trivial zeros of $Z(s)$, and then subtract the
poles at $s=1$ and at the $s_k \in (\frac{1}{2},1)$, $k=1,
\dots ,M$, corresponding to small eigenvalues. The problem one
then immediately faces is that because of the product over $n$
in the Euler product representation (\ref{Eulerprod}) of $Z(s)$
there is no convenient Dirichlet
series for the inverse of the Selberg zeta function.
The way out of
this problem will be to discard the $n$--product and to define a new
{\it Ruelle-type zeta function} $R(s):=\frac{Z(s)}{Z(s+1)}$,
which has for $Re\,s>1$ the Euler product representation
\beq
\label{EulerRuelle}
R(s)=\prod_{\{ \ga\}_p }(1-e^{-sl(\ga)})\ .
\eeq
This is easily obtained by inserting (\ref{Eulerprod}) into
the definition of $R(s)$. (\ref{EulerRuelle}) now is the exact
analogue to the inverse of the Euler product for $\ze (s)$.
Using the analytic properties of $Z(s)$ one can derive those
of $R(s)$. The latter is a meromorphic function of $s\in\C$
that is holomorphic for $Re\,s>0$. In this half-plane it has
the same zeros as $Z(s)$. Small eigenvalues lead to zeros at
$s_1 ,\dots,s_M$ in $(\frac{1}{2},1)$ and at $1-s_1 ,\dots,1-s_M$
in $(0,\frac{1}{2})$; the eigenvalue $E_0 =0$
produces a zero at $s_0 =1$. If the small eigenvalues are not
degenerate, $R(s)$ behaves in the vicinity of $s_k$, $k=0,
\dots, M$, like $R(s)=\frac{Z'(s_k )}{Z(s_k +1)}\,
(s-s_k )+O((s-s_k )^2 )$. Its logarithmic derivative hence
has a simple pole with residue one at the $s_k$'s. In order to
obtain a meromorphic function with poles at the non-trivial
zeros of $Z(s)$ on the critical line that is holomorphic
for $Re\,s>\frac{1}{2}$ one can define
\beq
\label{RuelleDiri}
f_R (s):= \frac{R'(s)}{R(s)}-\sum_{k=0}^M \frac{Z'(s_k )}
{Z(s_k +1)}\frac{1}{R(s)}\ .
\eeq
To simplify the notation it will henceforth be assumed that
there do not occur small eigenvalues, hence $M=0$. The only pole
that has to be subtracted then is the one at $s_0 =1$.

The next step now consists of finding a Dirichlet series
representation for $f_R (s)$. From the Euler product
(\ref{EulerRuelle}) for $R(s)$ one finds that for $Re\,s>1$
\beq
\label{logRuelle}
\frac{R'(s)}{R(s)}=\sum_{\{\ga\}_p }l(\ga)\,\frac{e^{-sl(\ga)}}
{1-e^{-sl(\ga)}}=\sum_{\{\ga\}_p }\sum_{k=1}^\infty
l(\ga)\,e^{-skl(\ga)}\ .
\eeq
In view of Beurling's theory of generalized prime numbers
(see e.g.\ \cite{BateDia}) one can
identify primitive hyperbolic conjugacy classes $\{\ga\}_p $
with primes $p$ \cite{Aurich92a}.
Comparing (\ref{logRuelle}) with the
Dirichlet series for $-\frac{\ze' (s)}{\ze (s)}$ then leads to the
definition of an analogue of the von Mangoldt function $\La (n)$,
see appendix A. What is lacking so far is an analogue of the positive
integers $\nz$. Recently it has become customary to introduce
such analogues as {\it pseudo-orbits} \cite{McKean,BerryKeating}.
These are the generalized integers in Beurling's theory and comprise
of formal combinations of powers of primitive conjugacy classes,
\beq
\label{pseudoorbit}
\ro := \{\ga_1^{k_1}\}_p \oplus\cdots\oplus \{\ga_n^{k_n}\}_p \ .
\eeq
On the surface $\GaH$ $\ro$ corresponds to a formal combination of the
primitive closed geodesics related to $\ga_1 ,\dots,\ga_n $ that are
traversed $k_1 ,\dots,k_n $ times, respectively. These are the
objects that were named pseudo-orbits \cite{BerryKeating}.
The corresponding ``lengths''
\beq
\label{pseudolength}
L_\ro =k_1 l(\ga_1)+\cdots +k_n l(\ga_n )
\eeq
are then called {\it pseudo-lengths}. Having defined analogues of
integers it is now possible to introduce a von Mangoldt
function for $R(s)$,
\beq
\label{MangoldtR}
\La_R (\ro) :=\left\{ \begin{array}{ccl} l(\ga) & , & \ro =
\{ \ga^k \}_p \\ 0 & , & \mbox{otherwise} \end{array} \right. \ .
\eeq
With the help of the above notions one can first introduce
a Dirichlet series for $R(s)^{-1}$,
\beq
\frac{1}{R(s)}=\prod_{\{\ga\}_p}\sum_{k=0}^\infty e^{-skl(\ga)}\,
=\sum_\ro e^{-sL_\ro }\ ,
\eeq
and then one for $f_R (s)$,
\beq
\label{fRDirichlet}
f_R (s) =\sum_\ro A_\ro \,e^{-sL_\ro }\ ,\ \ \ \ \ A_\ro =\La_R (\ro)
-\frac{Z'(1)}{Z(2)}\ .
\eeq
In appendix A the Chebyshev functions $\th (x)$ and $\psi (x)$
were used to relate the analytic properties of $\ze (s)$ to the
PNT. Trying to carry this over to the case of the Ruelle-type
zeta function and the PGT one is led to define analogues of the
Chebyshev functions as
\beqa
\label{ChebyshevR}
\th_R (L)  &:=& \sum_{l(\ga)\leq L}l(\ga)\ ,\nonumber \\
\psi_R (L) &:=& \sum_{L_\ro \leq L}\La_R (\ro) =\sum_{k\geq 1}
                \sum_{kl(\ga)\leq L}l(\ga) \\
           & =& \sum_{k\geq 1}\th_R (L/k) \ . \nonumber
\eeqa
(In the above notations the $l(\ga)$'s and the $L_\ro$'s are counted
with their respective multiplicities.)
As with the classical Chebyshev functions, see appendix A, an
estimate of the remainder in $\psi_R (L)=\th_R (L)+\cR_R (L)$
gives $\cR_R (L)=O(e^{\frac{1}{2}L}L^2 )$, $L\rto\infty$. Since
the analogue of the prime counting function $\pi (x)$ is the
counting function $\cN_p (l)$ for primitive closed geodesics,
exactly the same reasoning as in appendix A leads to
\beq
\label{psiPGT}
\cN_p (l)=\frac{\psi_R (l)}{l}+\int_{l_1}^l \frac{dl'}{l'^2}\
\psi_R (l') +O(e^{\frac{1}{2}l}l)\ ,
\eeq
where $l_1$ denotes the length of the shortest closed geodesic
on $\GaH$. Using the integral (\ref{Perron}) then allows to
express $\psi_R (L)$ through $\frac{R'(s)}{R(s)}$ by ($b>1$,
$L\not= L_\ro$)
\beq
\label{psiint}
\psi_R (L)=\sum_\ro \La_R (\ro)\,\frac{1}{2\pi i}\int_{b-i\infty}^{
b+i\infty}\frac{ds}{s}\ e^{s(L-L_\ro )}=\frac{1}{2\pi i}
\int_{b-i\infty}^{b+i\infty}\frac{ds}{s}\ e^{sL}\,\frac{R'(s)}{R(s)}\ .
\eeq
Since the analytic properties of the integrand on the very right
of (\ref{psiint}) are known one would expect to obtain an analogue
of the explicit formula (\ref{explicitformula}) for the classical
Chebyshev function $\psi (x)$. This would involve a sum
$\sum_{s_n}\frac{e^{s_n L}}{s_n}$ over the non-trivial zeros of
$Z(s)$ on the critical line. At this point, however, the present case
differs from that of the Riemann zeta function in that the sum
over the $s_n$'s diverges and thus no explicit formula in the
desired manner exists for $\psi_R (L)$. The reason for this
difference lies in the stronger growth of $N(E)$ as compared to the
counting function $N_\ze (p):=\{s_n =\be_n +i\ga_n ;\ 0<\ga_n \leq p\}$.
{}From (\ref{SelbergTF}) one can rederive Weyl's law (\ref{Weyl})
to yield $N(E(p))\sim\frac{area(\cF)}{4\pi}p^2 $, $p\rto\infty$,
whereas it is known \cite{Titchmarsh} that $N_\ze (p)\sim \frac{p}
{2\pi}\log p$, $p\rto\infty$. There are hence ``too many'' terms
in the sum over the $s_n$'s that prevent it from converging.

Hejhal \cite{Hejhal} proceeds in defining an integrated version
$\psi_{R,1}(L):=\int_0^L dl\,e^l \,\psi_R (l)$ of the Chebyshev
function $\psi_R (L)$. Inspecting (\ref{psiint}) one observes
that one has to deal with the sum $\sum_{s_n}\frac{e^{s_n L}}{s_n
(s_n +1)}$ instead, which is conditionally convergent. A
tedious analysis then leads to the PGT ({\it Theorem 6.19}
in \cite{Hejhal})
\beqa
\label{generalPGT}
\psi_R (L) &=& e^L +\sum_{k=1}^M e^{s_k L}+P_R (L)\ ,\ \ \ \ \
               P_R (L)=O(e^{\frac{3}{4}L}L^{\frac{1}{2}})\ ,
               \nonumber \\
\cN_p (l)  &=& Ei(l)+\sum_{k=1}^M Ei(s_k l)+Q_R (l)\ ,\ \ \ \ \
               Q_R (l)=O(e^{\frac{3}{4}l}l^{-\frac{1}{2}})\ ,
\eeqa
where the contributions of small eigenvalues have been reintroduced
explicitly.

Another way to obtain the PGT is to restrict the contour of
integration in (\ref{psiint}) to a finite interval and then to
estimate the resulting sum. Employing methods that can be
found in \cite{Titchmarsh}, pp.\,60, and in \cite{Iwaniec}
one can show that ($b>1$, $0<T<\infty$, $L\not= L_\ro$)
\beq
\label{psiintfinite}
\psi_R (L)=\frac{1}{2\pi i}\int_{b-iT}^{b+iT}\frac{ds}{s}\ e^{sL}\,
\frac{R'(s)}{R(s)}+O(L^2 T^{-1} e^L )+O(T^{-1}(b-1)^{-1}e^{bL})\ .
\eeq
The proof for the remainder terms on the r.h.s.\ is the same as
for the lemma of section 4.5,
where $Z(s)$ replaces $\frac{R'(s)}{R(s)}$. We therefore
postpone its explicit discussion to section 4.5. Deforming the
contour of integration in (\ref{psiintfinite}) and using estimates
for $\frac{R'(s)}{R(s)}$ on the contour Iwaniec succeeded to
estimate $Q_R (l)$ for the modular group as \cite{Iwaniec}
\beq
\label{Qmod}
Q_{R,mod}(l)=O(e^{(\frac{35}{48}+\ve)l})\ \ \ \ \forall \ve >0\ ,
\eeq
which is slightly better than the general result (\ref{generalPGT}).
This is also the best upper bound for the remainder term in a
PGT available. It seems that it is very hard to ``break the
barrier'' at $e^{\frac{3}{4}l}$ for the upper bound of $Q_R (l)$.

This difficulty has a consequence for estimating the abscissa
of conditional convergence for the Dirichlet series representation
(\ref{fRDirichlet}) of $f_R (s)$. Since $f_R (s)$ has been designed
in complete analogy to the function $f(s)$ introduced in
appendix A to study the PNT, the analysis applied to the latter
function will now be repeated for the former one. In appendix A
also the convergence properties
of general Dirichlet series have been reviewed,
according to which the series (\ref{fRDirichlet}) converges for
$Re\,s>\si_c $ and converges absolutely for $Re\,s>\si_a$,
$\si_a \geq\si_c$. Arranging the pseudo-orbits in ascending order
of their lengths, $L_1 \leq L_2 \leq L_3 \leq \dots$, one finds that
\beqa
\si_a &=& \lsup\ \frac{1}{L_N}\log \sum_{n=1}^N \left| \La_R (n)-
          \frac{Z'(1)}{Z(2)}\right|\ , \nonumber \\
\si_c &=& \lsup\ \frac{1}{L_N}\log \left| \sum_{n=1}^N \left(
          \La_R (n)-\frac{Z'(1)}{Z(2)}\right)\right| \\
      &=& \lsup\ \frac{1}{L_N}\log \left| \psi_R (L_N )-\cN^{(P)}
          (L_N )\frac{Z'(1)}{Z(2)}\right| \ , \nonumber
\eeqa
where $\cN^{(P)}(L):=\#\{\ro ;\ L_\ro \leq L\}$ denotes the
counting function for pseudo-orbits. In \cite{Aurich92a} it was
shown that
\beq
\label{pseudoPGT}
\cN^{(P)}(L)=\frac{Z(2)}{Z'(1)}\ e^L +Q^{(P)}(L)\ ,
\ \ \ \ \ Q^{(P)}(L)=O(e^{L-c_1 L^\al})\ ,
\eeq
with some constants $c_1 >0$ and $0<\al<\frac{1}{3}$.
As for $f(s)$ (see appendix A) one also concludes here, using
(\ref{pseudoPGT}), that $\sum_{n=1}^N |\La_R (n)-\frac{Z'(1)}{Z(2)}|
\sim const.\cdot e^{L_N}$. The Dirichlet series (\ref{fRDirichlet})
therefore converges absolutely for $Re\,s>\si_a =1$. Inserting
$\psi_R (L)=e^L +P_R (L)$ one observes that
\beq
\label{sic}
\si_c =\lsup\ \frac{1}{L_N}\log \left| P_R (L_N )-\frac{Z'(1)}{Z(2)}
Q^{(P)}(L_N) \right| \ .
\eeq
The only conclusion one can now draw from (\ref{sic}), and from the
PGT (\ref{generalPGT}), $P_R (L)=O(e^{\frac{3}{4}L}L^{\frac{1}{2}})$,
is that $\si_c \leq 1$. The analytic properties of $f_R (s)$, however,
suggest that $\si_c =\frac{1}{2}$, since $f_R(s)$ is a holomorphic
function for $Re\,s>\frac{1}{2}$ and has poles on the critical
line. The analogous function $f(s)$ built from the Riemann
zeta function indeed shows this behaviour: its Dirichlet series
converges (conditionally) on the maximal half-plane $Re\,s>\si_0$
where $f(s)$ is still holomorphic.

The weakness of the upper bound $\si_c \leq 1$ hinges on the estimate
$Q^{(P)}(L)=O(e^{L-c_1 L^\al})$, which was obtained in \cite{Aurich92a}
from the theory of generalized prime numbers, see e.g.\ \cite{BateDia}.
However, this upper bound only requires the
rather weak estimate $Q_R (l)=
O(e^{l-cl^{\be}})$ for some constants $c>0$ and $0<\be <1$.
Since $Q_R (l)$ is known to be smaller than required, see
(\ref{generalPGT}), the true magnitude of $Q^{(P)}(L)$ could be much
smaller than just being below $e^L$. To support this idea a function
similar to $f_R (s)$ will be introduced. Starting with the Dirichlet
series (\ref{fRDirichlet}) for $R(s)^{-1}$ one obtains for $Re\,s>1$
through an integration by parts
\beqa
\frac{1}{R(s)}=\int_0^\infty d\cN^{(P)}(L)\ e^{-sL}
  &=& \left. \cN^{(P)}(L)\,e^{-sL}\right|_0^\infty +s\int_0^\infty
      dL\ \cN^{(P)}(L)\,e^{-sL} \nonumber \\
  &=& \frac{Z(2)}{Z'(1)}\frac{s}{s-1}+s\int_0^\infty dL\ Q^{(P)}(L)\,
      e^{-sL}\ .
\eeqa
This yields an integral representation
\beq
\label{Rreg}
\frac{1}{R(s)}-\frac{Z(2)}{Z'(1)}\frac{1}{s-1}=\frac{Z(2)}{Z'(1)}+
s\int_0^\infty dL\ Q^{(P)}(L)\,e^{-sL}\ ,
\eeq
of a function that is holomorphic for $Re\,s>\frac{1}{2}$ and has
poles on the critical line. The estimate for $Q^{(P)}(L)$,
however, only permits to use the r.h.s.\ of (\ref{Rreg}) for
$Re\,s\geq 1$. Again, the domain of holomorphy of the l.h.s.\
suggests that the integral might exist for $Re\,s>\frac{1}{2}$,
yielding the estimate $Q^{(P)}(L)=O(e^{\frac{1}{2}L})$.
Assuming now that indeed $\si_c =\frac{1}{2}$ for the
Dirichlet series of $f_R (s)$, then by (\ref{sic}) one concludes
that $P_R (L)=O(e^{(\frac{1}{2}+\ve)L})$ $\forall \ve>0$,
leading to the estimate $Q_R (l)=O(e^{(\frac{1}{2}+\ve)l})$
$\forall \ve >0$. The desired result (\ref{QRexpect}) thus
would follow if one assumed that the two representations
(\ref{fRDirichlet}) and (\ref{Rreg}) converged in the maximal
half-planes where the functions they define are still holomorphic.

Certainly, (\ref{Rreg}) gives the lower bound $Q^{(P)}(L)=\Om
(e^{\frac{1}{2}L})$. This estimate
already accounts for the lower bound $\si_c \geq\frac{1}{2}$,
that arises from the fact that $f_R (s)$ has poles on the
critical line,
by inserting $Q^{(P)}(L)$ into (\ref{sic}).
For $P_R (L)$ Hejhal could prove the lower bound
$P_R (L)=\Om_\pm (e^{\frac{1}{2}L}
(\log L)^{\frac{1}{2}})$ \cite{Hejhal}, implying for the
remainder term in the PGT $Q_R (l)=\Om_\pm (e^{\frac{1}{2}l}l^{-1}
(\log l)^{\frac{1}{2}})$. But unfortunately no further rigorous
conclusion can be drawn from (\ref{sic}) and (\ref{Rreg}). The
analytic theory seems to be stuck at this point. The question one
might ask now is whether the problem is a technical one or whether
there lurks some yet undiscovered phenomenon behind it.

One idea that could come to one's mind is that $|Q_R (l)|=e^{\frac{1}
{2}l}\om(l)$ is indeed true for generic, i.e.\ non-arithmetic,
Fuchsian groups. The arithmetic case should then be treated
separately and might violate the expected behaviour of $Q_R (l)$.
It seems to be quite natural to distinguish arithmetic from
non-arithmetic groups, after having discussed the exceptional
structure of length spectra in the arithmetic case. $\cN_p (l)$
is a staircase function with steps of width $\De l_n =l_{p,n+1}
-l_{p,n}$ and of height $g_p (l_{p,n})$ at $l=l_{p,n}$.
It is thus the interplay of fluctuations of lengths and multiplicities
that results in fluctuations of the staircase function $\cN_p (l)$.
This is in contrast to the fluctuation properties of the spectral
staircase $N(E)$, for which only fluctuations in the quantum
energies $E_n$ are responsible (since we require the systems
under consideration to be completely desymmetrized and thus
being void of degeneracies in their energy spectra).
The interferences of the two contributions to $Q_R (l)$
describing the fluctuations of $\cN_p (l)$ are involved, but
different multiplicities of lengths in the arithmetic and the
non-arithmetic cases clearly lead to different kinds of
fluctuations. For arithmetic groups the mean step heights
are $<g_p (l)>\sim c_\Ga \frac{e^{l/2}}{l}$, $l\rto\infty$.
These alone give a contribution of the order $l^{-1}e^{l/2}$
to $Q_R (l)$, since the mean behaviour $Ei(l)$ cannot follow
the step structure of $\cN_p (l)$. In addition to its mean
behaviour fluctuations of $g_p (l)$
can give further contributions to $Q_R (l)$, let alone the
fluctuations of the lengths themselves. For non-arithmetic
groups the mean multiplicities, i.e.\ the mean step heights,
do not give an exponential contribution to $Q_R (l)$. Since
in any case the lower bound $Q_R (l)=\Om_\pm (e^{\frac{1}{2}l}
l^{-\frac{1}{2}}(\log l)^{\frac{1}{2}})$ requires
exponentially large oscillations about $Ei(l)$ (in the
positive and the negative direction), in the
non-arithmetic case
these have to come
from length fluctuations.

In the following arithmetic and non-arithmetic Fuchsian groups
will therefore be treated separately. At first for the generic,
non-arithmetic case an approach of {\it inverse quantum
chaology} will be employed. The latter notion stands for drawing
conclusions on the classical properties of a chaotic system
from its quantum energy spectrum
using the (Selberg or Gutzwiller) trace formula. As mentioned
in chapter 2 Berry's theory of spectral rigidity predicts
for the energy dependence of the saturation value $\De_\infty
(E)$ of the rigidity a logarithmic behaviour, $\De_\infty (E)
\sim\frac{1}{2\pi^2}\log E$, $E\rto\infty$, if the classical
system is chaotic and time-reversal invariant. It has also been
mentioned in chapter 2 that this result cannot be applied to
arithmetic systems because of their exponentially
degenerate length spectra. Since $\De_\infty (E)$ is related
to $N_{fl}(E)$ via (\ref{deltainf}), one can conclude that
$|N_{fl}(E)|\sim \frac{1}{\sqrt{2}\pi}\sqrt{\log E}$,
$E\rto\infty$. Hejhal now proved a theorem that yields an
estimate for the remainder term in the PGT depending on an
upper bound for $N_{fl}(E)$, therefore being truely a result
in the spirit of inverse quantum chaology. In the mathematical
literature on the Selberg trace formula
it has become customary to introduce
the notation $S(p):=N_{fl}(E(p))=\frac{1}{\pi}\,\mbox{arg }
Z(\frac{1}{2}+ip)$. Hejhal's {\it Theorem 14.18}
\cite{Hejhal} then states
that for $0<\de <\infty$ the estimate $|S(p)|=O((\log p)^\de )$
implies $P_R (L)=O(e^{\frac{1}{2}L}L^{2+\de})$. In our case,
$\de =\frac{1}{2}$, the upper bound $Q_R (l)=O(e^{\frac{1}{2}l}
l^{\frac{3}{2}})$ follows. Comparing this with the lower bound
$Q_R (l)=\Om_\pm (e^{\frac{1}{2}l}l^{-1}(\log l)^{\frac{1}
{2}})$ one concludes that $|Q_R (l)|=e^{\frac{1}{2}l}\om (l)$,
where the asymptotics of $\om (l)$ lie somewhere between
$\frac{\sqrt{\log l}}{l}$ and $l^{\frac{3}{2}}$.
It should be stressed that this conclusion has not been drawn
rigorously because it hinges on Berry's non-rigorous theory
for $\De_3 (L;E)$. It is not clear either what effect the
unboundedness of the multiplicities in the length spectra
even for non-arithmetic groups will have on the applicability
of Berry's theory. However, together with everything else
discussed above, this gives a further hint supporting
the expectation that the remainder term in the PGT grows
like (\ref{QRexpect}) for generic Fuchsian groups.

Considering arithmetic groups the inverse quantum chaology reasoning
does not yield the same result as in the generic case
since the upper bound for $S(p)$ assumed in {\it Theorem 14.18}
of \cite{Hejhal} is violated here. For a certain class
of arithmetic groups (derived from quaternion algebras defined
over $\Q$) Hejhal obtains in his {\it Theorem 18.8} and {\it Remark
18.14} of \cite{Hejhal} the lower bound
$S(p)=\Om_\pm (\sqrt{p}/\log p)$. (Originally this was
an unpublished result of Selberg.) Since the proof only requires
the exponential increase of the multiplicities of lengths present
for all arithmetic groups, the $\Om$--result for $S(p)$
applies to any arithmetic group. This lower bound is quite close
to the general upper bound $S(p)=O(p/\log p)$,
see \cite{Hejhal}. In order to deal with the arithmetic
case one then has to use Hejhal's {\it Theorem 15.13}
\cite{Hejhal} which states that if $|S(p)|=O(p^\al )$,
$0<\al <1$, then $P_R (L)=O(e^{\frac{1+2\al}{2+2\al}L}
L^{\frac{1}{1+\al}})$. The general upper bound for $S(p)$
requires to take the limit $\al\rto 1$, yielding $Q_R =O(e^{
\frac{3}{4}l}l^{-\frac{1}{2}})$, which is exactly like in the PGT
(\ref{generalPGT}). In section 4.4, however, it will be argued
that $|S(p)|\sim\sqrt{\frac{\db}{\pi}}\,\sqrt{\frac{p}{\log p}}$,
$p\rto\infty$,
so that $\al =\frac{1}{2}$ can be chosen. This yields the
result $Q_R (l)=O(e^{\frac{2}{3}l}l^{-\frac{1}{3}})$. Thus,
using inverse quantum chaology in conjunction with reasonable
assumptions on the fluctuations in the respective quantum
energy spectra, the upper bounds on $Q_R (l)$ could be
improved. But only in the non-arithmetic case the expectation
(\ref{QRexpect}) could be supported.
In the arithmetic case it was merely possible to bring the
bound down to $e^{\frac{2}{3}l}l^{-\frac{1}{3}}$.

The methods of inverse quantum chaology did not work as effective
as desired when applied to arithmetic Fuchsian groups.
It will thus now be tried to support the assumption (\ref{QRexpect})
on the magnitude of the remainder term in the PGT by numerical
investigations of some arithmetic groups. Again, there are
two possible approaches one could choose: a direct calculation
of $Q_R (l)$ from length spectra, or an indirect one in the
spirit of inverse quantum chaology. The latter way was employed by
Aurich and Steiner in \cite{Aurich92.1}. To explain this approach
one has to go back to the starting point (\ref{psiintfinite})
of a modified explicit formula for $\psi_R (L)$. Iwaniec
succeeded in showing from (\ref{psiintfinite}) \cite{Iwaniec}
that for $1\leq T\leq e^{\frac{1}{2}L}L^{-2}$
\beq
\label{modexplicit}
\psi_R (L)=e^L +\sum_{|p_n |\leq T}\frac{e^{s_n L}}{s_n}
+O(T^{-1}L^2 e^L ) \ .
\eeq
The (finite) sum runs over pairs of non-trivial zeros $s_n =\frac{1}
{2}\pm ip_n$ of $Z(s)$. The optimal choice for T to resolve
a given $L$ then is $T=L\,e^{\frac{1}{4}L}$ \cite{Iwaniec}.
Inserting (\ref{modexplicit}) into (\ref{psiPGT}) thus yields
\beq
\label{IwaniecPGT}
\cN_p (l)=Ei(l)+\sum_{|p_n |\leq T}Ei(s_n l)+O(T^{-1}le^l )\ .
\eeq
The optimal choice for $T$ results in a remainder term of
$O(e^{\frac{3}{4}l})$ in (\ref{IwaniecPGT}). Aurich and Steiner,
however, derived the formula
\beq
\label{AurichPGT}
\cN_p (l)=Ei(l)-\frac{1}{2}Ei(l/2)+\sum_{|p_n |\leq T}Ei(s_n l)
+\dots \
\eeq
by employing the Selberg trace formula with
a special test function.
They hence added an additional contribution of $-\frac{1}{2}Ei(l/2)$
to the r.h.s.\ of (\ref{IwaniecPGT}). In the modified explicit
formula (\ref{modexplicit}) this extra term does not appear
explicitly because it is hidden in the remainder. In the
numerical evaluation of (\ref{AurichPGT}) this, however, appears
to be needed. For the (arithmetic) regular octagon group
Aurich and Steiner calculated the first 200 eigenvalues,
leading to a cut-off at $T=14.2$. Relating this to $l$ through
the optimal choice $T=l\,e^{\frac{1}{4}l}$ yields $l\approx 4.6$.
Indeed, the numerical calculations presented in \cite{Aurich92.1}
show a good approximation of $\cN_p (l)$ in the interval
$3\leq l\leq 6$. And furthermore, the r.h.s.\ of (\ref{AurichPGT})
reveals a reasonable approximation to the actual staircase
even at larger values of $l$, which, however, cannot follow
the step structure properly. But the approximation does
not ``leave'' the staircase. This can be taken as an indication
that the omitted remainder on the r.h.s.\ of (\ref{AurichPGT})
does not exceed the mean step height $<g_p (l)>\sim 8\sqrt{2}
\frac{e^{l/2}}{l}$. Thus it seems that the complete remainder
on the r.h.s.\ of (\ref{IwaniecPGT}) is of the order of magnitude
of $\frac{e^{l/2}}{l}$.

In order to investigate the fluctuation properties of arithmetic
length spectra numerically
in detail, first fluctuations of the multiplicities
will be studied. The first example chosen is given by Artin's
billiard \cite{Matthies}. This is a system that is derived from
a desymmetrization of the modular surface $\Ga_{mod}\backslash\cH$.
The latter possesses an orientation reversing symmetry commuting
with the hyperbolic Laplacian. Dividing this symmetry out leads
to a billiard problem in a triangle on the hyperbolic plane $\cH$
first discussed by Artin \cite{Artin}. For a detailed presentation of
the desymmetrization procedure see \cite{Venkov78}. The regular
octagon group $\Ga_{reg}$ then serves as the second example.
A third arithmetic group that is included in the numerical
studies presented here is the {\it Gutzwiller octagon group}
$\Ga_G$. This is a group commensurable with $\Ga_{reg}$
leading to a compact surface $\Ga_G \backslash\cH$ of genus two.
The intersection of $\Ga_{reg}$ and $\Ga_G$ is a subgroup of index
two in $\Ga_{reg}$ and of index five in $\Ga_G$. The mean
multiplicity of lengths of closed geodesics on $\Ga_G \backslash\cH$
is thus asymptotically smaller by a factor of $\frac{2}{5}$ (see
(\ref{multcommensurable})) than that for the regular octagon group.
For a detailed description of $\Ga_G$ see \cite{Ninnemann}.
The numerical data of the lengths and multiplicities in the three
examples have been kindly placed at disposal by the authors of
\cite{Matthies,AurichBogo,Ninnemann}.
\begin{figure}[hbt]
\label{fig1}
\epsfbox[0 0 557 390]{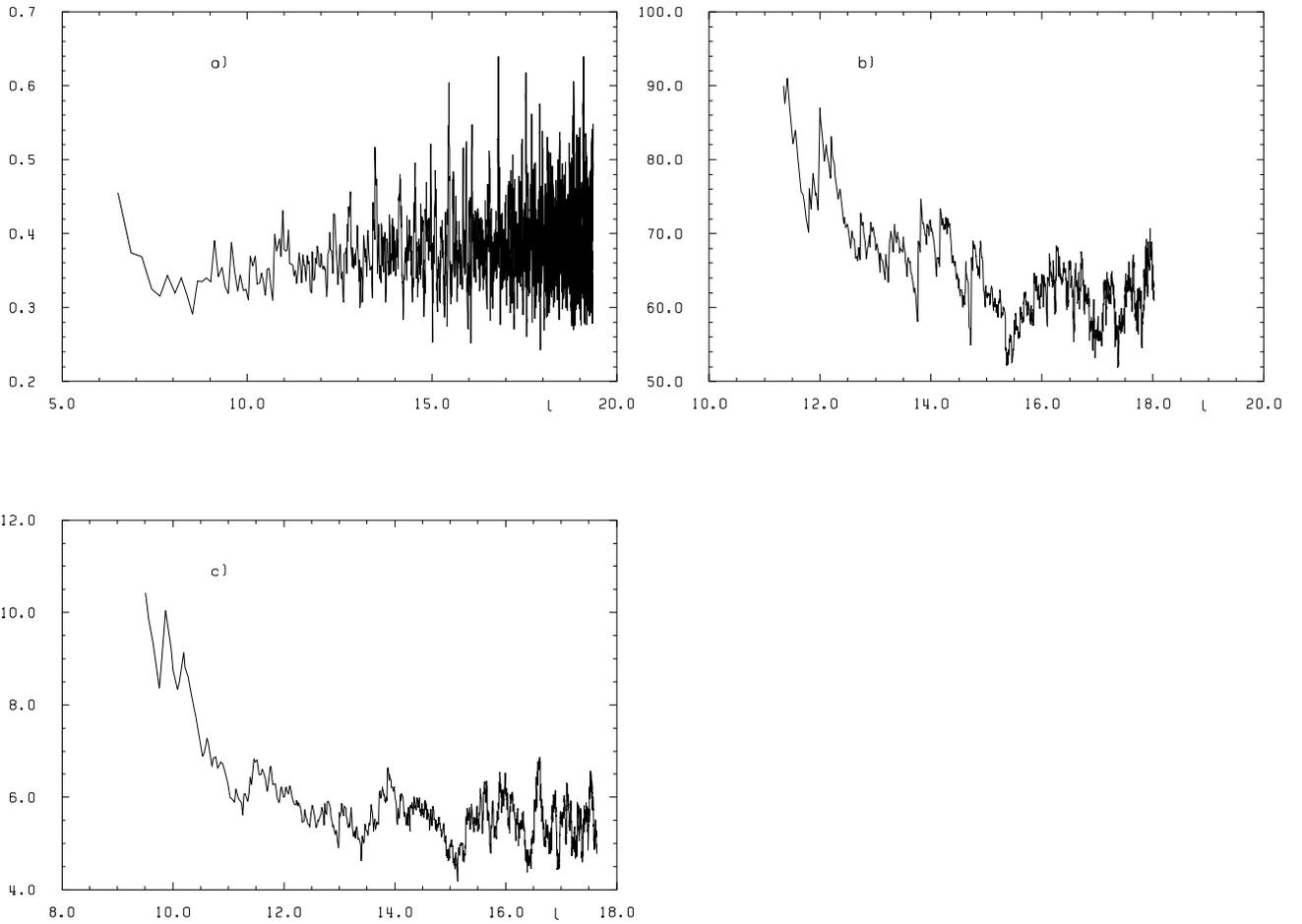}
\caption{For a) Artin's billiard, b) the regular octagon and c)
the Gutzwiller octagon the quantity $\si_g^2 (l)\,l^2 \,e^{-l}$
is shown.}
\end{figure}

As a numerical expression to calculate the mean multiplicities
\beq
\label{nummult}
\frac{1}{2N}\sum_{k=-N}^{+N}g_p (l_{p,n+k})=\frac{1}{2N}\left[
\cN_p (l_{n+N})-\cN_p (l_{n-N-1})\right]
\eeq
is taken. Asymptotically, for $l_n \rto\infty$, this expression
approaches $<g_p (l_n )>$, compare (\ref{multdef}). Analogously,
$<g_p (l_n )^2 >$ is approximated by $\frac{1}{2N}\sum_{k=-N}^{
+N} g_p (l_{p,n+k})^2 $. Employing Chebyshev's inequality, stating
that $(\sum_{n=1}^N a_n )^2 \leq N\sum_{n=1}^N a_n^2 $ if all
$a_n \geq 0$, one merely ends up with the lower bound
\beq
\label{varmult}
\si_g^2 (l):=<g_p (l)^2 >-<g_p (l)>^2 \ \geq 0\ .
\eeq
To find out about the asymptotic $l$--dependence of the fluctuations
$\si_g^2 (l)$ numerical calculations were performed in the three
systems mentioned above. In fig.1
$\si_g^2 (l)\,l^2 \,e^{-l}$ is plotted for the three systems mentioned
above. The respective primitive length spectra are completely known
up to $l_{max}=19.360$ for Artin's billiard, $l_{max}=18.092$ for
the regular octagon, and $l_{max}=17.680$ for the Gutzwiller
octagon. The averaging has been performed over $2N=100$ lengths
in all three examples. One notes that asymptotically $\si_g^2 (l)
\,l^2\,e^{-l}$ seems to fluctuate about a constant value so that
the asymptotic behaviour of the fluctuations
appears to be $|\si_g (l)|\sim const.\
\frac{e^{l/2}}{l}$, $l\rto\infty$. It thus turns out that the
fluctuations of the multiplicities about their mean values do
not give stronger contributions to $Q_R (l)$ than the mean
multiplicities themselves.

\begin{figure}[hbt]
\label{fig2}
\epsfbox[0 0 557 390]{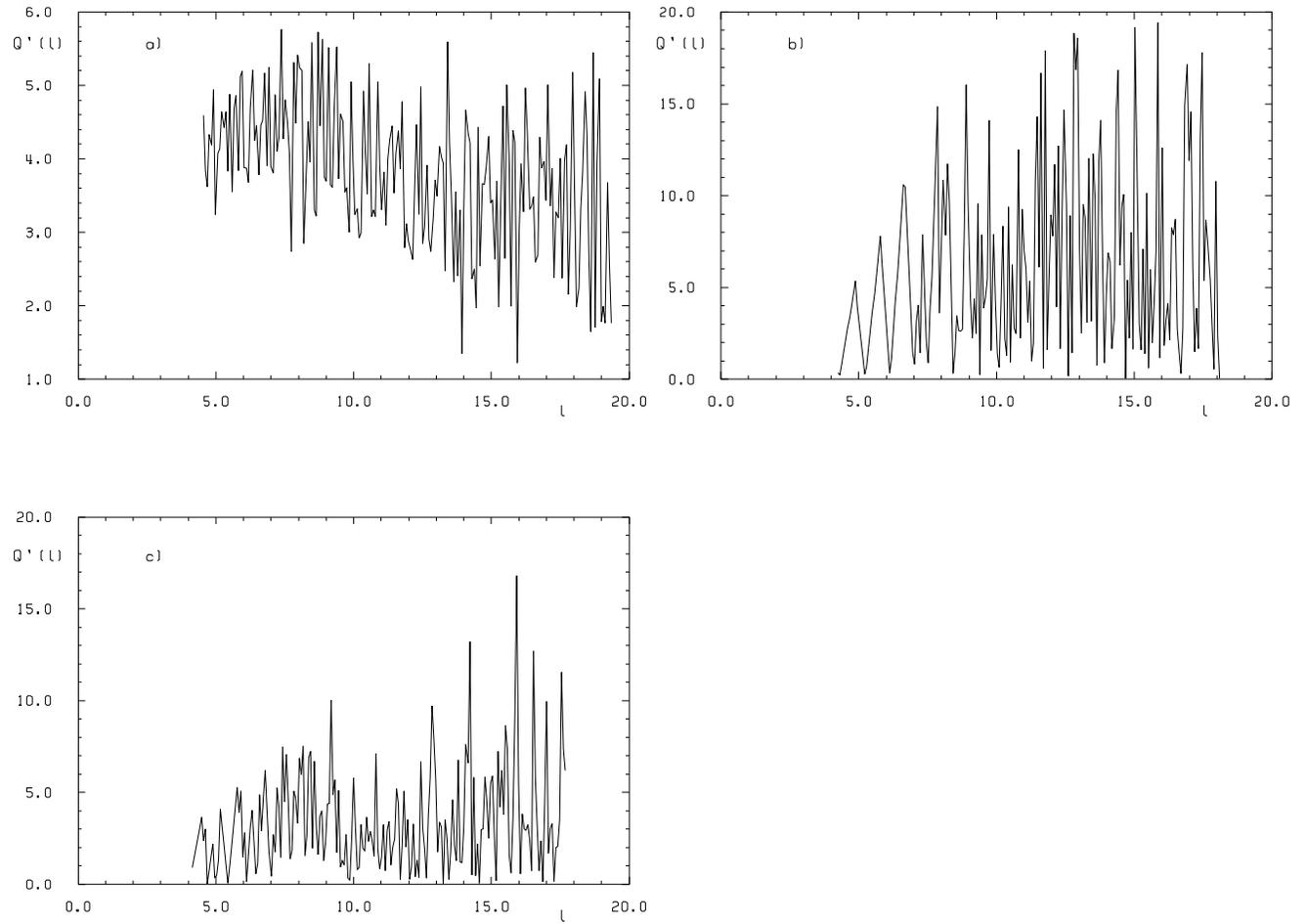}
\caption{The quantity $Q'(l)=|Q_R (l)|\,l\,e^{-\frac{l}{2}}$ is
shown for a) Artin's billiard, b) the regular octagon and c)
the Gutzwiller octagon.}
\end{figure}
One might now be interested in seeing the behaviour of $Q_R (l)$
directly. Since the contributions from the mean of the multiplicities
and from their fluctuations are of the order of magnitude of
$\frac{e^{l/2}}{l}$, fig.2 presents a numerical
calculation of $Q'(l):=|Q_R (l)|\,l\,e^{-l/2}$. As it is observed
from fig.2 that $Q'(l)$ fluctuates in some bounded strip
one can conclude that $Q_R (l)$ appears to be of the order of
magnitude of $\frac{e^{l/2}}{l}$ in the computed range of
$l$--values. It should, however, be noted that a slightly different
power of $l$ in the asymptotics of $Q_R (l)$ can hardly be
excluded by the numerical results. Therefore, the expectation
(\ref{QRexpect}) is supported by numerical evidence to hold
also for the arithmetic Fuchsian groups studied here.
Numerical calculations for the modular group as well as for
a non-arithmetic group also supporting (\ref{QRexpect})
were performed in \cite{BogoSch}.

In summary, there are several reasons, both from an analytical
as well as from a numerical point of view, that the remainder
term $Q_R (l)$ in the PGT is asymptotically for $l\rto\infty$
given by $|Q_R (l)|=e^{\frac{l}{2}}\om (l)$, where $\om (l)$ is
a combination of powers and logarithms of $l$, both for
arithmetic and non-arithmetic Fuchsian groups. But since the
mechanisms of fluctuations are different for these two classes
of systems, the methods employed to support the expected behaviour
of $Q_R (l)$ had to be different ones. It seems that the
non-arithmetic case behaves like other ``generic'' chaotic
systems (see also \cite{SieberPhD}). But for those the analytic
tools used here are not applicable because the Gutzwiller trace
formula is not an exact identity and hence no such detailed
information about the analytic properties of the associated
dynamical zeta functions are available like for the Selberg zeta
function. Inspecting (\ref{GutzwillerTF}), however, there seems
to be no fundamental difference between the general case and the
case studied here. It is only that the powerful machinery of the
Selberg trace formula is available for analytic investigations.
The only peculiarity of the arithmetic systems that played a role
was their exceptional structure of their respective length
spectra including the exponential degeneracies, a property
not shared by generic systems. One would therefore expect a
behaviour like the one for non-arithmetic Fuchsian groups
also for general, generic chaotic systems.

\xsection{Quantum Aspects of Arithmetical Chaos}
The item of this chapter is a discussion of the quantized
versions of geodesic flows on hyperbolic surfaces derived from
arithmetic Fuchsian groups. The main objective thereby is
to understand the statistical properties of arithmetic energy
spectra in contrast to those of non-arithmetic systems. The
Selberg trace formula will play a decisive role because it allows
to trace back the exceptional spectral statistics of the
arithmetical systems to the peculiarities of their respective
classical limits, namely to the exponentially growing degeneracies
in their classical length spectra.

The first section presents a discussion of pseudosymmetries
in the quantum mechanical context and introduces Hecke operators.
The discussion and interpretation of these leads to a discovery
of constraints on the arithmetical quantum energy spectra that
are taken as indications for exceptional statistical
properties for the latter. Then the empirical observations about
arithmetic quantum energy spectra are reviewed and the role
of the form factor for the spectral statistics is discussed.
Sections 4.3 and 4.4 present a model to describe the level
spacings distribution and the number variance, respectively.
The final section of this chapter is devoted to an
investigation of the convergence properties of the Selberg
zeta function both for arithmetic and non-arithmetic Fuchsian
groups.

\subsection{Hecke Operators}
In section 3.5 pseudosymmetries on arithmetic surfaces have
been introduced as generalizations of symmetries. In quantum
mechanics the latter ones manifest themselves as being
represented unitarily on the wave functions. The {\it Hecke
operators} are generalizations of the representation operators of
symmetries to pseudosymmetries. For
arithmetic groups they form an infinite algebra of
self-adjoint operators commuting with the Hamiltonian $H=-\De$.

As in section 3.5 we will now open the discussion of the
realization of pseudosymmetries in quantum systems by briefly
reviewing the case of symmetries. Let therefore be $\Ga$ a
Fuchsian group of the first kind that is a normal subgroup
of index $N$ in $\Ga'$. $\Ga' /\Ga \cong\Si =\{\unmat, g_1 ,\dots,
g_{N-1}\}$ then is the symmetry group of the surface $\GaH$.
$\Si$ shall be unitarily represented on the (finite dimensional)
vector space $V_\chi$ by $\chi\in\mbox{End}(V_\chi )$. Since
$\chi$ may be decomposed into irreducible components,
henceforth only unitary irreducible representation will be discussed;
$\Si^*$ then denotes the set of these (the unitary dual of $\Si$).
The relevant quantum mechanical Hilbert space is the space of
$V_\chi$--valued square-integrable functions $L^2 ({\GaH})\otimes
V_\chi$, which may be realized as follows. Let $\psi\,:\,\cH \rto
V_\chi$ transform under $\Ga'$ via the unitary representation
$\chi$ that is defined on $\Ga'$ by extending it trivially onto $\Ga$,
i.e.\ $\chi$ is viewed as a representation of $\Ga'$ with
$\Ga\subseteq \mbox{ker}(\chi)$. Expressed in explicit terms:
if $\ga'\in\Ga'$, then $\ga'\in g_i \Ga$ for some $g_i \in \Si$, and,
with some $\ga\in\Ga$, $\psi(\ga'z)=\psi(g_i \ga z)=\chi(g_i )\psi (z)$.
One is hence considering the spectral problem of the hyperbolic
Laplacian on $\Ga'$--automorphic functions transforming under
unitary representations of $\Ga'$ that act trivially on $\Ga$
such that they yield irreducible representations of $\Si \cong
\Ga'/\Ga$. The representation operators $T_{g_i}$ for
$g_i \in\Si$ act as
\beq
\label{symrep}
T_{g_i}\psi(z):=\psi(\ga' z)=\chi(g_i )\psi(z) \ .
\eeq
On the level of the Selberg trace formula and the Selberg zeta
function the desymmetrization procedure can be carried through
by decomposing the representation of $\Ga'$ that is induced from
the trivial representation of $\Ga\subset\Ga'$ into irreducible
components, see \cite{Venkov,VenkovZograf}. The Selberg
zeta function $Z_\Ga (s)$ for the group $\Ga$ then factorizes as
\beq
\label{zetafactor}
Z_\Ga (s)=\prod_{\chi\in\Si^*}Z_{\Ga'}(s,\chi)^{\mbox{dim V}_\chi}\ ,
\eeq
where for $Re\,s>1$
\beq
\label{zetachi}
Z_{\Ga'}(s,\chi)=\prod_{\{\ga'\}_p}\prod_{n=0}^\infty \det\left(
1-\chi(\ga')\,e^{-(s+n)l(\ga')}\right)
\eeq
is the Selberg zeta function referring to $\Ga'$ and incorporating
the representation $\chi$. (The determinant is defined on the
representation space $V_\chi$.) (\ref{zetafactor}) clearly
shows that an eigenvalue $E_n^\chi =(p_n^\chi )^2 +\frac{1}{4}$
of the hyperbolic Laplacian related to the symmetry class $\chi$
occurs with multiplicity $\mbox{dim }V_\chi$, since this is
the multiplicity of the corresponding zero $s_n =\frac{1}{2}-i
p_n^\chi$ of $Z_\Ga (s)$. The explicit treatment of an example
for such a desymmetrization will be presented in appendix B.

According to the algebraic setting of pseudosymmetries
reviewed in section 3.5 the latter ones are related to
(non-trivial) elements of the commensurator $\overline{\Ga}$
of the arithmetic Fuchsian group $\Ga$ under consideration.
For $g\in\overline{\Ga}$ the double cosets $\Ga g\Ga$ are the
basic objects the Hecke ring is constructed from, see
\cite{Shimura,Miyake} as general references. Let now be
$\De$ a semi-group, $\Ga\subset\De\subseteq\overline{\Ga}$,
and form the free $\gz$--module $\cR (\Ga,\De)$ generated
by all $\Ga g\Ga$ for $g\in\De$, i.e.\
\beq
\label{Heckering}
\cR (\Ga,\De):=\left\{ \sum_{g\in\De}c(g)\,\Ga g\Ga ;\ c(g)\in\gz,\
c(g)\not= 0\mbox{ for finitely many }g\right\}\ .
\eeq
On $\cR(\Ga,\De)$ a multiplication will be introduced. To this end
consider the decompositions (\ref{doublecoset})
\beqa
\label{cosets1}
\Ga g_1 \Ga &=& \Ga\al_1 \cup\dots\cup\Ga\al_k \ ,\nonumber \\
\Ga g_2 \Ga &=& \Ga\be_1 \cup\dots\cup\Ga\be_l \ ,
\eeqa
corresponding to two elements $g_1 ,g_2 \in\De$. For every
$g\in\De$ define the integer
\beq
\label{mvong}
m(g):= \#\,\{(i,j);\ \Ga\al_i \be_j =\Ga g\}\ .
\eeq
The product of the cosets $\Ga g_1\Ga$ and $\Ga g_2\Ga$ is then
defined as
\beq
\label{product}
\Ga g_1 \Ga\cdot\Ga g_2 \Ga :=\sum_{g\in\De}m(g)\Ga g\Ga \ .
\eeq
One can now show \cite{Shimura,Miyake} that this product does
not depend on the choice of the representatives $\{\al_i \}$
and $\{\be_j \}$ in (\ref{cosets1}), and that it is associative.
Furthermore, $\unmat\in\De$ yields the identity element
$\Ga\unmat\Ga$ for this multiplication. Extending the law of
multiplication (\ref{product}) linearly to all of $\cR(\Ga,\De)$
turns this module into an associative ring with identity, which
is called the {\it Hecke ring of $\Ga$ with respect to} $\De$.
Choosing $\De =\overline{\Ga}$, $\cR(\Ga):=\cR(\Ga,\overline{\Ga})$
is named the {\it Hecke ring} of $\Ga$. In general, the Hecke
ring need not be commutative, but there exists a sufficient
criterion for its commutativity, see \cite{Shimura,Miyake}:
if there exists a one-to-one mapping $\io :\,\De\rto\De$
such that $(g_1 g_2)^\io =g_2^\io g_1^\io $, $ \Ga g^\io \Ga =
\Ga g\Ga$ for all $g\in\De$, and $\Ga^\io =\Ga$, then the Hecke
ring $\cR(\Ga,\De)$ is commutative. Several examples
of arithmetic groups have been
dealt with in the literature that actually do result in
commutative Hecke rings, the most prominent ones, as always,
being the modular group and its congruence subgroups,
see \cite{Shimura,Terras,Miyake}.
Also the Hecke rings of arithmetic groups that are obtained
from unit groups of maximal orders \cite{Shimura61} or
of orders of the Eichler type of level $N$ \cite{Miyake}
in indefinite quaternion algebras over $\Q$ are
known to be commutative.

The Hecke ring $\cR(\Ga,\De)$ can be represented on various linear
spaces; see \cite{Hecke,Shimura,Miyake} for examples not treated here.
The one we are interested in is the Hilbert space $L^2 (\GaH)$.
Let therefore be $\psi\in L^2 (\GaH)$ and define for
$g\in\De$
\beq
\label{Heckeoperator}
T(g)\,\psi(z):=\sum_{i=1}^n \psi (\al_i z)\ ,
\eeq
where the fractional linear transformations
$\al_1 ,\dots,\al_n $ are yielded from the decomposition
(\ref{doublecoset}). The bounded linear
operator $T(g)$ on $L^2 (\GaH)$ is called a {\it Hecke operator},
see \cite{Venkov} for proofs of its properties. Since $T(g)$
acting on the function $\psi$ results in a linear combination
of $\psi$ taken at points translated by operations of
$GL^+ (2,\rz)$, the Hecke operator commutes with the hyperbolic
Laplacian. In \cite{Venkov} one also finds that $T(g^{-1})=
T(g)^*$, where ``$*$'' denotes the adjoint
with respect to the scalar
product of $L^2 (\GaH)$. Thus, $\widetilde{T}(g):= T(g)+T(g^{-1})$
is self-adjoint. Moreover, the $T(g)$, $g\in\De$, form a
representation of $\cR(\Ga,\De)$ on $L^2 (\GaH)$, i.e.\ for
$g_1 ,g_2 \in\De$, $\Ga g_1 \Ga\cdot\Ga g_2 \Ga=\sum_{g\in\De}
m(g)\Ga g\Ga$, one can show that $T(g_1)T(g_2)\,\psi(z)=
\sum_{g\in\De}m(g)T(g)\,\psi(z)$.

In case $g\in\overline{\Ga}$ is a symmetry, i.e.\ a
pseudosymmetry of order $n=1$, it is an element of the group $\Ga'$
that contains $\Ga$ as a normal subgroup such that $\Ga'/\Ga$
is isomorphic to the symmetry group. By (\ref{decompGa})
then only $\ga_1 =\unmat$ occurs in the decomposition of
$\Ga$ into cosets of $\Ga'(g)$ ($=\Ga$). Thus the Hecke operator
related to $g$ only involves $\al_1 =g$, see (\ref{Heckeoperator}).
One now observes that this Hecke operator is nothing else
than the representation operator of the symmetry $g$, see
(\ref{symrep}),
\beq
T(g)\,\psi(z)=\psi(gz)=\chi(g)\psi(z)\ .
\eeq
Hecke operators therefore yield proper generalizations of
symmetry operators, which is in accordance with the assertion that
pseudosymmetries are generalizations of symmetries.

At this point one might seek for a geometrical interpretation of
Hecke operators. In section 3.5 pseudosymmetries of order $n$
were demonstrated to result in $n$--sheeted coverings of the
surface $\GaH$. The $\ga_1 ,\dots,\ga_n$ in the decomposition
(\ref{decompGa}) of $\Ga$ into cosets of $\Ga'(g)$ were
found to define the mappings that interchange the $n$ sheets of
the covering. Exactly these $\ga_i$'s now occur in the definition
(\ref{Heckeoperator}) of the Hecke operator $T(g)$ via
$\al_i =g\ga_i$, which therefore in some sense averages the wave
function $\psi(z)$ defined on $\GaH$ over the $n$ sheets of
the covering $\Ga'(g)\backslash\cH \rto\GaH$.
Looking at the diagram
in section 3.5 one observes that the $\vp_g (\ga_i z)$
are the $n$ points on $\Ga'(g)\backslash\cH$ that lie over
$\vp (z)\in\GaH$. Transforming these $n$ points by $g$,
$T(g)$ averages $\psi$ over the resulting $n$ images.

As in section 3.5 the modular group $\Ga_{mod}=SL(2,\gz)$
will serve as an example to illustrate the general procedure
just discussed, see also \cite{BoStSt}. An exhaustive
treatment of this particular case may be found in e.g.
\cite{Shimura,Terras,Miyake}. The semi-group $\De$ needed to define
the Hecke ring $\cR(\Ga,\De)$ has been introduced in
(\ref{Delta}) as $\De =\{g\in M(2,\gz);\ \det g >0\}$.
According to the convention (\ref{mnzdecomp}) to look at all
$g\in\De$ with $\det g=n$, $n\in\nz$, simultaneously,
the Hecke operators for $\Ga_{mod}$ are defined as
\beq
\label{Heckemod}
T_n :=\frac{1}{\sqrt{n}}\sum_{g\in\De\atop \det g=n} T(g)\ ,
\ \ \ \ \ n\in\nz\ ,
\eeq
after normalizing them as it is commonly done in the literature,
see e.g.\ \cite{Terras}. The decomposition (\ref{mnzdecomp})
then yields the explicit form
\beq
\label{Heckemodexplicit}
T_n \psi(z)=\frac{1}{\sqrt{n}}\sum_{ad=n\atop0\leq b<d}
\psi \left( \frac{az+b}{d}\right) \ ,
\eeq
for $\psi (z)$ fulfilling $\psi (\ga z)=\psi (z)$, $\ga\in\Ga_{mod}$.
One can prove several properties for the Hecke operators
(\ref{Heckemodexplicit}), see e.g.\ \cite{Terras}: they are
self-adjoint on $L^2 (\GaH)$ and form a commutative algebra. Their
law of multiplication can be drawn from (\ref{mvong}) and
(\ref{product}) as
\beq
\label{HeckerelationT}
T_n \,T_m =\sum_{d|(n,m)}T_{\frac{nm}{d^2}}\ ,\ \ \ \ \
n,m\in\nz\ .
\eeq
Since the $T_n$'s, for $n\in\nz$, and the hyperbolic Laplacian
all commute with one another, they can be simultaneously
diagonalized. The square-integrable eigenfunctions of the
Laplacian for $\Ga_{mod}$, the so-called {\it Maa\3 cusp forms},
can be expanded on $\cH$ as
\beq
\label{cuspFourier}
\psi (z)=N\sum_{k\neq 0}c(k)\sqrt{y}K_{ip}(2\pi |k|y)\,e^{
2\pi ikx}\ ,
\eeq
see \cite{Terras}. $K_\nu (t)$ denotes the modified Bessel
function, and $p$ is the momentum related to the eigenvalue $E$,
$-\De\psi (z)=E\psi (z)$, by $E=p^2 +\frac{1}{4}$. The $c(k)$'s
are the Fourier expansion coefficients, and $N$ is a
normalization factor. The surface $\Ga_{mod}
\backslash\cH$ possesses one symmetry that can be realized
in the coordinates employed to derive (\ref{cuspFourier})
as $z\mapsto -\bar{z}$. Accordingly, the eigenfunctions of the
type (\ref{cuspFourier}) can be distinguished as having
positive or negative parity under this symmetry operation,
$\psi_\pm (z)=\pm \psi_\pm (-\bar{z})$. Thus the expansion of
$\psi_+ (z)$ contains $\cos (2\pi kx)$ and the one of $\psi_-
(z)$ contains $\sin (2\pi kx)$ replacing $e^{2\pi ikx}$ in
(\ref{cuspFourier}). In addition, the sum then extends only over
positive $k$'s. The coefficients $c(k)$ will now be chosen
in order to yield simultaneous eigenfunctions of all the Hecke
operators $T_n$, $n\in\nz$, $T_n \psi(z) =t_n \psi(z)$,
$t_n \in\rz$. Choosing the normalization $N$ such that
$c(1)=1$, which is possible since it is known that $c(1)\neq 0$,
results in the identification $c(k)=t_k $, see \cite{Terras}.
One thus observes the interesting property that the coefficients
of the Fourier expansions (\ref{cuspFourier}) of eigenfunctions
of the Laplacian can be chosen to be the eigenvalues of
the Hecke operators. This realization immediately yields
from (\ref{HeckerelationT}) the constraints
\beq
\label{Heckerelation}
c(n)\,c(m)=\sum_{d|(n,m)} c\left(\frac{nm}{d^2}\right) \ ,
\eeq
which are named {\it Hecke relations},
for the Fourier coefficients of the eigenfunctions.
There are further properties of these coefficients and of
their statistics either known, conjectured, or
numerically investigated, see \cite{Hejhal,Terras,HejhalArno,Steil}
for further references.

It is also possible to derive a trace formula for Hecke operators
in close analogy to the Selberg trace formula. The former already
occurs in Selberg's original article \cite{Selberg} and
has been further explored in \cite{Hejhal}. The first numerical
computations with this trace formula that yield the lower
Fourier coefficients for some cusp forms (\ref{cuspFourier})
of definite parity have appeared in \cite{BoGeGiSch2}.

The Hecke relations allow to express the Fourier coefficients
$c(k)$ as polynomials in the coefficients $c(p)$ for primes
$p\leq k$. Hence the coefficients are not independent, as it
would be required by RMT for wave functions of classically chaotic
systems \cite{Brody}. There it is asserted that the expansion
coefficients of wave functions in a generic basis are Gaussian
random variables. The correlations induced by (\ref{Heckerelation})
are, however, known to be weak in the sense that \cite{Hejhal92}
\beq
C_N (l):=\frac{1}{N}\sum_{k\leq N}c(k+l)\,c(k) =O(N^{-\frac{1}{3}
+\ve})\ ,
\eeq
for all $\ve >0$ and every $l\in\nz$, leading to
$\lim_{N\rto\infty}C_N (l)=0$.
Although the expansion
(\ref{cuspFourier}) uses a special basis, the Hecke relations
manifest themselves in the wave functions irrespective of any
basis because the values of $\psi (z)$, being an eigenfunction
of every $T_n$, are related by $T_n$ at different points (see
(\ref{Heckemodexplicit})) for every $n\in\nz$. Berry's
{\it random wave conjecture} \cite{Berry77} now asserts that
the values of wave functions for classically chaotic systems
become, in the semiclassical limit, Gaussian random variables.
The existence of Hecke operators inducing relations on the
values of wave functions seems to indicate a violation of this
randomness assumption. Hejhal and Rackner \cite{Hejhal92}, however,
computed numerically several eigenfunctions of the hyperbolic
Laplacian on $\Ga_{mod}\backslash\cH$ and found good agreement
with the random wave conjecture. It therefore seems that the
correlations in the eigenfunctions of arithmetic systems induced
by the presence of the infinitely many pseudosymmetries do
not suffice to violate the conjectured random character for
wave functions in generic classically chaotic systems. This
observation is in contrast to the case of symmetries, where
it is known that in order to obtain a generic behaviour one
has to desymmetrize the systems first.

Although the presence of the Hecke operators seemingly does not
cause an exceptional behaviour of the eigenfunctions, the
pseudosymmetries they have been derived from could have an
influence on the distribution of the eigenvalues. Therefore
a final remark concerning the spectrum of the hyperbolic
Laplacian on an arithmetic  surface shall conclude this section.
The infinitely many non-trivial pseudosymmetries
of the surface $\GaH$
lead to infinitely many constraints for the eigenvalue
spectrum $\si_p (\Ga)=\{E_0 <E_1 \leq E_2 \leq\dots \}$ of $-\De$
on $\GaH$. For
every $g\in\overline{\Ga}$ the Fuchsian group $\Ga'(g)$ is a
subgroup of finite index in $\Ga$ to which there is
related the eigenvalue problem of $-\De$ on $\Ga'(g)\backslash\cH$.
Denoting these spectra by $\si_p (\Ga'(g))$, the inclusions
$\Ga'(g)\subset\Ga$ of Fuchsian groups result in inclusions
$\si_p (\Ga)\subset\si_p (\Ga'(g))$ of the respective
spectra. In order to proof these inclusions one recalls that
$E\in\si_p (\Ga)$ is equivalent to the existence of a Maa\3
waveform $\psi :\,\cH\rto\C$ fulfilling
\begin{enumerate}
\item $-\De \psi(z)=E\psi(z)$,
\item $\psi (\ga z)=\psi(z)$ for all $\ga\in\Ga$,
\item $\int\limits_{\GaH}\frac{dx\,dy}{y^2}\ |\psi (z)|^2 <\infty$.
\end{enumerate}
The assertion that then also
$E\in\si_p (\Ga'(g))$ is proved once one can
show that $\psi (z)$ obeys $1.-3.$ with $\Ga'(g)$ replacing
$\Ga$ for all $g\in\overline{\Ga}$: $1.$ is trivial, and $2.$
is also obviously true, since $\Ga'(g)\subset\Ga$. The requirement
that $\Ga'(g)$ is of finite index $n$ in $\Ga$ results in  $\mbox
{area}(\Ga'(g)\backslash{\cH})=n\cdot\mbox{area}({\GaH})<\infty$,
so that the integral in $3.$ gets multiplied by $n$ when replacing
$\Ga$ through $\Ga'(g)$ and thus remains finite. In conclusion,
the discrete spectrum $\si_p (\Ga)$ for an arithmetic group $\Ga$
is a subspectrum of infinitely many spectra $\si_p
(\Ga'(g))$,
\beq
\si_p (\Ga) \subseteq \bigcap_{g\in\overline{\Ga}}\si_p (\Ga'(g))\ .
\eeq
The groups $\Ga'(g)$ are themselves arithmetic, since they are
commensurable with the (arithmetic) group $\Ga$. Arithmetic
spectra are thus constrained by obeying infinitely
many inclusions. The question now is whether these constraints are
strong enough in order to yield exceptional statistical
properties of discrete energy spectra related to arithmetic groups.
The following sections of the present chapter are devoted to
trying to answering this question in the affirmative. The methods
employed there, however, do not rely on these constraints but
rather use the classical properties of arithmetical systems
as starting points, since it seems to be difficult to
formulate the constraints in a way that enables one to use
them in a quantitative analysis.

\subsection{Spectral Statistics and the Form Factor}
Following the general belief on the existence of a
universal classification for the statistical properties of
discrete quantum energy spectra according to the characters
of the corresponding classical systems it is expected that
time-reversal invariant systems with chaotic classical limits
possess spectra that can be described by the GOE random matrix
ensemble up to a maximal scale $L_{max}$. Berry's theory
for the spectral rigidity supporting this assumption on scales
$1\ll L\ll L_{max}$, however, presupposes that the
multiplicities of lengths of periodic orbits asymptotically
approach two for long orbits. The discussion of geodesic
flows on hyperbolic surfaces with arithmetic Fuchsian groups
$\Ga$ in chapter 3, however, revealed that for these systems
the mean multiplicities of lengths grow exponentially, $<g_p (l)>
\sim c_\Ga \frac{e^{l/2}}{l}$, $l\rto\infty$. This property
will certainly influence the spectral rigidity and therefore
the medium- and long-range correlations in the respective
quantum energy spectra. Conjecturally, then also the short-range
correlations, in particular the level spacings distributions,
will be affected, since the universal behaviour of the spectral
statistics for generic systems holds, according to empirical
observations, on all scales up to $L_{max}$. The question then
remains what spectral statistics the arithmetical systems
share?

To get an idea what the answer might look like one can recall
what is known about the function $S(p)=\frac{1}{\pi}\arg\,Z(\frac{1}
{2}+ip)=N_{fl}(E(p))$ for arithmetic as well as for non-arithmetic
Fuchsian groups, and compare this with the corresponding results
for classically integrable systems. For general cocompact
Fuchsian groups the best known asymptotic upper bound is
$N_{fl}(E)=O(\sqrt{E}/\log E)$, whereas the best lower bound
is only $N_{fl}(E)=\Om_\pm (\sqrt{\frac{\log E}{\log \log E}})$,
see \cite{Hejhal} and especially {\it Theorem 7.10} in
\cite{HejhalDuke}. One hence observes a rather large gap
between the upper and the lower bound. This can be understood
once one consults the lower bound valid for arithmetic groups.
In section 3.6 this lower bound was already employed and it was
remarked there that although in \cite{Hejhal} this was only
proved for a certain class of arithmetic groups, the result
extended to every arithmetic group, since it were the
exponential degeneracies in the respective geodesic length spectra
that were responsible for this lower bound of $N_{fl}(E)=\Om_\pm
(E^{\frac{1}{4}}/\log E)$. There hence remains for arithmetic groups
only a much more modest gap to the upper bound. Berry's result
on the spectral rigidity for generic systems, $\De_\infty (E)
\sim \frac{1}{2\pi^2}\log E$, now suggests that in the non-arithmetic
case $|N_{fl}(E)|\sim\frac{1}{\sqrt{2}\pi}\sqrt{\log E}$,
thus being close to the general lower bound. It therefore
appears that the non-arithmetic groups obstruct the lower
bound to be improved considerably, whereas the arithmetic groups
are responsible for the upper bound. Certainly, the lower
bound for arithmetic groups shows that $\De_\infty (E)\geq const.
\frac{\sqrt{E}}{(\log E)^2}$ asymptotically for $E\rto\infty$,
which clearly violates Berry's result, reflecting the fact that
the presuppositions to apply the latter are not met. Actually, the
discussion in section 4.4 yields that $\De_\infty (E)\sim \frac{2\db}
{\pi}\frac{\sqrt{E}}{\log E}$, which is only by a factor of
$\log E$ larger than the lower bound. Therefore the
arithmetic case reminds more of the saturation value of the rigidity
for classically integrable systems, $\De_\infty (E)\sim const.
\sqrt{E}$.

The integrable case also indicates that it is rather the lower
bound for $N_{fl}(E)$ that comes closer to its true magnitude
than the upper one.
Rigorous results for $N_{fl}(E)$ are known for the quantization
of the geodesic flow on a torus $T=(\rz/2\pi\gz)^2$. The associated
spectral problem is that of minus the euclidean Laplacian $\De_E
=\partial_x^2 +\partial_y^2$ acting on doubly periodic wave
functions on $\rz^2$, $\psi (x,y)=\psi (x+2\pi ,y)=\psi (x,y+2\pi )$.
The spectrum is then given by $E_{nm}=n^2 +m^2$, $n,m\in\gz$.
Since $N(E)$ is the number of points of $\gz^2$ inside a
circle of radius $\propto\sqrt{E}$, the estimation
of $N(E)$ is the classical {\it circle problem},
see \cite{HejhalDuke} for a review. It is known
that $N_{fl}(E)=O(E^{\frac{1}{3}-\de})$ for some small values of
$\de$, and $N_{fl}(E)=\Om_\pm (E^{\frac{1}{4}})$. The rigidity
result is therefore in accordance with the lower bound for $N_{fl}
(E)$.

Everything discussed so far has been concerned with $\De_3 (L;E)$
for $L\rto\infty$, that is with correlations in the spectra
on very large scales. It seems that in this regime the arithmetical
systems behave more like classically integrable ones than
like generic classically chaotic ones. It would now be interesting
to learn whether or not this similarity to the integrable case
pertains also to smaller scales, especially for the level
spacings. The first numerical calculations of quantum energies
for arithmetical systems are due to Schmit, who considered a
special symmetry class for the spectral problem related to the
regular octagon group. In \cite{Cuernavaca,BalazsVoros} he
obtained a level spacings distribution that revealed a level
attraction somewhat weaker than for a Poissonian $P(s)$, but the
--at that time expected-- Wigner surmise was clearly ruled out.
In addition it seemed that the computed $P(s)$ would the more
approach a Poissonian distribution the higher in energy one
went. Aurich and Steiner \cite{Aurich89} then calculated
eigenvalues in all symmetry classes for the regular octagon
group corresponding to one dimensional representations of the
symmetry group for $\Ga_{reg}\backslash\cH$, and obtained the
same findings as Schmit did. Hejhal was the first to compute
a considerable number of eigenvalues for the modular group
\cite{HejhalHua} and it was observed \cite{SteinerSantaBa}
that the corresponding level spacings behaved like the ones
for the regular octagon group. Later, Schmit \cite{Schmit91}
could calculate more eigenvalues for the odd symmetry class
on $\Ga_{mod}\backslash\cH$
and further confirmed the results for the modular group.
At that time, however, it remained unclear how the observed
violation of the RMT hypothesis could come about and what class
of systems would share alike properties.
Steil then computed \cite{Steil,BoStSt} 3167 eigenvalues for the
even symmetry class and 3475 eigenvalues for the odd symmetry class
comprising the complete spectrum up to $p=300$, i.e.\ in energy
up to $E=90\,000.25$, revealing that $P(s)$ can be rather well
described for high energies, corresponding to
$250\leq p\leq 300$, by a Poissonian distribution.
In \cite{BoStSt,BoGeGiSch} then the explanation was given that
the arithmetic properties of the Fuchsian groups $\Ga_{reg}$ and
$\Ga_{mod}$ involved were responsible for the exceptional
statistical properties observed, and that consequently all
arithmetical systems would share alike spectral statistics.
In contrast, non-arithmetic systems were considered in
\cite{Aurich90,Aurich92.1,Schmit91,BoGeGiSch} and found to be in good
agreement with a RMT behaviour of their energy spectra. Up to
now, however, no rigorous argument or quantitative heuristics
could be given that explains the observed phenomena.

It is our aim to present in the following two sections a simple
model that should account for the observed spectral
properties of the arithmetical systems. The key quantity
studied there in order to determine the level spacings
distribution $P(s)$ and the number variance $\Si^2 (L;x)$
is the {\it spectral form factor} $K(\tau;x)$
\cite{BerryA400} for the unfolded spectrum $\{x_i \}$.
It is defined as a Fourier transform of the pair correlation
of the energy fluctuations and will be introduced below. The
{\it pair correlation function} $g(t)$ is the two-point
correlation function of the spectral density $d(x)$,
\beq
\label{paircorr}
g(t):= <d(x-\frac{t}{2})\,d(x+\frac{t}{2})>\ ,
\eeq
where $<\dots>$ denotes a semiclassical averaging, as in section 2.
$g(t)$ then is the density function for the probability of finding
an energy level in the interval $(0,T)$, given one at $t=0$.
This can be used to construct the level spacings distribution
$P(s)$ as
\beq
\label{levelspacing}
P(s)=g(s)\,\exp\left[ -\int_0^s dt\ g(t)\right]\ ,
\eeq
see e.g.\ Porter's contribution in \cite{Porter}. Defining
$G(s):=\int_0^s dt\,g(t)$, one observes that $P(s)=-\frac{d}{ds}\,
e^{-G(s)}$, and therefore
\beq
\label{Pnorm}
\int_0^T ds\,P(s)=e^{-G(0)}-e^{-G(T)}\ ,\ \ \ \ \ T>0\ .
\eeq
Hence, as long as $g(s)$ is integrable on any finite interval
$[0,T]$, but with $\int_0^\infty dt\,g(t)=+\infty$,
$P(s)$ is a normalized probability density, $\int_0^\infty ds\,
P(s)=1$.

The interesting
contribution to $d(x)=1+d_{fl}(x)$ now comes from its
fluctuating part $d_{fl}(x)$. Inserting this splitting into
(\ref{paircorr}) for the pair correlation function, one is left
among others
with two terms of the form $<d_{fl}(x\pm\frac{t}{2})>$. It will
now be argued that these vanish in the semiclassical limit.
To support this idea one can go back to the regularization
(\ref{dflucreg}) for $d_{fl}(E)$. In section 2 it was demonstrated
that averaging over an interval of length $L$, $<d_{fl}(E)>$
vanishes like $L^{-1}$. Choosing now $\De x=x^a$
for the length of the interval involved in the semiclassical
averaging, which meets the prerequisit $1\ll \De x\ll x$ for
a small enough power $1>a>0$, $<d_{fl}(x\pm\frac{t}{2})>$ behaves
like $x^{-a}$ and thus vanishes in the semiclassical
limit $x\rto\infty$. Hence, in this limit,
\beq
\label{paircorrsc}
g(t)\sim 1+<d_{fl}(x-\frac{t}{2})\,d_{fl}(x+\frac{t}{2})> \ .
\eeq
$K(\tau;x)$ is now defined as the Fourier transform of the
correlation function on the r.h.s.\ of (\ref{paircorrsc}),
\beq
\label{Kvontau}
K(\tau;x):=\int_{-\infty}^{+\infty}dt\ e^{2\pi it\tau}<d_{fl}( x-
\frac{t}{2})\,d_{fl}( x+\frac{t}{2})>\ .
\eeq
Since this definition involves only $d_{fl}(x)$, which is related
via (\ref{dflucreg}) to the periodic orbits of the classical
system, the form factor is especially suited to be used in
periodic-orbit theory. It only remains to reexpress the pair
correlation function $g(t)$ in terms of $K(\tau;x)$ in order
to be prepared for a periodic-orbit analysis of the level
spacings distribution $P(s)$. To this end one inserts
$\de (x)=\int_{-\infty}^{+\infty}dz\ e^{2\pi ixz}$ into the
r.h.s.\ of the identity
\beq
\label{eq6}
<d_{fl}(x-\frac{t}{2})\,d_{fl}(x+\frac{t}{2})>
=\frac{1}{2t}\int_{-\infty}^{+\infty}dy\ y\,<d_{fl}(x-\frac{y}
{2})\,d_{fl}(x+\frac{y}{2})>\left[ \de (y-t)-
\de (y+t)\right] \ ,
\eeq
yielding after some simple manipulations
\beqa
\label{eq7}
<d_{fl}(x-\frac{t}{2})\,d_{fl}(x+\frac{t}{2})> &=&
-\frac{1}{2\pi t}\int_{-\infty}^{+\infty}d\tau\ \sin(2\pi t\tau)\
\cdot \nonumber \\   & & \cdot
\frac{\partial}{\partial\tau}\,\int_{-\infty}^{+\infty}dy\ e^{2\pi
i\tau y}\,<d_{fl}(x-\frac{y}{2})\,d_{fl}(x+\frac{y}{2})>\ .
\eeqa
Using (\ref{paircorrsc}) and (\ref{Kvontau}) then leads to
\beq
\label{paircorrK}
g(t)\sim 1-\frac{1}{\pi t}\,\int_0^{\infty}d\tau\ \sin(2\pi t\tau)\,
\frac{\partial}{\partial\tau}K(\tau;x)\ ,
\eeq
which is a fundamental expression to be used in the following
section.

In order to derive a periodic-orbit expression for the form
factor one has to rescale the unfolded spectrum by $d_{fl}(x)=
\frac{dE}{dx}\frac{d}{dE}N_{fl}(E)=\frac{1}{\db(E)}d_{fl}(E)$.
One can then use (\ref{dflucreg}) and insert it into the definition
of $K(\tau;x)$ expressed in terms of $E$,
\beq
\label{KvontauE}
K(\tau;E)=\frac{1}{\db(E)}\int_{-\infty}^{+\infty}d\la\ e^{2\pi i\tau
\db(E)\la}<d_{fl}(E-\frac{\la}{2})\,d_{fl}(E+\frac{\la}{2})>\ .
\eeq
Introducing the momentum variable $p$ for convenience, (\ref{dflucreg})
for the hyperbolic Laplacian on a surface $\GaH$ reads
\beqa
\label{dflucA}
d_{\ve,fl}(p)&=& \frac{1}{\pi}\sum_{\{l_n \}}\sum_{k=1}^\infty
                 A_{n,k}\,\cos(pkl_n)\,e^{-\frac{\ve^2}{4}k^2 l_n ^2}
                 \ , \nonumber \\
A_{n,k}      &:=& \frac{l_n}{2p}\frac{\chi_n^k\ g_p (l_n)}{|e^{\frac{k}
                 {2}l_n}-\si_n^k e^{-\frac{k}{2}l_n}|}\ .
\eeqa
The outer sum runs over all distinct lengths $l_n$ of primitive
closed geodesics, whereby their respective multiplicities have
been incorporated in the amplitude factors
$A_{n,k}$. The possibility to include {\it inverse hyperbolic orbits}
has been left open by allowing for $\si_n =-1$. As may be drawn
from the Selberg trace formula (\ref{SelbergTF}) ordinary closed
geodesics corresponding to hyperbolic $\ga\in \Ga$ are {\it
hyperbolic orbits}, i.e.\ possessing $\si_\ga =1$.
The consideration of inverse hyperbolic orbits is necessary
in order to be able to treat Artin's billiard, which is obtained
from $\Ga_{mod}\backslash\cH$ by dividing out the orientation
reversing symmetry $z\mapsto -\bar z$, thus yielding a billiard
system on the hyperbolic plane $\cH$. Choosing the standard
fundamental domain $\cF_{mod}=\{ z\in{\cH};\ |z|\geq~1,\ -\frac{1}{2}
\leq x\leq\frac{1}{2}\}$ for the modular group,
Artin's billiard takes place on $\cF_A =
\{ z\in{\cF_{mod}};\ x\geq 0\}$. Closed geodesics on $\Ga_{mod}
\backslash\cH$ that are invariant under the reflection $z\mapsto
-\bar z$ then result in inverse hyperbolic orbits on the billiard
domain, which are reflected an odd number of times on $\partial
\cF_A$, see \cite{Matthies} for further information. In addition,
representations $\chi:\ \Ga \rto\{\pm 1\}$ for the Fuchsian groups
$\Ga$ are admitted in (\ref{dflucA}). For Artin's billiard it turns
out that $\chi_n =\si_n$. In (\ref{dflucA}) it has further been
assumed that all $g_p (l_n )$ geodesics $\ga$ that share
the same length $l_n$ have identical $\chi_\ga =\chi_n$ and
$\si_\ga =\si_n$, respectively. In Artin's billiard this
requirement is indeed met \cite{Matthies}.

Using the regularized fluctuating part of the spectral density
(\ref{dflucA}) in (\ref{KvontauE}), and performing the limit
$\ve\rto 0$ at the end of the calculation, leads to
\beq
\label{Ksemicl}
K(\tau;E)\sim\frac{1}{(2\pi\db(E))^2}<\sum_{\{l_n\},\{l_m\}}
\sum_{r,s} {'} A_{n,r}A_{m,s}\,e^{ip(rl_n -sl_m)}\,\de (\tau-
\frac{rl_n +sl_m}{8\pi p\db(E)})>\ ,
\eeq
see \cite{BerryA400} and Berry's contribution in \cite{LesHouches}.
The inner sums extend over all non-zero integers, the prime
indicating the omission of $r,s=0$.

The r.h.s.\ of (\ref{Ksemicl}) consists of two double sums, each
running over the primitive closed geodesics and their repetitions.
The diagonal contribution of these two double sums is
\beq
\label{Kdiagonal}
K_D (\tau;E)=\frac{1}{(2\pi \db(E))^2}\sum_{\{l_n \}}\sum_{k=1}^\infty
A_{n,k}^2\,\de (\tau-\frac{kl_n}{4\pi p\db(E)})\ .
\eeq
Naively one would expect the non-diagonal terms in (\ref{Ksemicl})
to be washed out by the semiclassical averaging. This to work
would require large enough ``random'' phases $e^{ip(rl_n -sl_m)}$.
However, by going to longer orbits different lengths $l_n$
lie closer and closer due to their exponentially increasing density.
This effect can be drawn from the exponential growth of $\hat\cN_p (l)$.
Thus, for long orbits, the random phase argument fails, and
the diagonal approximation (\ref{Kdiagonal}) is only reasonable
up to some $\tau^*$. For small values of $\tau$, $\tau\leq\tau^*$,
only short orbits contribute to (\ref{Ksemicl}), and these are
well separated in length. In \cite{BerryA400,LesHouches}
Berry demonstrates that for $\tau\rto\infty$ the form factor
saturates, $K(\tau;E)\rto 1$. He further claims that even for
$\tau >1$ one obtains $K(\tau;E)\approx 1$. (Notice that by
(\ref{Kdiagonal}) already $\tau =1$ corresponds in the semiclassical
limit to long orbits, $l=4\pi p\db(E)$, and thus $\tau^* \ll 1$.)
It should be remarked, however, that very recently
Aurich and Sieber \cite{Aurich92x} found a violation
of this saturation for the form factor in cases where
there exists an eigenvalue $E_0 =0$ for the hyperbolic Laplacian.
They obtain an exponentially increasing $K(\tau;E)$ for $\tau
\rto\infty$ instead. In such cases the contribution to (\ref{Ksemicl})
coming from the zero mode has to be subtracted using the trace
formula, see \cite{Aurich92x}.

For small values of $\tau$, $\tau\leq\tau^*$, the diagonal approximation
(\ref{Kdiagonal}) indicates that $\de$--spikes determined by
individual lengths $l_n$ will characterize the form factor.
In this regime $K(\tau;E)$ therefore behaves non-universally.
The intermediate range $\tau^* <\tau\leq 1$ is then governed by
the contribution of rather long orbits. According to Berry
\cite{BerryA400,LesHouches} a sum rule of Hannay and Ozorio de
Almeida \cite{Hannay}, which exploits the uniform exploration of
phase space by long orbits, yields $K(\tau;E)\approx\overline{g}
\tau$ for $\tau^* <\tau\leq 1$. Thereby it is assumed that the
multiplicities of lengths of primitive orbits approach
$\overline{g}=const.$ asymptotically for long orbits. For
generic time-reversal invariant systems thus $\overline{g}=2$,
whereas generic systems without time-reversal symmetry show
$\overline{g}=1$. Obviously this sum rule cannot be applied to
arithmetic systems because the exponentially increasing
multiplicities of lengths prohibit $<g_p (l)>$ to approach a
constant. One would, however, expect from the sum rule that
$K(\tau;E)$ grows much faster for the arithmetic systems
than generically, see also \cite{BoGeGiSch}. This observation
lies at the heart of the model that will be presented in the
next two sections.

\subsection{A Model for the Level Spacings Distribution}
It will now be attempted to set up a model describing the
statistical properties of the eigenvalues of hyperbolic
Laplacians on surfaces with arithmetic Fuchsian groups. The
main tool to be employed will be the spectral form factor
$K(\tau;E)$. Then, as explained in the preceding section,
the relations (\ref{levelspacing}) and (\ref{paircorrK})
allow to determine the level spacings distribution $P(s)$
once $K(\tau;E)$ is known in sufficient detail. Following
Berry's reasoning as reviewed in section 4.2, the form
factor can be substituted by its diagonal approximation
$K_D (\tau;E)$ (\ref{Kdiagonal}) for $\tau\leq\tau^* \ll 1$.
{}From now on $\tau^*$ will be fixed at some value so that
$K_D (\tau;E)$ approximates the complete form factor sufficiently
well. Then only $K_D (\tau;E)$ will be used for $\tau\leq\tau^*$.

The semiclassical limit $E\rto\infty$ can for fixed $\tau$
also be viewed as the limit of long orbits, $l\rto\infty$, as
can be drawn e.g.\ from (\ref{Kdiagonal}). Thus the sum in
(\ref{Kdiagonal}) will be evaluated in the asymptotic regime
$l\rto\infty$. In particular the multiplicities $g_p (l_n )$
appearing in the amplitude factors $A_{n,k}$ (\ref{dflucA})
will be replaced by their asymptotic values $c_\Ga \frac{e^{l_n /2}}
{l_n}$. Therefore one obtains
\beq
\label{ampasymptotic}
A_{n,k}\sim\frac{c_\Ga}{2p}\,\chi_n^k \,e^{\frac{1}{2}(1-k)l_n}\,
[1+O(e^{-kl_n})]\ ,\ \ \ \ \ l_n \rto\infty\ .
\eeq
As always when dealing with periodic-orbit sums one observes
also here that the $(k=1)$--contribution to the sum over
repetitions of primitive orbits is the leading one for $l_n \rto
\infty$. Thus, asymptotically in the semiclassical limit,
one finds that
\beq
\label{KDdelta}
K_D (\tau;E)\sim \frac{c_\Ga^2}{(4\pi p\db(E))^2}\sum_{\{l_n\}}
\de(\tau-\frac{l_n}{4\pi p\db(E)})\ .
\eeq
To get rid of the hardly tractable Dirac--$\de$'s one integrates
(\ref{KDdelta}),
\beq
\label{KNhat}
\int_0^\tau dt\ K_D (t;E)\sim\frac{c_\Ga^2}{(4\pi p\db(E))^2}\,
\hat\cN_p (4\pi p\db(E)\tau)\ .
\eeq
Introducing the asymptotic behaviour $\hat\cN_p (l)\sim\frac{2}
{c_\Ga}\,e^{l/2}$, $l\rto\infty$, and differentiating the result
with respect to $\tau$ yields
\beq
\label{KDexp}
K_D (\tau;E)\sim\frac{c_\Ga}{4\pi p\db(E)}\,e^{2\pi p\db(E)\tau}\ ,
\eeq
compare also \cite{BoGeGiSch}. Relation (\ref{KDexp}) shows that
the diagonal approximation $K_D (\tau;E)$ grows exponentially and
already at a value of $\tau_0 := \frac{1}{2\pi p\db(E)}\log
(\frac{4\pi}{c_\Ga}p\db(E))$ it has reached the value one.
$\tau_0$ being a function of $E$ decreases in the semiclassical
limit and above some energy $E=\hat E$ it is smaller than the fixed
value $\tau^*$. Hence $K_D (\tau;E)$ can be taken as an
approximation for the complete form factor $K(\tau;E)$ in the
whole range $[0,\tau_0]$ once the energy is chosen
sufficiently large, $E\geq \hat E$. In conclusion, the
exponential behaviour (\ref{KDexp}) semiclassically describes
the complete form factor in the mean up to $\tau_0$.

According to Berry's investigation of the form factor
\cite{BerryA400,LesHouches} this approaches one for $\tau\rto
\infty$. Since even for $\tau =1$ mainly long orbits
contribute, he argues that $K(\tau;E)\approx 1$ for $\tau\geq 1$,
and then takes $K(\tau;E)\equiv 1$ to model the actual
form factor in this domain, leading to his result for the spectral
rigidity. In the sequel we will proceed analogously and
define a model form factor
\beq
\label{Kmodel}
K_M (\tau;E):=\left\{ \begin{array}{ccl} \frac{c_\Ga}{4\pi p\db(E)}
\,e^{2\pi p\db(E)\tau} &,& \tau\leq\tau_0 \\1 &,& \tau>\tau_0
\end{array}\right. \ .
\eeq
In addition to Berry's reasoning $K_M (\tau;E)$ is defined
to be one also on the interval $[\tau_0 ,1]$. This may be justified
by the finding that $K_D (\tau;E)$  already reaches the value one
at $\tau_0$, whereas in the generic case Berry considers the
form factor has to be interpolated on the interval $[\tau^* ,1]$
by the result obtained from the sum rule of \cite{Hannay}, as
described in section 4.2. Put in sloppy terms, the exponential increase
(\ref{KDexp}) somehow exempts one from the need to discuss the
interval $[\tau^* ,1]$ separately. Inspecting (\ref{KDdelta})
one notes that the definition (\ref{Kmodel}) of $K_M (\tau;E)$
results in cutting-off the periodic-orbit sum at $l_{max}=4\pi
p\db(E)\tau_0 =2\log(\frac{4\pi}{c_\Ga}p\db(E))$. In the semiclassical
limit the cut-off is therefore being removed automatically.
Certainly, the actual form factor
will oscillate about the mean value described by (\ref{Kmodel}).
This fine structure is left out in the model and thus one cannot
really expect the resulting model distribution $P_M (s)$ to describe
the actual level spacings in full detail. However, the model should
reproduce the facts at least qualitatively.

Inserting (\ref{Kmodel}) into (\ref{paircorrK}) yields the
model pair correlation function
\beqa
\label{paircorrmodel}
g_M (t)&=& 1-\frac{c_\Ga}{2\pi t}\int_0^{\tau_0}d\tau\ \sin(2\pi
           t\tau)\,e^{2\pi p\db(E)\tau} \nonumber \\
       &=& 1-\frac{1}{\pi t}\frac{p\db(E)}{(p\db(E))^2 +t^2 }\left\{
           p\db(E)\sin(2\pi t\tau_0)-t\cos(2\pi t\tau_0)+\frac{c_\Ga}
           {4\pi}\frac{t}{p\db(E)}\right\} \\
       &=& 1-\frac{1}{\pi}\frac{p\db(E)}{(p\db(E))^2 +t^2 }\left\{
           \log(\frac{4\pi}{c_\Ga}p\db(E))-1+\frac{c_\Ga}{4\pi}
           \frac{1}{p\db(E)}+O((\frac{\log p\db(E)}{p\db(E)})^2 )
           \right\} \ . \nonumber
\eeqa
Integrating this result, $G_M (s)=\int_0^s dt\,g_M (t)$, yields
\beq
\label{Gmodel}
G_M (s)\sim s-\frac{1}{\pi}\left\{ \log(\frac{4\pi}{c_\Ga}p\db(E))
-1+\frac{c_\Ga}{4\pi}\frac{1}{p\db(E)}\right\}\arctan(\frac{s}{p\db(E)})
\ .
\eeq
The model level spacings distribution is then obtained as $P_M (s)=
g_M (s)\,e^{-G_M (s)}$,
\beqa
\label{spacemodel}
P_M (s)&\sim& \left\{ 1-\frac{1}{\pi}\frac{p\db(E)}{(p\db(E))^2
              +s^2}\left[ \log(\frac{4\pi}{c_\Ga}p\db(E))-1
              +\frac{c_\Ga}{4\pi}\frac{1}{p\db(E)}\right]\right\}
              \cdot\nonumber \\
          & & \cdot\exp\left\{ -s+\frac{1}{\pi}\left[ \log(\frac{4\pi}
              {c_\Ga}p\db(E))-1+\frac{c_\Ga}{4\pi}\frac{1}{p\db(E)}
              \right] \arctan(\frac{s}{p\db(E)})\right\}\ .
\eeqa
Consulting (\ref{Pnorm}), and $G_M (s)
\rto\infty$ for $s\rto\infty$, one obtains that $P_M (s)$
is a normalized distribution. Moreover, for $E\rto\infty$
one finds that $g_M (s)\rto 1$ and $G_M (s)\rto s$, leading to the
observation that in the semiclassical limit $P_M (s)$ approaches
a Poissonian distribution. Since $P_M (s)$ is designed to reproduce
the actual level spacings distribution for $E\rto\infty$, also
the latter is expected to converge to a Poissonian.

The model shows a level attraction, which is for finite $E$ weaker
than that one of a pure Poissonian spectrum, since for $s\rto 0$
one finds
\beqa
\label{modelrepulsion}
P_M (s)&\sim& g_M (0)+\left[ g_M '(0)-g_M (0)^2 \right] s\nonumber\\
       &\sim& 1-\frac{1}{\pi p\db(E)}\left[ \log(\frac{4\pi}{c_\Ga}
              p\db(E))-1+\frac{c_\Ga}{4\pi}\frac{1}{p\db(E)}\right] \\
       &    & -\left\{ 1-\frac{1}{\pi p\db(E)}\left[ \log(\frac{4\pi}
              {c_\Ga}p\db(E))-1+\frac{c_\Ga}{4\pi}\frac{1}{p\db(E)}
              \right]\right\}^2 s \ . \nonumber
\eeqa
Thus $P_M (0)<1$ for finite $E$, and also $0>P_M '(0)>-1$. In the
semiclassical limit the strength of the level attraction then approaches
the Poissonian result $P(s)\sim 1-s$. The model hence qualitatively
reproduces the numerical findings in
\cite{Cuernavaca,BalazsVoros,Aurich89,BoStSt,BoGeGiSch}
correctly.

\begin{figure}[tb]
\epsfbox[0 0 557 390]{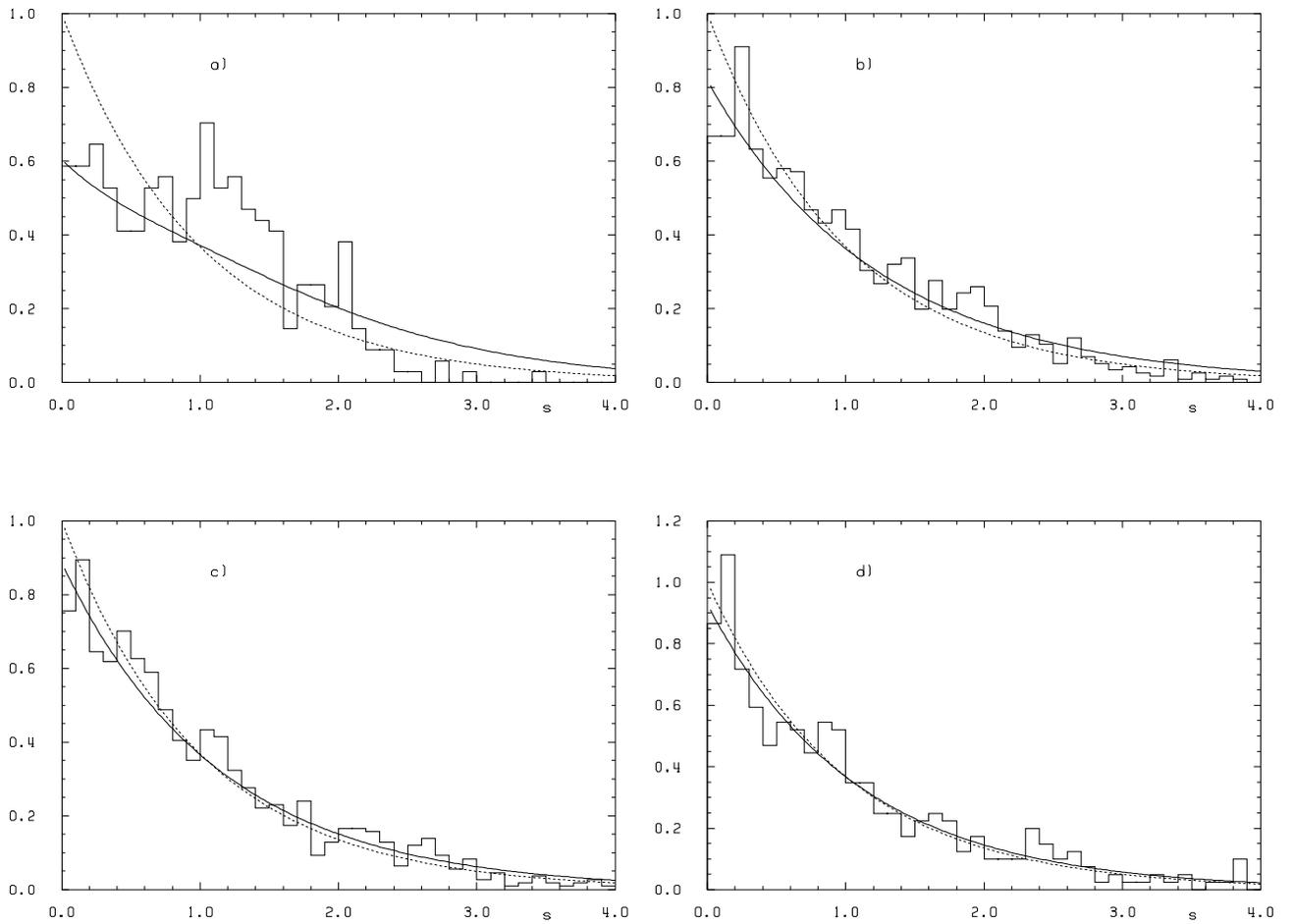}
\caption{\label{fig3}
The model $P_M (s)$ applied to the odd symmetry class
of Artin's billiard is shown as the full curves
compared to the actual level spacings distributions in the
intervals a) $0\leq p\leq 100$, b) $100\leq p\leq 200$,
c) $250\leq p\leq 300$, and
d) $500\leq p\leq 510$. The dotted curves show Poissonian
distributions.}
\end{figure}
Using the modular group as an example it will now be studied
how well the model describes the actual level spacings
distribution of an arithmetical system quantitatively. The
energy eigenvalues for the two symmetry classes occurring for the
modular group have been kindly placed at disposal by Gunther Steil,
see also \cite{Steil}. The mean spectral density for the odd
symmetry class of Artin's billiard reads \cite{Matthies}
\beq
\db_- (E)\sim \frac{1}{24}-\frac{1}{8\pi}\frac{\log E}{\sqrt{E}}
-\frac{3\log 2}{8\pi}\frac{1}{\sqrt{E}}\ ,
\eeq
and for the even symmetry class
\beq
\db_+ (E)\sim \frac{1}{24}-\frac{3}{8\pi}\frac{\log E}{\sqrt{E}}
-\frac{\log2 -4\log \pi}{8\pi}\frac{1}{\sqrt{E}}\ .
\eeq
As a final input the constant $c_\Ga =1$ of $<g_p (l)>\sim c_\Ga
\frac{e^{l/2}}{l}$ is needed. Fig.3 presents histograms of the
level spacings in the odd symmetry class
for the four momentum-intervals $0\leq p\leq 100$, $100\leq p
\leq 200$, $250\leq p\leq 300$, and $500\leq p\leq 510$,
comprising of $341$, $1157$, $1093$, and $409$ levels, respectively.
The momenta $p$ that have been used in the model (\ref{spacemodel}),
which is shown as the full curves,
\begin{figure}[bt]
\epsfbox[0 0 557 390]{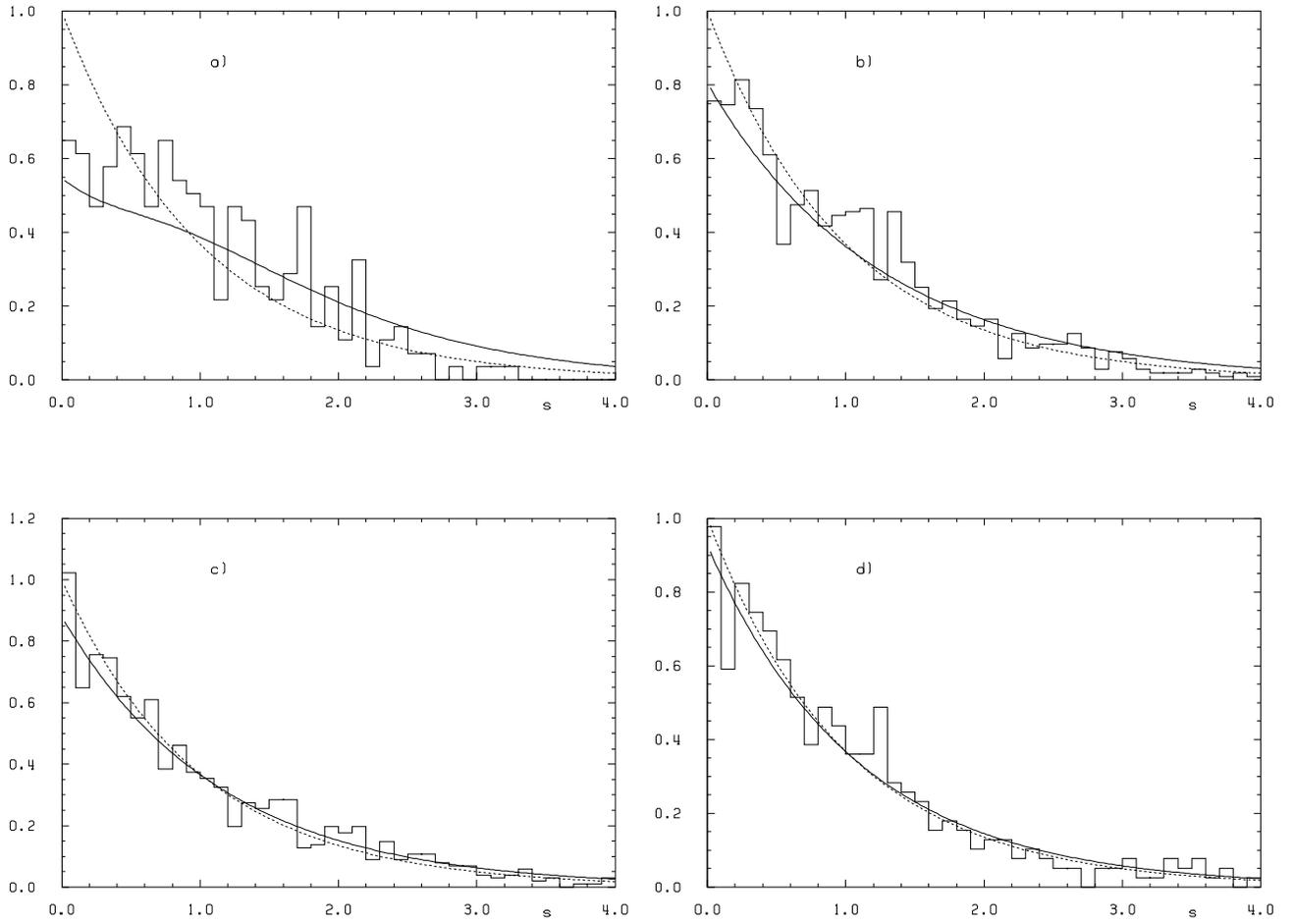}
\caption{\label{fig4}
The same as in fig.3, but for the even symmetry class.}
\end{figure}
have been chosen in the middle of each interval, i.e.\ $p=50$,
$p=150$, $p=275$, and $p=505$, respectively. One now observes from
fig.3 that the model $P_M (s)$ approximates the actual distributions
reasonably well, and that the quality of the approximation grows
with energy. Especially, the strength $P(0)$ of the level attraction
is reproduced rather well, even in the low energy regime.
Apparently the histograms as well as the model the more approach
a Poissonian distribution the higher in energy one goes.
Fig.4 contains the same information as fig.3, but for the even symmetry
class. Here the respective momentum-intervals contain $277$,
$1040$, $1026$, and $395$ levels. From fig.4
one observes the same level of agreement of the model with the
data as for the odd symmetry class.

\subsection{A Model for the Number Variance}
The discussion in the preceding section has revealed that the
statistical properties of quantum energy spectra in arithmetical
chaos can on small scales and in the semiclassical limit be described
by those of a Poissonian spectrum. The present section now provides
a continuation of that study to medium- and long-range
correlations in arithmetical quantum energy spectra. In
chapter 2 the spectral rigidity $\De_3 (L;x)$ has been introduced
as a means to investigate spectral statistics on scales $L>1$
that take several levels into account. For the following, however,
it proves useful to study a different quantity that principally
provides the same information on spectral correlations;
this is the {\it number variance} $\Si^2 (L;x)$,
defined as the variance of the distribution of the numbers
$n(L;x)=N(x+L)-N(x)$ of levels in intervals $[x,x+L]$,
\beq
\label{defvariance}
\Si^2 (L;x):=<[n(L;x)-L]^2 >\ ,
\eeq
where $<\dots >$ as usual denotes a semiclassical averaging over
$x$. A truely Poissonian spectrum excels by a linear number variance,
$\Si^2 (L;x)=L$, whereas the GOE in RMT possesses a $\Si^2 (L;x)$
that is asymptotically given by $\Si^2 (L;x)\sim L$ for $L\rto 0$,
and by $\Si^2 (L;x)\sim \frac{2}{\pi^2}[\log(2\pi L)+\ga +1-\frac{\pi^2}
{8}]$ for $L\rto\infty$; here $\ga$ denotes Euler's constant. Further
details can be found e.g.\ in Bohigas' contribution in
\cite{LesHouches}.

In \cite{BerryNL,LesHouches} Berry presents a semiclassical treatment
of the number variance very much in the spirit of his considerations
of the spectral rigidity \cite{BerryA400}. In the semiclassical
limit he expresses the number variance through the form factor
by
\beq
\label{varianceform}
\Si^2 (L;E)\sim \frac{2}{\pi^2}\int_0^\infty \frac{d\tau}{\tau^2}
\sin^2 (\pi L\tau)\,K(\tau;E)\ .
\eeq
Using the conclusions Berry draws for the functional form of
$K(\tau;E)$ one obtains in the intermediate $L$-range
$1\ll L\ll L_{max}$ that the number variance is given by the GOE result,
whereas for $L\gg L_{max}$ it oscillates non-universally about
a saturation value $\Si^2_\infty (E)$. These oscillations as well
as the value of $\Si^2_\infty (E)$ are determined by the
contributions of the short periodic orbits to the form factor.
As can be drawn from the relation
\beq
\label{rigvar}
\De_3 (\frac{L}{2};E)=\frac{2}{L^4}\int_0^L dr\ \Si^2 (r;E)\,
[ L^3 -2L^2 r+r^3 ]
\eeq
of the spectral rigidity to the number variance, the saturation
values of both quantities satisfy $\Si^2_\infty (E)=2\De_\infty
(E)$, see \cite{Aurich90}. Hence the number variance can also
be used to find out about the asymptotic energy dependence
of $<N_{fl}(E)^2 >=\De_\infty (E)=\frac{1}{2}\Si^2_\infty (E)$
for $E\rto\infty$.

The relation (\ref{varianceform}) now easily allows for an application
of the model from the preceding section also to the number variance.
It turns out, however, that it is possible to improve the form factor
for the model a little in that the complete diagonal approximation
is taken into account for $\tau\leq\tau_0$. Therefore
\beq
\label{modmodel}
\hat K_M (\tau;E):=\left\{ \begin{array}{ccl} K_D (\tau;E) &,&
\tau\leq\tau_0 \\ 1 &,& \tau >\tau_0 \end{array}\right.\ ,
\eeq
with the diagonal term $K_D (\tau;E)$ taken from (\ref{Kdiagonal}),
will be used in (\ref{varianceform}) to yield a model $\Si^2_M (L;
E)$ for the number variance in arithmetical chaos. $\Si_M^2 (L;E)$
then consists of two contributions $\Si^2_{M,1}(L;E)$ and $\Si^2_{M,2}
(L;E)$, the first one being derived from the integration in $\tau$
along $[0,\tau_0]$, and the second one resulting
from the respective integration along the remaining interval
$[\tau_0 ,\infty]$. These integrals can be performed exactly,
yielding
\beqa
\label{Si1and2}
\Si^2_{M,1}(L;E) &=& \frac{8p^2}{\pi^2}\sum_{\{l_n \}}\sum_{k\geq 1
\atop kl_n \leq l_{max}}\frac{A_{n,k}^2}{(kl_n)^2}
\sin^2 \left( \frac{kl_n L}{4p\db(E)} \right)\ ,\nonumber \\
\Si^2_{M,2}(L;E) &=& \frac{1}{\pi^2 \tau_0}-\frac{\cos (2\pi L\tau_0)}
{\pi^2 \tau_0}-\frac{2L}{\pi}\mbox{Si}\,(2\pi L\tau_0) +L \ .
\eeqa
$l_{max}:=4\pi p\db(E)\tau_0$, and
$\mbox{Si}\,(x)=\int_0^x dt\,\frac{\sin t}{t}$ denotes the
sine integral, see e.g.\ \cite{Bateman}. The amplitude factor
$A_{n,k}$ is given by (\ref{dflucA}). Because of the exponential
vanishing of $A_{n,k}$ for $k\geq 2$ and $l_n \rto\infty$
(\ref{ampasymptotic}) only the $(k=1)$--contribution from the
summation over $k$ in the periodic orbit sums will be considered
for the further analytic investigations; for the numerics, however,
the complete double summation of (\ref{Si1and2}) will be used.

In the limit $L\rto 0$ one can expand the sine in (\ref{Si1and2})
and then use the asymptotic value (\ref{ampasymptotic}) for
$A_{n,k}$, yielding
\beq
\label{Si1L0}
\Si^2_{M,1}(L;E)\sim \frac{c_\Ga^2}{8\pi^2 p^2 \db(E)^2}\,L^2 \,
\hat\cN_p (4\pi p\db(E)\tau_0 )+O(L^4 )\ .
\eeq
Employing the (semiclassical) asymptotics $\hat\cN_p (l)\sim\frac{2}
{c_\Ga}\,e^{l/2}$, $l\rto\infty$, and expanding $\Si^2_{M,2}(L;E)$
for $L\rto 0$ leaves one with
\beq
\label{SiL0}
\Si^2_M (L;E)\sim L-\frac{1}{\pi p\db(E)}\left[ \log(\frac{4\pi}
{c_\Ga}p\db(E))-1\right]\,L^2+O(L^3 )\ .
\eeq
Thereby the definition $\tau_0 =\frac{1}{2\pi p\db(E)}
\log(\frac{4\pi}{c_\Ga}p\db(E))$ has been employed. Thus, for $L\rto
0$, the model $\Si^2_M (L;E)$ follows in lowest order a Poissonian
number variance. For finite values of $L$ and large enough energies,
however, it is smaller than the latter.

\begin{figure}[hbt]
\epsfbox[0 0 557 390]{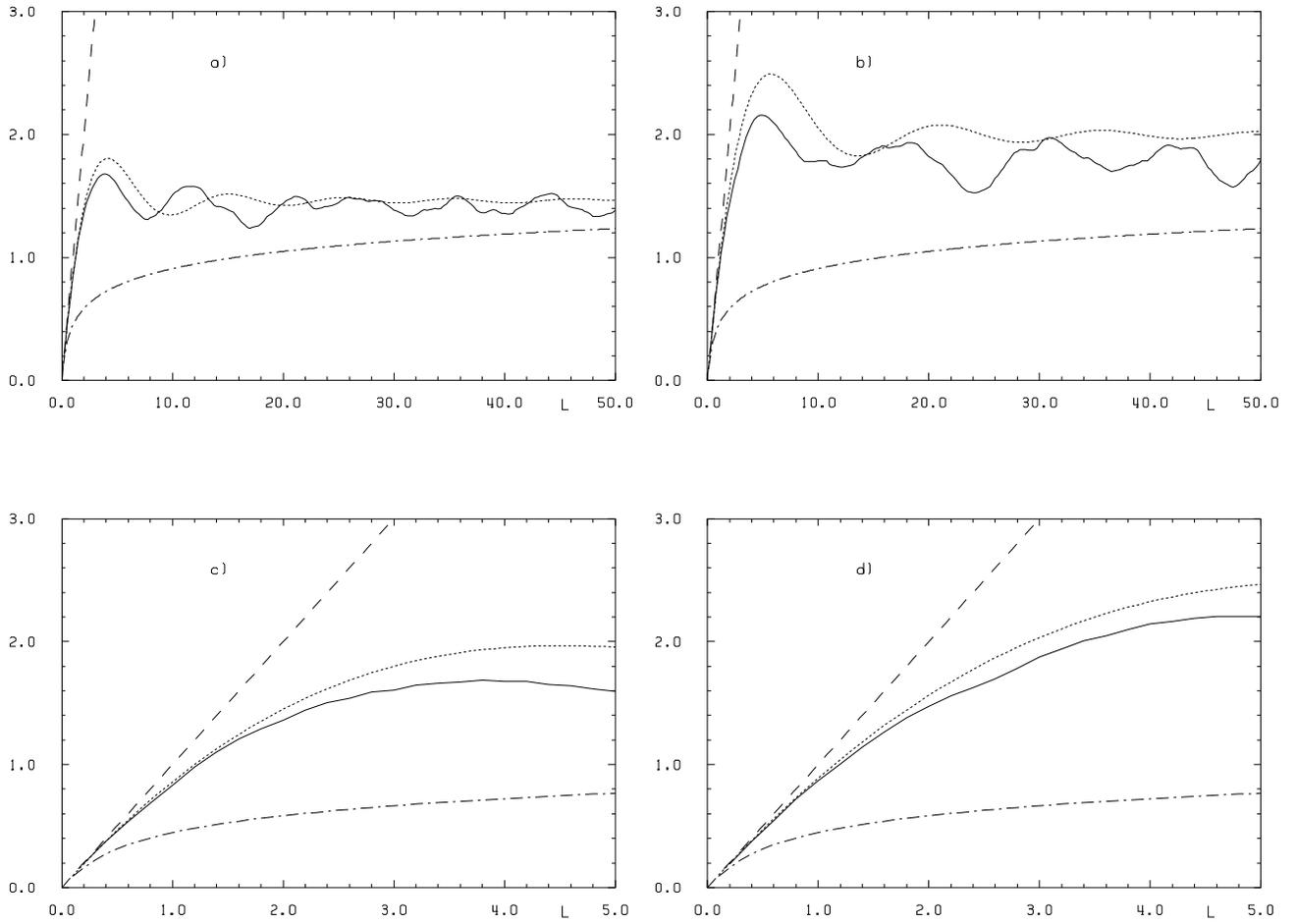}
\caption{\label{fig5} The number variances obtained from the
eigenvalues are given as the full curves in comparison with the model,
which is shown as the dotted curves. The Poissonian
(dashed curves) and GOE (dashed-dotted curves) results are added.
a) and c) refer to the interval $164.92\leq p\leq 229.70$, b) and d),
however, to $250.86\leq p\leq 296.83$.}
\end{figure}
Fig.\ref{fig5} presents as the dotted curves
a numerical evaluation of (\ref{Si1and2})
for the odd symmetry class of Artin's billiard, and compares
these to the number variance obtained from the
quantum energies computed by Steil \cite{Steil}. The full curves
in \ref{fig5}a) and c) refer to a sample of eigenvalues between
the $1000^{th}$ and the $2000^{th}$ one, whereas \ref{fig5}b)
and d) are obtained from the $2400^{th}$ up to the $3400^{th}$
eigenvalue. The respective momentum intervals are $164.92 \leq p
\leq 229.70$ and $250.86 \leq p\leq 296.83$. The model
(\ref{Si1and2}) has been evaluated using momentum values from
the middle of each interval, namely $p=200$ and $p=270$,
respectively. The dashed and the dashed-dotted curves provide a
comparison with Poissonian and GOE number variances, respectively.
One observes from fig.\ref{fig5} that the model reproduces the
small--$L$ behaviour as well as the saturation of the actual
number variance reasonably well. It fails, however, to describe
the oscillations properly; but due to the simplicity of the
assumed form factor (\ref{modmodel}) no better agreement of the
model with reality could actually be expected. A further
conclusion that can be drawn from fig.\ref{fig5} is that
$\Si^2 (L;E)$ leaves the Poissonian form factor already for
rather small values of $L$. The small--$L$ asymptotics
(\ref{SiL0}) of $\Si^2_M (L;E)$, however, reveals that the
(negative) coefficient of $L^2$ vanishes for $E\rto\infty$.
Thus the small--$L$ behaviour of the model is Poissonian-like
on the larger $L$--intervals the higher in energy one goes.
This Poissonian behaviour of the number variance on small scales
is in accordance with the findings about the level spacings
distributions in the preceding section. Moreover, the rate of
approaching a Poissonian distribution observed there is the same as the
respective rate here, compare (\ref{SiL0}) with (\ref{spacemodel})
and (\ref{modelrepulsion}).

A further question that e.g.\ arises in the context of inverse
quantum chaology as employed in sections 3.6 and 4.5 is the
one for the energy dependence of the saturation value $\De_\infty
(E)=\frac{1}{2}\Si^2_\infty (E)$. In the following therefore
the limit $L\rto\infty$ will be studied for the number variance
$\Si^2_M (L;E)$ derived from the model introduced above. The second
contribution $\Si_{M,2}^2 (L;E)$ immediately yields
\beq
\label{Si2Linf}
\Si^2_{M,2} (L;E)=\frac{1}{\pi^2 \tau_0}+O(L^{-2})\ ,\ \ \ \ \
L\rto\infty\ ,
\eeq
when using the asymptotic behaviour $\mbox{Si}\,(x)=\frac{\pi}{2}
-\frac{\cos x}{x}+O(\frac{1}{x^2})$ for $x\rto\infty$.
The periodic-orbit term $\Si_{M,1}^2 (L;E)$ contributes oscillations
to the large--$L$ asymptotics of the model number variance
caused by the $\sin^2$'s. For fixed $E$ the sum represents a
superposition of finitely many oscillations of incommensurable
wave lengths. In order to obtain the average value about which
this superposition of oscillations fluctuates, one replaces
each $\sin^2$ by its mean value $\frac{1}{2}$. Therefore
\beqa
\label{Si1Linf}
<\Si_{M,1}^2 (L;E)> &\sim& \frac{c_\Ga^2}{\pi^2}\sum_{l_n \leq l_{max}}
                       \frac{1}{l_n^2} \nonumber \\
                & =  & \frac{c_\Ga^2}{\pi^2}\int_{l_1}^{l_{max}}
                       \frac{d\hat\cN_p (l)}{l^2} \ .
\eeqa
Since $l_{max}=4\pi p\db(E)\tau_0 =2\log(\frac{4\pi}{c_\Ga}p\db(E))\rto
\infty$ in the semiclassical limit, one can introduce the asymptotics
$\hat\cN_p (l)\sim\frac{2}{c_\Ga}\,e^{l/2}$, $l\rto\infty$, on
the r.h.s.\ of (\ref{Si1Linf}),
\beqa
\label{Si2saturation}
<\Si_{M,1}^2 (L;E)>&\sim&\frac{c_\Ga}{2\pi^2}\int_{\frac{l_1}{2}}^{
                    \frac{l_{max}}{2}}\frac{dt}{t^2}\ e^{-t}\nonumber \\
               &=& \frac{c_\Ga}{2\pi^2}\,Ei\left( \log\left(\frac{4\pi}
                   {c_\Ga}p\db(E)\right)\right)-\frac{2}{\pi}\,
                   \frac{p\db(E)}{\log(\frac{4\pi}{c_\Ga}p\db(E))}
                   +C(l_1 )\ .
\eeqa
$C(l_1 ):=\frac{c_\Ga}{2\pi^2}[2\frac{e^{l_1 /2}}{l_1}-Ei(\frac{l_1}
{2})]$ is an energy independent constant determined by the shortest
primitive length $l_1$ on $\GaH$.
The contribution (\ref{Si2Linf}) to the saturation
value coming from $\Si_{M,2}^2 (L;E)$ now exactly cancels the
second term on the r.h.s.\  of (\ref{Si2saturation}),
\beqa
\label{Sisaturation}
\Si_{M,\infty}^2 (E) &\sim& \frac{c_\Ga}{2\pi^2}\,Ei\left(\log\left(
                            \frac{4\pi}{c_\Ga}p\db(E)\right)\right)
                            +C(l_1) \nonumber \\
                     &\sim& \frac{2}{\pi}\,\frac{p\db(E)}{\log(
                            \frac{4\pi}{c_\Ga}p\db(E))}\ ,\ \ \ \ \
                            E\rto\infty\ .
\eeqa
The energy dependence of $\De_\infty (E)$ derived from the
model thus is
\beq
\label{deltainfmodel}
\De_{M,\infty} (E)\sim \frac{2\db}{\pi}\,\frac{\sqrt{E}}{\log E}
\ ,\ \ \ \ \ E\rto\infty\ .
\eeq
\begin{figure}[bt]
\epsfbox[0 0 557 290]{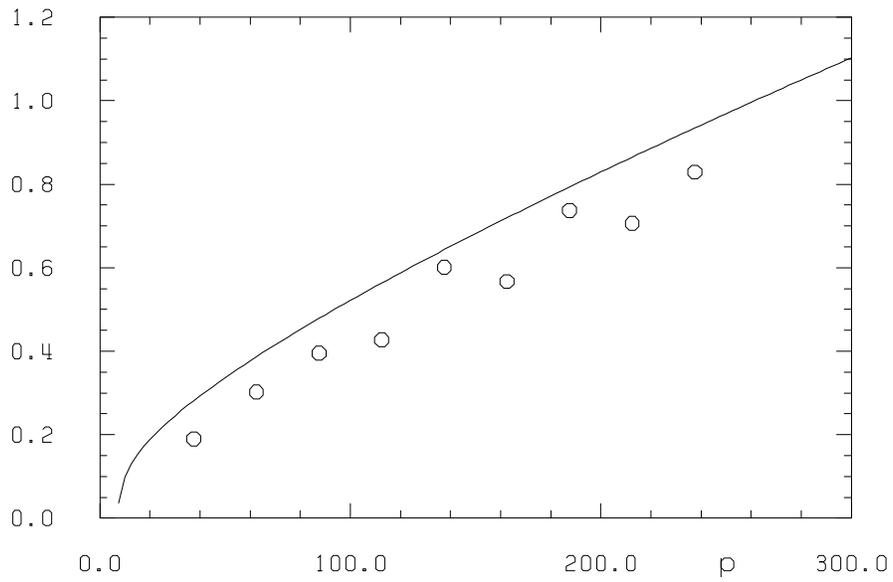}
\caption{\label{fig6} The saturation value $\frac{1}{2}\Si^2_{M,\infty}
(p)$ of the model number variance for the odd symmetry class of
Artin's billiard is shown as the full curve. The dots mark
$\frac{1}{2}\Si^2_\infty (p)$ obtained from the eigenvalues, see
\protect\cite{Steil}.}
\end{figure}
This result should be compared with the rigorous lower bound
$N_{fl}(E(p))=S(p)=\Om_\pm (\frac{\sqrt{p}}{\log p})$ \cite{Hejhal}
for arithmetic groups. Via (\ref{deltainf}) the latter yields
\beq
\label{deltainflb}
\De_\infty (E)=\Om\left( \frac{\sqrt{E}}{(\log E)^2} \right)\ .
\eeq
This being a lower bound is well in accordance with the result
(\ref{deltainfmodel}) obtained from the model. Given the latter
describes the actual saturation value of $\Si^2 (L;E)$
correctly in the semiclassical limit, this means that the lower
bound (\ref{deltainflb}) is off the true magnitude only by a
factor of $\log E$. The upper bound $S(p)=O(\frac{p}{\log p})$,
yielding $\De_\infty (E)=O(\frac{E}{(\log E)^2 })$, is therefore
much less sharp than (\ref{deltainflb}).

In order to test the model, the r.h.s.\ of the first line of
(\ref{Sisaturation}) has been evaluated numerically and
multiplied by $\frac{1}{2}$ for the odd symmetry class of Artin's
billiard. Fig.\ref{fig6} presents the result as the full
curve. Steil has computed $\De_\infty (E)$ from his numerically
obtained quantum energies, see also \cite{Steil}. These values
are given as dots in fig.\ref{fig6}.
One notes from fig.\ref{fig6} that the model appears to reproduce the
functional form of $\De_\infty (E)$ correctly, but that the constant
$C(l_1)=0.075$ seems to come out a little bit too large.

Finally the above observation should be compared to very recent
rigorous results of Sarnak and Luo \cite{SarnakLuo}. They
define the number variance as
\beq
\label{SiSL}
\Si^2_{SL}(L;x):=\frac{1}{x}\int_x^{2x}dx'\ \left[ N(x'+L)
-N(x')-L\right]^2
\eeq
for $x\rto\infty$, and obtain the following \\

\noindent {\sc Theorem:} In the range $\frac{\sqrt{x}}{\log x}
\ll L\ll L_{max}$, $L_{max}\propto \sqrt{x}$, the estimates
\beqa
\label{theoremSL}
\Si^2_{SL}(L;x) &=& \Om \left( \frac{L}{\log L}\right) \nonumber \\
\Si^2_{SL}(L;x) &=& \Om \left( \frac{\sqrt{x}}{\log x}\right)
\eeqa
hold for all arithmetic Fuchsian groups.\\

The $L$--range these bounds refer to is the upper part of
Berry's universal regime $1\ll L\ll L_{max}$. The first line
bounding the $L$--dependence is, if taken as sharp,
in accordance with the numerical
observations as well as with our model in that in any case
$\Si^2 (L;E)\leq L$ on the interval alluded to above. A direct
comparison of the model (\ref{SiL0}) and the first line of
(\ref{theoremSL}) is not possible because (\ref{SiL0}) is
only valid for $L\rto 0$. Since $\Si_M^2 (L;E)$ is certainly
positive, an approximation using the r.h.s.\ of (\ref{SiL0})
up to $O(L^2 )$ is restricted to
\beq
\label{Lorelei}
L\leq \frac{\pi p\db (E)}{\log (\frac{4\pi}{c_\Ga}p\db(E))-1}\ .
\eeq
Thus there is no overlap with the range of validity of the
Theorem. Those two results should rather be viewed as being
complementary.
The second line of the Theorem estimating the $E$--dependence
at finite $L$ is in accordance with the above findings about
$\Si^2_{M,\infty}(E)$ and, of course, with the rigorous estimate
(\ref{deltainflb}), although a direct comparison is also, strictly
speaking, prohibited by the restriction $L\ll L_{max}$ in the Theorem.

Summarizing the results on the number variance of arithmetical
systems obtained in this section one notices that $\Si^2 (L;E)$
is for small values of $L$ reasonably well approximated by a Poissonian
behaviour. The $L$--range on which this agreement takes place
grows in the semiclassical limit. For larger values of $L$
$\Si^2 (L;E)$ deviates slowly from being linear as first described
by the model (\ref{SiL0}) and then by the Theorem of Sarnak
and Luo. At some $L_{max}\propto\sqrt{E}$ the number variance
saturates and oscillates non-universally beyond that value. The
oscillations and their mean value are determined by the short
closed geodesics on the surface
$\GaH$. The number of distinct lengths contributing,
however, is $\hat\cN_p (l_{max})\sim \frac{8\pi}{c_\Ga^2}p\db(E)$
and therefore tends to infinity for $E\rto\infty$. The saturation
value $<N_{fl}(E)^2 >=\De_\infty (E)=\frac{1}{2}\Si^2_\infty (E)
\sim \frac{2\db}{\pi}\,\frac{\sqrt{E}}{\log E}$, $E\rto\infty$,
grows slightly less (by a factor of $\frac{1}{\log E}$) than
the corresponding value for classically integrable systems.
It is, however, certainly well beyond the one for generic
classically chaotic systems, $\De_\infty (E)\gg \frac{1}{2\pi^2}
\log E$.

Finally a remark on the class of systems that are being described
by the findings of the present chapter will be added.
The reason why the model works is provided by the exponentially
increasing multiplicities of lengths of closed geodesics for
arithmetic Fuchsian groups, since the exponential behaviour
(\ref{KDexp}) of the form factor is caused by the compensation
of the exponential damping present in the amplitude factor
$A_{n,k}$ (\ref{dflucA}) through $g_p (l_n)$. In order this
to work it has been assumed that all $g_p (l_n )$ closed
geodesics of the same length $l_n$ shared alike factors $\chi(\ga)$.
Otherwise the sum over closed geodesics could not have been
rewritten as a sum over distinct primitive lengths, see
(\ref{dflucA}). Assuming the simplest case of $\gz_2$--valued
representations $\chi :\,\Ga \rto\{\pm 1\}$, one can group
the geodesics in classes of alike signs and define
$g_p (l_n )=g_n^+ +g_n^- $, where $g_n^\pm$ denotes the number of
geodesics of length $l_n$ with $\chi(\ga)=\pm 1$, respectively.
The diagonal term (\ref{Kdiagonal}) then reads
\beq
\label{Kdiagonalchi}
K_D (\tau;E)=\frac{1}{(4\pi p\db(E))^2}\sum_{\{l_n \}}\sum_{k=1}^%
\infty l_n^2\,e^{-kl_n}\,[ g_n^+ +(-1)^k g_n^- ]^2\,\de(\tau -
\frac{kl_n}{4\pi p\db(E)})\,[ 1+O(e^{-kl_n})]\ .
\eeq
The leading $(k=1)$--contribution thus contains the difference
$[g_n^+ -g_n^- ]^2 $ of the multiplicities referring to $\chi(\ga)
=+1$ and $\chi(\ga)=-1$. Only if this difference grows like
$|g_n^+ -g_n^- |\sim const.\,\frac{e^{l_n /2}}{l_n}$, $l_n
\rto\infty$, the model of sections 4.3 and 4.4 is applicable.
Once $g_n^+$ and $g_n^-$ are of the same order of magnitude
the leading contribution comes from $k=2$. Since then $[ g_n^+
+(-1)^2 g_n^- ]^2 =g_p (l_n)^2 \sim c_\Ga^2 \,\frac{e^{l_n}}
{l_n^2}$, this term is of a similar form as the analogous one
for $k=1$ in the non-arithmetic case. Thus the statistical
properties are expected to be generic, i.e.\ the level
spacings should be close to the GOE behaviour and the medium- and
long-range correlations should be described by Berry's theory.

As an example for a non-trivial representation $\chi$ take
an arithmetic Fuchsian group $\Ga_1$ leading to a symmetric
surface $\Ga_1\backslash\cH$. Then $\Ga_1$ is a normal subgroup
of index $N$ in another, also arithmetic, group $\Ga_2$.
According to (\ref{multcommensurable}) thus $<g_p^{(2)}(l)>
\sim\frac{1}{N}<g_p^{(1)}(l)>$, $l\rto\infty$. The symmetry group
$\Ga_2 /\Ga_1$ is represented via $\chi:\,\Ga_2 \rto\mbox{End}\,
(V_\chi)$ with $\mbox{ker}\,\chi \supseteq\Ga_1$. Since
therefore $\chi(\ga)=+1$ for $\ga\in\Ga_1$, one concludes that
$g_n^+ \geq g_p^{(1)}(l_n)$. Again only $\gz_2$--valued symmetry
classes shall be considered for simplicity. Then $\chi(\ga)=-1$
is only possible for $\ga\in\Ga_2$, hence $<g_n^- >\,\leq\, <g_p^{
(2)}(l_n)>\sim\frac{1}{N}<g_p^{(1)}(l_n)>$, yielding
$<g_n^+ >-<g_n^- >\,\geq\, (1-\frac{1}{N})<g_p^{(1)}(l_n)>\sim
const.\ \frac{e^{l_n /2}}{l_n}$ for $l_n\rto\infty$.
Therefore the model is still applicable to this case.

The more general situation of arbitrary representations
of a symmetry group can be treated analogously. In view
of the possible cancellations of multiplicities in the
$(k=1)$--term of (\ref{Kdiagonalchi}) the case of a $\gz_2$--valued
representation,
however, is the worst possible. Hence the Laplacian on a
symmetric arithmetic surface is always expected to share statistical
properties as discussed in the preceding sections. Artin's
billiard may serve as an example that has already been
studied above. Although the symmetry on the modular surface
is orientation reversing and thus $\Ga_2$ is not a subgroup
of $SL(2,\rz)$ but rather of the full group of isometries
of $\cH$, the above reasoning extends also to Artin's billiard,
since inverse hyperbolic orbits occurring due to orientation
reversing symmetries can as well be dealt with in the present
framework, see \cite{Venkov78,Venkov,Matthies}.

There may of course exist more general representations of
arithmetic Fuchsian groups $\Ga$ than the ones being
derived from symmetries. The latter excel by their triviality
on a subgroup of finite index, which leads to the observation just
made that a large enough fraction of closed geodesics is equipped
with positive $\chi$'s. Examples for the former may be provided
by the presence of Aharonov-Bohm type magnetic fluxes on
arithmetic surfaces $\GaH$. Depending on the strength of such
a flux the spectral properties of the respective Laplacian
are expected to deviate from the findings of the present
chapter. Once the phases $\chi(\ga)$ that arise when a
wave function $\psi (z)$ is carried along the geodesics related
to the $\ga\in\Ga$ and enclosing the Aharonov-Bohm flux line
``mix'' sufficiently among those geodesics being degenerate in length,
one could even retain generic spectral statistics like those
for non-arithmetic groups.

\subsection{Convergence Properties of the Selberg Zeta Function}
The final topic of the present chapter now again deals with
general Fuchsian groups of the first kind. However, it is observed
in the course of the following discussion that arithmetic groups
play a special role. As in section 3.6 methods of inverse
quantum chaology are applied and once again it turns out that
the strong fluctuations present for arithmetic quantum energy
spectra violate the prerequisits to apply the formalism
developed below to the arithmetic case. Referring to the heuristic
reasoning presented first in \cite{entropy} one can, however,
understand the reason for the obstruction occurring for arithmetic
groups.

The item of this section lies at the foundation of one of the
major objectives of quantum chaology, namely the derivation of
certain quantization rules that allow to determine the quantum
energies of a classically chaotic system in a semiclassical
approximation. Recently such quantization rules involving
dynamical zeta functions have been introduced and successfully
applied to a variety of different chaotic systems, see e.g.\
\cite{SieberPRL,Matthies,Tanner,Bogomolny,Keating,BerryKeating2,%
AurichProc}.

All these methods make strong use of the fact that the
semiclassical quantum energies are directly related to the
zeros of the dynamical zeta function on the critical line.
The problem one immediately faces when trying to compute
the non-trivial zeros explicitly is that the Euler product
(\ref{zetafct}) defining the dynamical zeta function in
general does not converge on the critical line. This
phenomenon is also referred to as the existence of an
{\it entropy barrier}, since it is the topological entropy
$\tau$ that determines the half-plane of convergence $Re\,s
>\tau$ for the Euler product. Concerning geodesic flows on
hyperbolic surfaces and the Selberg zeta function it was mentioned
earlier that the topological entropy universally is $\tau =1$,
and that the critical line is located at $Re\,s=\frac{1}{2}$.
The entropy barrier to be overcome hence has a width of
$\frac{1}{2}$.

In order to find a quantization rule one therefore has to
develop a consistent procedure to calculate the non-trivial
zeros of the zeta function other than searching for the
zeros of the Euler product. It was McKean \cite{McKean} who apparently
first noticed the possibility to rewrite the Euler product
(\ref{Eulerprod}) of the Selberg zeta function as a
Dirichlet series, but made no use of this. Berry and
Keating \cite{BerryKeating} then were the first to
introduce this Dirichlet series in order to obtain a
quantization rule from it, leading to their {\it Riemann-Siegel
look alike formula}. Sieber and Steiner \cite{SieberPRL}, Matthies
and Steiner \cite{Matthies} and Aurich and Steiner \cite{AurichProc}
then investigated the convergence properties of the Dirichlet series in
several examples and used it to calculate non-trivial
zeros. After that in \cite{entropy} a statistical model was
developed that predicts the domain of conditional
convergence for the Dirichlet series of the Ruelle-type
zeta function (\ref{EulerRuelle}), which as well can be
used to set up a quantization rule. This model can also be
extended to the Selberg zeta function, see e.g.\ \cite{AurichBolte}.

The above remarks stress the importance of investigating
the convergence properties of the Dirichlet series representing
the Selberg zeta function. Very much alike the discussion
in section 3.6 methods from analytic number theory will be
used below in conjunction with inverse quantum
chaology to derive statements about the abscissa of
convergence for the Dirichlet series. Again, as in section
3.6, for simplicity only cocompact Fuchsian groups $\Ga$
will be considered. But as also has been stated earlier, the
results do not depend on this restriction, since it is only
the contributions of the hyperbolic $\ga\in\Ga$ that are relevant.

Before entering the detailed studies just announced
the model introduced in \cite{entropy} will
be reviewed. In order to convert the Euler product
(\ref{Eulerprod}) into a Dirichlet series one has to
transform the product over $n\in\nz_0$ with the help of
Euler's identity
\beq
\label{Eulerident}
\prod_{n=0}^\infty (1-y\,x^n )=1+\sum_{m=1}^\infty \frac{(-1)^m
\,y^m \,x^{\frac{m}{2}(m-1)}}{\prod_{k=1}^m (1-x^k )}
\eeq
first; $|x|<1$ and $y\in\C$ suffice for the product and the
sum to converge. One then obtains that for $Re\,s>1$
\beq
\label{ZDirichlet}
Z(s)=\sum_\ro A_\ro \,e^{-sL_\ro}\ ,
\eeq
where the sum extends over all pseudo-orbits $\ro$, see
(\ref{pseudoorbit}), and the $L_\ro$'s denote the pseudo-lengths
(\ref{pseudolength}). The coefficients $A_\ro$ of the generalized
Dirichlet series (\ref{ZDirichlet}) are given by
\beq
\label{ai}
A_\ro =\prod_{i=1}^r a_i \ ,\ \ \ \ \ \ \
a_i :=\frac{(-1)^{m_i}\,e^{-\frac{m_i}{2}(m_i -1)l_{\ga_i}}}{\prod_{k=1
}^{m_i}(1-e^{-kl_{\ga_i}})}\ ,
\eeq
for $\ro =\{\ga_1^{m_1}\}_p \oplus\dots\oplus\{\ga_r^{m_r}\}_p$.
Recall that a pseudo-orbit $\ro$ is a formal combination of
finitely many (not necessarily primitive) closed geodesics, the
integers $m_1,\dots,m_r$ denoting the numbers of respective
repetitions of the primitive closed geodesics corresponding
to $\{\ga_1\}_p ,\dots,\{\ga_r \}_p$. The asymptotic relation
$\cN (l)\sim\cN_p (l)$, $l\rto\infty$, which is an analogue
of (\ref{Nhatprim}), expresses the fact that the number of
all closed geodesics of lengths not exceeding $l$ is in the
asymptotic regime $l\rto\infty$ already
given by the respective number of primitive geodesics. In the
limit $L\rto\infty$ therefore the number $\cN^{(P)}(L)$
of pseudo-orbits with $L_\ro \leq L$
is dominated by those $\ro$ that are completely composed of
primitive closed geodesics traversed only once, i.e.\ $m_1 =
\dots =m_r =1$. Their respective coefficients $A_\ro$ look
like
\beq
\label{Aro}
A_\ro =\prod_{i=1}^r \frac{(-1)}{1-e^{-l_{\ga_i}}}\sim(-1)^{|\ro|}\ ,
\eeq
where $|\ro|:=r$ denotes the number of primitive geodesics the
pseudo-orbit $\ro$ consists of.

Since already the lengths of primitive closed geodesics are
at least twice degenerate because of the time-reversal symmetry,
the pseudo-lengths $L_n$ will in general also possess multiplicities
exceeding one. Assuming that these multiplicities $g^{(P)}(L_n)$
are exclusively caused by degenerate primitive lengths and not
by different combinations of primitive lengths yielding the
same pseudo-length, degenerate $\ro$'s have identical
coefficients $A_\ro$. Thus
\beq
Z(s)=\sum_{\{L_n\}}g^{(P)}(L_n)\,A_n\,e^{-sL_n}
\eeq
for $Re\,s>1$. Numerical calculations now hint at an exponential
increase of the multiplicities according to
\cite{entropy,AurichBolte}
\beq
\label{multalpha}
<g^{(P)}(L)>\sim d\,e^{\al L}\ ,\ \ \ \ \ L\rto\infty\ ;
\eeq
$d$ and $\al$ are constants to be determined numerically. Thus
\beq
\label{Dirichletalpha}
Z(s)=\sum_{\{L_n\}}d\,A_n\,\frac{g^{(P)}(L_n)}{de^{\al L_n}}\,
e^{-(s-\al)L_n}\ ,
\eeq
which can be considered as
a generalized Dirichlet series in the variable $s-\al$.
The theory of Dirichlet series \cite{Hardy} as
e.g.\ also briefly reviewed in
appendix A now allows to determine the convergence properties of the
Dirichlet series for $Z(s)$ from (\ref{Dirichletalpha}) once
the pseudo-orbits have been arranged in ascending order of their
respective pseudo-lengths, $0=L_0 <L_1 <L_2 <\dots$. According to this
theory (\ref{ZDirichlet}) converges for $Re\,s>\si_c$ and diverges for
$Re\,s<\si_c$; furthermore it converges absolutely for $Re\,s>\si_a$,
$\si_a \geq \si_c$. The abscissa of convergence $\si_c$ and
of absolute convergence $\si_a$ are determined by the formulae
(\ref{absconv}). The asymptotic growth (\ref{pseudoPGT}) of the number
$\cN^{(P)}(L)$ of pseudo-orbits with pseudo-lengths not exceeding $L$
then fixes the abscissa of absolute convergence to be $\si_a =1$.
In \cite{entropy} a statistical model was established that
yielded $\si_c =\frac{1+\al}{2}$. The model assumes that
after applying the approximation (\ref{Aro}) the
coefficients $A_n$ in (\ref{Dirichletalpha}) represent random signs.
This conjecture is based on the observation that by (\ref{Aro})
$A_\ro \sim \pm 1$, the sign depending on whether the number
of primitive closed geodesics comprising the pseudo-orbit $\ro$
is even or odd, respectively. Arranging the pseudo-lengths in
ascending order and taking into account a supposed irregularity
in the distribution of primitive lengths should make the
numbers $|\ro_n|$ and $|\ro_{n+1}|$ modulo two independent of
one another.

Taking this randomness hypothesis for granted
one can obtain the result for $\si_c$ also slightly
differently, although in the same spirit as in \cite{entropy}.
For it is known \cite{Kac} that a series
\beq
\label{randomseries}
\sum_{k=1}^\infty (-1)^{\om_k}c_k
\eeq
of positive coefficients $c_k$ and random signs
$(-1)^{\om_k}$ either converges or diverges, depending on whether
the series $\sum_{k=1}^\infty c_k^2$ converges or diverges,
respectively. Recalling (\ref{pseudoPGT}) and (\ref{multalpha}),
one observes that the number of distinct pseudo-lengths up to a value
of $L$ grows asymptotically proportional to $e^{(1-\al)L}$.
Thus the criterion for convergence stated after (\ref{randomseries})
requires $Re\,s>\frac{1+\al}{2}$ for the Dirichlet
series (\ref{Dirichletalpha}) to converge, reproducing hence the
outcome of the model in \cite{entropy}. In conclusion, one learns
from this model
that the Dirichlet series representing the Selberg zeta function
is not expected to converge on the critical line $Re\,s=\frac{1}{2}$.
The distance of $\si_c$ to the critical line is determined by
the growth of the multiplicities of pseudo-lengths.

For arithmetic systems $\al$ is expected to be large since already
the multiplicities of primitive lengths grow exponentially.
Indeed, by numerical calculations of pseudo-length spectra up
to some cut-off value $L_{max}$ it was observed in \cite{entropy}
that $\al =0.4658$ for the regular octagon group, and
$\al =0.279$ for Artin's billiard. However, non-arithmetic
and completely desymmetrized systems should possess a
considerably smaller value for $\al$. In \cite{AurichBolte}
an example for such a system was studied numerically and
$\al=0.0572$ was found. It even may be that in these cases the
multiplicities will not really show an exponential behaviour,
but will rather follow a power law, leading to an effective
vanishing of $\al$. Then the Dirichlet series would converge
for $Re\,s>\frac{1}{2}$ and diverge for $Re\,s<\frac{1}{2}$,
but it would not be known whether it converged on the
critical line. In any case it would be possible to evaluate
$Z(s)$ close to the critical line and to obtain the non-trivial
zeros as minima. One could even try to go onto the critical
line and hope that a divergence would not show up when using
the finitely many available pseudo-lengths in (\ref{ZDirichlet}).
An example of such a procedure is presented in \cite{AurichBolte}.

A remark on the (arithmetic) case of Artin's billiard seems to be
in place now.
It is expected that $\si_c$ will in this
case be well above $\frac{1}{2}$, keeping in mind the rather
large value of $\al=0.279$. However, a numerical evaluation
of the formula (\ref{absconv}) for $\si_c$ yields a value
below $\frac{1}{2}$ \cite{Matthies,entropy}, at least
in the finite range of available pseudo-lengths. An
explanation for this observation is that the randomness
hypothesis the statistical model is built upon is apparently
violated \cite{entropy} in the computed range of the pseudo-length
spectrum. Thus Artin's billiard can apparently not be understood by the
above considerations.
The question whether or not this phenomenon pertains to
higher values of $L$ remains open.

In the following the question for the location of the abscissa
of convergence $\si_c$ of the Dirichlet series (\ref{ZDirichlet})
will be approached from a different side, employing methods
from analytic number theory and inverse quantum chaology. As
guiding references \cite{Titchmarsh,Hejhal} may be consulted.
The idea to be pursued below is similar to the one how to
obtain the PNT from the analytic properties of $\frac{\ze'(s)}
{\ze(s)}$ (see (\ref{psirz})), or how to obtain the PGT from
$\frac{R'(s)}{R(s)}$ (see (\ref{psiint})).
To this end define the function
\beq
\label{psizeta}
\psi_Z (L):=\sum_{\ro ,L_\ro\leq L} A_\ro \ ,
\eeq
where the above notation should indicate that the pseudo-lengths
have to be counted with their respective multiplicities.
The abscissa of convergence is then according to
(\ref{absconv}) given by
\beq
\label{sigmapsi}
\si_c =\limsup_{L_\ro \rto\infty}\,\frac{1}{L_\ro}\,\log |\psi_Z
(L_\ro)|\ .
\eeq
If it were possible to derive an $O$--estimate for $\psi_Z (L)$,
this would yield an upper bound for $\si_c$, i.e.\ $\psi_Z (L)=
O(L^a e^{bL})$, $a\in\rz$, $b>0$, results in the bound $\si_c
\leq b$. Accordingly, an $\Om$--result for $\psi_Z (L)$ would
give a lower bound on $\si_c$. Thus it will be attempted to
estimate $\psi_Z (L)$ for $L\rto\infty$. The principle tools
to be employed have already been used in section 3.6 and in
appendix A.

Using the Dirichlet series (\ref{ZDirichlet}) and
the integral (\ref{Perron}) one easily obtains
\beq
\label{psiPerron}
\psi_Z (L)=\frac{1}{2\pi i}\int_{c-i\infty}^{c+i\infty}\frac{ds}{s}\
e^{sL}\,Z(s)\ ,\ \ \ \ \ c>1,
\eeq
in analogy to (\ref{psiint}) and (\ref{psirz}). However, it does
not proof particularly useful to extend the contour of integration
in (\ref{psiPerron}) from $c-i\infty$ to $c+i\infty$. In the
following the integral will therefore be restricted to the finite
interval $[c-iT,c+iT]$, $T>0$, and the remainder that has been left
out in comparison to (\ref{psiPerron}) will be estimated.
This is achieved by the following\\

\noindent {\sc Lemma:} With the notations introduced above, and
$c>1$, $L>0$, $L\neq L_\ro$,
\beq
\label{lemma}
\psi_Z (L)=\frac{1}{2\pi i}\int_{c-iT}^{c+iT}\frac{ds}{s}\ e^{sL}\,
Z(s)+O\left(\frac{e^{cL}}{(c-1)T}\right)+O\left(\frac{L}{T}e^L\right)\ .
\eeq \\

\noindent {\sc Proof:} The proof is a standard calculation
in analytic number theory, see e.g.\ \cite{Titchmarsh}, pp.60.
Due to the importance of the result for the further considerations
the main ideas shall, however, be reproduced here.

Inserting the Dirichlet series (\ref{ZDirichlet}) into the integral
on the r.h.s.\ of (\ref{lemma}) one is left with an integrand
of $\frac{1}{s}e^{s(L-L_\ro )}$. Depending on the sign of
$L-L_\ro$ one has to choose different contours to render
the following integrals finite:
\begin{enumerate}
\item $L>L_\ro :$
\beq
\label{int1}
\frac{1}{2\pi i}\left\{ \int_{-\infty -iT}^{c-iT}+\int_{c-iT}^{c+iT}
+\int_{c+iT}^{-\infty +iT}\right\} \frac{ds}{s}\ e^{s(L-L_\ro)}
=1\ ,
\eeq
because the pole of the integrand at $s=0$ is enclosed by the
contour.
\item $L<L_\ro :$
\beq
\label{int2}
\frac{1}{2\pi i}\left\{ \int_{\infty -iT}^{c-iT}+\int_{c-iT}^{c+iT}
+\int_{c+iT}^{\infty +iT}\right\} \frac{ds}{s}\ e^{s(L-L_\ro)}=0\ .
\eeq
\end{enumerate}
A typical integral to be estimated now can be treated by an
integration by parts ($L>L_\ro$),
\beqa
\int_{-\infty-iT}^{c-iT}\frac{ds}{s}\ e^{s(L-L_\ro)}&=& \left.
     \frac{e^{s(L-L_\ro)}}{s(L-L_\ro)}\right|_{-\infty -iT}^{
     c-iT}+\frac{1}{L-L_\ro}\int_{
     -\infty-iT}^{c-iT}\frac{ds}{s^2}\ e^{s(L-L_\ro)} \nonumber \\
 &=& \frac{e^{(c-iT)(L-L_\ro)}}{(c-iT)(L-L_\ro)}+\frac{e^{-iT(L-L_\ro)}}
     {L-L_\ro}\int_{-\infty}^c d\si \frac{e^{\si (L-L_\ro)}}
     {(\si-iT)^2} \ .
\eeqa
To obtain an upper bound for the absolute value of the above
expression one extracts the maximal value of the exponential
under the integral on the r.h.s.\ and obtains
\beqa
\left| \int_{-\infty -iT}^{c-iT}\frac{ds}{s}\ e^{s(L-L_\ro)}\right|
&\leq& \frac{e^{c(L-L_\ro)}}{|c-iT|(L-L_\ro)}+\frac{e^{c(L-L_\ro)}}
       {L-L_\ro}\int_{-\infty}^{+\infty}d\si\ \frac{1}{\si^2 +T^2}
       \nonumber \\
   &=& O\left( \frac{e^{c(L-L_\ro)}}{T(L-L_\ro)}\right) \ .
\eeqa
The remaining three integrals from (\ref{int1}) and (\ref{int2})
are of the same type and obey the same bounds. Thus
\begin{enumerate}
\item $L>L_\ro :$
\beq
\label{int3}
\frac{1}{2\pi i}\int_{c-iT}^{c+iT}\frac{ds}{s}\ e^{s(L-L_\ro)}
=1+O\left( \frac{e^{c(L-L_\ro)}}{T(L-L_\ro)}\right) \ ,
\eeq
\item $L<L_\ro :$
\beq
\label{int4}
\frac{1}{2\pi i}\int_{c-iT}^{c+iT}\frac{ds}{s}\ e^{s(L-L_\ro)}
=O\left( \frac{e^{c(L-L_\ro)}}{T(L-L_\ro)}\right)\ .
\eeq
\end{enumerate}
Using these bounds one finds with the help of (\ref{psizeta}) that
\beq
\label{psifinite}
\frac{1}{2\pi i}\int_{c-iT}^{c+iT}\frac{ds}{s}\ e^{sL}\,Z(s)=
\psi_Z (L) +O\left( \frac{e^{cL}}{T}\sum_{\ro}|A_\ro |\frac{e^{-c
L_\ro}}{|L-L_\ro|}\right) \ .
\eeq
One is therefore left with the task of bounding the sum on the r.h.s.\
of (\ref{psifinite}). Somewhat tedious but straightforward
calculations that may e.g.\ be found in \cite{Titchmarsh}, pp.60,
and will not be reproduced here yield
\beq
\label{sumbound}
\sum_\ro |A_\ro |\frac{e^{-cL_\ro}}{|L-L_\ro|}=O\left( \frac{1}{c-1}
\right) +O(L\,e^{(1-c)L})\ ,
\eeq
finally proving the lemma.  \\

The next task in order to derive a bound for $\psi_Z (L)$ is to
estimate the integral that is left on the r.h.s.\ of
(\ref{lemma}). Because of the factor of $e^{sL}$ under the
integral it proofs useful to move the contour from $Re\,s=c>1$
to the left in the complex $s$-plane as far as possible. It
turns out that this can be achieved up to directly before the
critical line. What is missing yet is an estimate of $Z(s)$
on the contour. It is at this point where an inverse quantum
chaology argument enters the game. Namely, Hejhal can prove
an estimate for $Z(s)$ in the half-plane $Re\,s>\frac{1}{2}$
depending on an upper bound for $S(p)=N_{fl}(E(p))=\frac{1}{\pi}
\mbox{arg}Z(\frac{1}{2}+ip)$. Define
\beq
\label{deltap}
\De (p):=p^\mu (\log p)^\nu (\log\log p)^\la \ ,
\eeq
where the exponents $\mu,\nu,\la$ are chosen such that $\De (p)$
tends to infinity for $p\rto\infty$. Using the notation $s=\si +ip$,
$\si\in\rz$, $p>0$, {\it Theorem 10.10}
in \cite{Hejhal} then states that $|S(p)|=O(\De (p))$ implies
\beq
\label{logZestimate}
\log Z(s)=O\left(\De (p)^{2\max [0,1-\si]}\log \De(p)\right) \ ,
\eeq
for $\si =Re\,s\geq \frac{1}{2}+\frac{1}{\log \De(p)}$ and $p$ large
enough, $p\geq p_0 (\De)$. It will henceforth be assumed that
$p_0 (\De)$ is chosen that large that $\De (p)$ is monotonically
increasing for $p\geq p_0 (\De)$. From now on $\si$ shall be restricted
to the domain $\frac{1}{2}+\frac{1}{\log\De (p)}<\frac{1}{2}+\frac{1}
{\log\log\De (p)}\leq \si\leq 1$, on which $2\max [0,1-\si] =1-
\frac{2}{\log\log\De (p)}$. Thus $\log Z(s)=O( \De (p)\,e^{-
\frac{2\log\De (p)}{\log\log\De (p)}+\log\log\De (p)})=
O(\De (p)\,e^{-\frac{\log\De (p)}{\log\log\De (p)}})$. Since therefore
$|\log Z(s)|\leq \eta\,\De(p)^{1-\frac{1}{\log\log\De (p)}}$ for some
constant $\eta >0$ and $p\geq p_0 (\De)$, which again must be chosen
large enough, one obtains that $|\log Z(s)|\leq \ve\De (p)$ for all
$\ve >0$ and $p$ large enough. Thus
in the domain alluded to above
\beq
\label{Zbound}
|Z(s)|\leq e^{\ve\De (p)}\ ,\ \ \ \ \ \forall \ve >0\ .
\eeq
The integration contour in (\ref{lemma}) can by Cauchy's
theorem now be moved to $Re\,s=\frac{1}{2}+\de$, $\de :=\frac{1}
{\log\log\De (T)}$, without loosing control on the
magnitude of $Z(s)$,
\beq
\label{contourmove}
\frac{1}{2\pi i}\int_{c-iT}^{c+iT}\frac{ds}{s}\ e^{sL}\,Z(s)=
\frac{1}{2\pi i}\left\{ \int_{c-iT}^{\frac{1}{2}+\de -iT}+
\int_{\frac{1}{2}+\de -iT}^{\frac{1}{2}+\de +iT}+\int_{\frac{1}{2}
+\de +iT}^{c+iT}\right\} \frac{ds}{s}\ e^{sL}\,Z(s)\ .
\eeq
The first and the third integral on the r.h.s.\ behave alike,
\beqa
\label{int13}
\left| \int_{c\pm iT}^{\frac{1}{2}+\de\pm iT}\frac{ds}{s}\ e^{sL}\,
Z(s)\right| &\leq& \int^c_{\frac{1}{2}+\de}d\si\ \frac{e^{\si L}}
{|\si\pm iT|}|Z(\si\pm iT)| \nonumber \\
   &=& O\left( \frac{1}{T}\,e^{\ve \De (T)}\,e^{cL}\right) \ .
\eeqa
The second integral, however, can be bounded according to
\beqa
\label{scdint}
\left| \int_{\frac{1}{2}+\de -iT}^{\frac{1}{2}+\de +iT}\frac{ds}{s}
\ e^{sL}\,Z(s)\right| &\leq& e^{(\frac{1}{2}+\de)L}\int_{-T}^{+T}dt\
\frac{|Z(\frac{1}{2}+\de+it)|}{|\frac{1}{2}+\de +it|}\nonumber \\
  &=& O\left( e^{(\frac{1}{2}+\de)L}\,e^{\ve \De (T)}\int_0^T
      dt\,\frac{1}{1+t}\right) \\
  &=& O\left( e^{(\frac{1}{2}+\de)L}\,e^{\ve \De (T)}\log T
      \right) \ . \nonumber
\eeqa
Combining the estimates (\ref{int13}) and (\ref{scdint}) with
the lemma one obtains the following \\

\noindent {\sc Proposition:} The function $\psi_Z (L)$ can be
estimated for $L\rto\infty$ using the notations introduced
above as
\beq
\label{prop}
\psi_Z (L)=O\left(\frac{1}{T}e^{\ve \De(T)}e^{cL}\right)+O\left(
e^{(\frac{1}{2}+\de)L}e^{\ve \De(T)}\log T\right)+O\left(
\frac{1}{T}e^{cL}\right)+O\left(\frac{L}{T}e^L \right) \ .
\eeq  \\

Recall that one seeks for an upper bound of the type $\psi_Z (L)
=O(L^a e^{bL})$, $1>b>0$, in order to bound the abscissa of
convergence by $\si_c\leq b<1$. The third and the fourth term
on the r.h.s.\ of (\ref{prop}) therefore require to take
$T=e^{dL}$ for some appropriate $d>c-1>0$.
Then, however, $\De(T)=e^{\mu
dL}(dL)^\nu (\log dL)^\la$, and the first two terms
prohibit to obtain the desired form of the estimate unless
$\mu =0$. Once, however, $\De (p)=(\log p)^\nu (\log\log p)^\la$
for $\nu< 1$ or $\nu =1$, $\la \leq 0$, and thus $|S(p)|=O(\log p)$,
one observes with the choice $d=c>1$
\beq
\label{psiZestimate}
\psi_Z (L)=O( e^{\ve cL})+O( L\,e^{(\frac{1}{2}+\de+\ve c)L})+O( L\,
e^{-(c-1)L})
\eeq
for $L\rto\infty$ and for all $\ve >0$. Since $\de =\frac{1}{\log\log
\De (T)}$ vanishes for $L\rto\infty$ and $\ve$ can be made as small
as required, one draws from (\ref{sigmapsi})
determining the abscissa of convergence the bound
\beq
\label{sicbound}
\si_c \leq \frac{1}{2}+\ve' \ \ \ \ \ \ \mbox{for all }\ve'>0\ .
\eeq
Therefore, the Dirichlet series (\ref{ZDirichlet}) for the Selberg
zeta function converges (conditionally) for all $s$ with
$Re\,s>\frac{1}{2}$, since one can then always choose $\ve'$
as small as desired.

This being a conditional result, the question for the
validity of the input $|S(p)|=| N_{fl}(E(p))|$ $=O(\log p)$
immediately arises. For arithmetic groups the lower bound
already employed in sections 3.6 and 4.2, $S(p)=\Om_\pm (\frac{
\sqrt{p}}{\log p})$, forces to chose $\mu\geq \frac{1}{2}$ in
(\ref{deltap}), therefore ruling out an application of (\ref{prop})
to obtain an upper bound for $\si_c$. This negative observation
comes in accordance with the result obtained from the
statistical model, $\si_c =\frac{1+\al}{2}>\frac{1}{2}$,
where $\al$, describing the growth of the multiplicities of
pseudo-lengths, is rather large for arithmetic groups, and thus
$\si_c$ violates the lower bound (\ref{sicbound}).
The lower bound for general (cocompact) Fuchsian groups, $S(p)=\Om_\pm
(\sqrt{\frac{\log p}{\log\log p}})$, still allows for
expecting $\mu =0$ and $\nu\leq 1$. Now suppose
that $S(p)=O((\log p)^\nu )$, i.e.\ $N_{fl}(E)=O((\log E)^\nu )$.
In order to obtain the saturation value $\De_\infty (E)$ of the
spectral rigidity in the semiclassical limit (see (\ref{deltainf}))
one has to evaluate
\beq
\frac{\db}{2L}\int_{E-\frac{L}{\db}}^{E+\frac{L}{\db}}
dE'\ [N_{fl}(E')]^2
= O((\log E)^{2\nu})+O\left( (\log E)^{2\nu -1}
\frac{L^2}{E^2}\right)
\eeq
in the limit $L\rto\infty$ and $E\rto\infty$. The interval
of length $2\frac{L}{\db}$ to be integrated over has to be kept
small compared to $E$. Choosing $L=E^a$, $0<a<1$, then yields the
semiclassical asymptotics
\beq
\label{rigidityasympt}
\De_\infty (E)\sim O((\log E)^{2\nu})\ .
\eeq
Berry's semiclassical theory for the spectral rigidity yielding
for generic classically chaotic systems $\De_\infty (E)\sim
\frac{1}{2\pi^2}\log E$ now implies $\nu =\frac{1}{2}<1$. Once
one believes in the applicability of this heuristic theory
to non-arithmetic Fuchsian groups one has to draw the conclusion
that the Dirichlet series (\ref{ZDirichlet}) for the
respective Selberg zeta functions converge conditionally
for $Re\,s>\frac{1}{2}$. The above reasoning can be supported
by the numerical evaluation of the formula (\ref{absconv})
in an example of a non-arithmetic group in \cite{AurichBolte},
which yielded a result in accordance with the bound (\ref{sicbound}).
{}From the part of the pseudo-length spectrum calculated in
\cite{AurichBolte} one can, however, not draw a clear-cut
conclusion on the precise value of $\si_c$ because the plotted
curve still oscillates rather strongly.

A further confirmation of the above results may be provided by
considering the Riemann zeta function $\ze(s)$. Assuming the
Riemann hypothesis its non-trivial zeros are given by
$s_n =\frac{1}{2}\pm i\ga_n$, $\ga_n \geq 0$.
Supposing that either $\ga_n$ or $\ga_n^2$ correspond to
quantum energies of a yet unknown physical
system, the spectral statistics show a behaviour as if
the classical limit of this system were chaotic without
time-reversal invariance, i.e.\ the level spacings can be
well described by the GUE random matrix ensemble, and the
spectral rigidity and the number-variance saturate for
$L\rto\infty$, see e.g.\ \cite{Odlyzko,BerryNL}. The function
$S(p)=\frac{1}{\pi}\,\mbox{arg}\,\ze(\frac{1}{2}+ip)$ is known to obey
the same lower bound $S(p)=\Om_\pm (\sqrt{\frac{\log p}
{\log\log p}})$ \cite{MontgomeryHelv} as the analogous quantity
for a generic Selberg zeta function. Its
upper bound, however, is given by $S(p)=O(\frac{\log p}{\log\log p})$,
see e.g.\ \cite{HejhalDuke}. Berry's theory of the rigidity
stating that $\De_\infty (E)\sim\frac{1}{4\pi^2}\log E$ would
predict that $|S(p)|$ is asymptotically given by $\sqrt{\log p}$,
possibly times some power of $\log\log p$, therefore clearly
being within the rigorous bounds. This is also in accordance with
the belief that the $\Om$--estimate is ``likely to be best
possible'' \cite{MontgomeryHelv} and possibly sharp.
The upper bound for $S(p)$
leads to an analogue of (\ref{logZestimate}), namely $\log \ze
(s)=O((\log p)^{2\max [0,1-\si]}\log\log p)$ for $\si =Re\,s
\geq \frac{1}{2}+\frac{1}{\log\log p}$. The estimate corresponding
to (\ref{Zbound}), $\ze (s)=O(p^\ve )$, $\forall\ve >0$, is then
equivalent to the Lindel\"of hypothesis, see \cite{Titchmarsh}
for details.

In conclusion it appears that as long as the geodesic flow
on a hyperbolic surface $\GaH$ is generic in the sense that
the spectral statistics can be described by RMT on small
scales and follow Berry's prediction for the rigidity, especially
if the prescribed saturation occurs with the predicted
energy dependence, the Dirichlet series
for the Selberg zeta function converges for $Re\,s>\frac{1}{2}$.
The arithmetic systems once again drop out of this general
scheme by reasons that seem to be understood: sticking to
the statistical model they are provided by the exponential
degeneracies of the (pseudo-) length spectra, whereas consulting
inverse quantum chaology it is the exceptional spectral
statistics, showing much less correlations, that cause the trouble.
Hence the arithmetic case once again exemplifies the duality
of classical and quantum properties and demonstrates the
mechanisms of their interplay.

It is tempting now to express the expectation that generic
systems with a chaotic classical limit might show an analogous
behaviour regarding the convergence properties of their
dynamical zeta functions. As long as their phase spaces are
compact and Pesin's theorem \cite{Pesin} on the equality of metric and
topological entropy holds, the Dirichlet series for the
dynamical zeta functions should converge conditionally for
$Re\,s>\frac{\lb}{2}=\frac{\tau}{2}$, i.e.\ up to immediately
before the critical line. The expectation is based on the fact that
although the above inverse quantum chaology argument required detailed
knowledge about the analytic properties of the Selberg zeta
functions, the mechanism seems to be universal as it dwells on
the magnitude of $N_{fl}(E)$. One only needs a ``rigid''
spectrum with a saturating spectral rigidity that can be
described by Berry's general scheme. For generic chaotic systems
all evidence is for this to be satisfied, and
thus, although the knowledge about the associated dynamical
zeta functions is much poorer than
for the Selberg zeta function, at least there seems to be
no obvious obstacle around against the expectation to hold.
One is, however, far from proving this since the technical
problems are enormous in the general case.

\xsection{Summary}
This investigation contained a discussion of the semiclassical
quantization for a class of strongly chaotic systems. The relevant
aspects of classical and quantum mechanics for the unconstrained
motions of single particles on hyperbolic surfaces with
arithmetic fundamental groups were studied. The main body of
this text consisted of two major parts: chapter 3 discussed
classical mechanics, i.e.\ the geometry of the surfaces the
systems are defined on; chapter 4 then was devoted to an
investigation of the quantum mechanical energy spectra of the
arithmetical systems.
It was worked out that the arithmetic nature of the fundamental
groups involved had consequences for the geometry of the
respective surfaces. In particular the length spectra of closed
geodesics reveal high degrees of degeneracies.

In the context of
the semiclassical quantization of classically chaotic systems
the philosophy of studying ``generic'' systems includes the
requirement of only dealing with completely desymmetrized
systems. Symmetries may lead to unwanted effects that
superimpose the structures one tries to explore, and in many
cases they can rather easily be removed. Once the systems
have been desymmetrized one can compare them irrespective
of their differences in detail. Those quantities that
appear to share common properties can then be used to
characterize the class of chaotic dynamical systems. One
manifestation of a symmetry in a classical system is the
presence of degeneracies in the spectrum of periodic orbits
with respect to their actions. After removing symmetries
two classes of generic systems remain. The first one comprises of
time-reversal invariant systems, whereas the second one consists
of systems without time-reversal invariance. The philosophy
referred to above continues in assuming that then periodic
orbits generically can at most be twofold degenerate in
action due to a time-reversal invariance. Further multiplicities
would be considered as accidental. The discussion of the
arithmetical systems, however, revealed that there exist
perfectly chaotic Hamiltonian dynamical systems with
multiplicities of lengths of periodic orbits that even grow
exponentially with length. These multiplicities are not really
accidental since they can be traced back to the structure of
the set out of which the geodesic lengths are allowed to be
taken. The arithmetic structure inherent in this set forces the
lengths of closed geodesics not to cluster too strongly for
$l\rto\infty$. Since the total number of closed geodesics
with lengths up to $l$ has to grow according to the universal
prime geodesic theorem, $\cN(l)\sim\frac{e^l}{l}$, $l\rto\infty$,
the low number of distinct lengths up to~$l$, $\hat\cN (l)\sim
\frac{2}{c_\Ga}\,e^{l/2}$, $l\rto\infty$, has to be compensated
by exponentially increasing multiplicities, $<g(l)>\sim c_\Ga
\frac{e^{l/2}}{l}$, $l\rto\infty$. In chapter 3 details
on the mechanism resulting in the exponential law for the
mean multiplicities occurring for arithmetic fundamental
groups have been worked out. Since there exist infinitely many
arithmetic Fuchsian groups these exceptional systems form a
whole class of strongly chaotic dynamical systems that cannot
be neglected. In addition, there is a further (discrete)
dynamical system known with similar properties. This is the
so-called {\it cat map}, whose classical and quantum
properties have been discussed in detail by Keating \cite{KeatingNL}.

It has long been known that the arithmetical systems excel by
a further property, namely by the existence of infinitely
many pseudosymmetries. Although their definition includes
geometric symmetries, non-trivial pseudosymmetries cannot be removed
in a kind of desymmetrization procedure. It might appear that
the impossibility to ``de-pseudosymmetrize'' the arithmetical
systems suffices to consider them as generic, but
the discussion of multiplicities in their length spectra showed
that the algebraic and geometric structures induced by
pseudosymmetries are important enough to result in considerable
effects. On the quantum mechanical side of the problem, which
was discussed in chapter 4, these structures affect quantities
that are commonly be considered as characteristic for a distinction
of classically chaotic and integrable systems. In particular
statistical properties of quantum mechanical energy spectra were
discussed. Regarding the latter, symmetries manifest themselves as
independent superpositions of spectra referring to individual
symmetry classes. Thus the total quantum energy spectrum of a
system possessing a discrete and finite symmetry group
contains finitely many subspectra, which can each be viewed as
generic, since they are spectra of desymmetrized systems.
It was observed that regarding non-trivial pseudosymmetries
the situation is somehow reversed. The eigenvalue spectrum of
an arithmetical system is a subspectrum of infinitely many
other, also arithmetic, spectra. If it were an independent
superposition of infinitely many spectra, one could immediately
identify the result as showing Poissonian fluctuations, see
e.g.\ \cite{Mehta}. However, the fact that it is only a
subspectrum in such an infinite superposition complicates
the use of this point of view to draw conclusions on the
spectral statistics for arithmetical systems.

This is one reason for chapter 4 to proceed differently in its
investigation of arithmetic energy spectra. It appeared to be more
convenient to employ the exponential growth of multiplicities
in the geodesic length spectra. The spectral form factor
turned out to be a useful means for a periodic-orbit
investigation of spectral statistics. The two quantities that
were picked out to be studied were the level spacings distribution
and the number variance. The former yields information
on short-range correlations, whereas the latter takes
medium- and long-range correlations into account. The
exponential increase of multiplicities of lengths allowed for the
development of a simplified model for the form factor. This model
was essentially only based on the obtained exponential increase
for small $\tau$ and the saturation for $\tau\rto\infty$
of the form factor. Applied to the level spacings distribution
and to the number variance the model was found to describe
the numerically observed phenomena qualitatively correctly.
Quantum energy spectra of arithmetical systems are reminiscent
of those for classically integrable systems. Their fluctuations
are much stronger than those for generic classically chaotic
systems. They show a level attraction that grows with increasing
energy and the level spacings approach a Poissonian behaviour
for $E\rto\infty$. This finding is in contrast to
the integrable case that seems to yield stationary distributions
already at finite energies. On larger scales the correlations
in arithmetical spectra appear to be slightly stronger than
those observed for integrable systems. This is reflected in
the energy dependence of the saturation value of the spectral
rigidity. The latter was found by Berry \cite{BerryA400}
to be $\De_\infty (E)\sim const.\ \sqrt{E}$, $E\rto\infty$,
whereas the model for the number variance of arithmetical
systems yielded $\De_\infty (E)\sim\frac{2\db}{\pi}\,\frac{\sqrt{E}}
{\log E}$, $E\rto\infty$. However, the spectral statistics
in arithmetical quantum chaos are much more similar to those
of classically integrable systems than to the ones of generic
classically chaotic systems.

Sections 3.6 and 4.5 on fluctuations in geodesic length spectra
and on convergence properties of the Selberg zeta function,
respectively, had to take the different spectral statistics
for the hyperbolic Laplacian on arithmetic and non-arithmetic
surfaces into account. It turned out that in both sections
an application of inverse quantum chaology proved useful. The
desired results, however, could only be obtained in the
non-arithmetical case. There Berry's observation on the
saturation value of the spectral rigidity, $\De_\infty (E)
\sim\frac{1}{2\pi^2}\log E$, $E\rto\infty$, sufficed as an
input to apply Hejhal's theorems \cite{Hejhal} of inverse
quantum chaology. In section 3.6 the remainder term to the
leading asymptotics in the PGT thus followed to be of the form
$Q_R (l)=e^{\frac{1}{2}l}\,\om (l)$, with $\om(l)$ denoting some
unknown function containing powers and logarithms of $l$.
The result
of section 4.4 on $\De_\infty (E)$ for arithmetical systems
could only bring down the upper bound of $e^{\frac{3}{4}l}\,\om (l)$
for $Q_R (l)$ to $e^{\frac{2}{3}l}\,\om(l)$. Numerical evidence
obtained from three arithmetic groups, however, suggested
that the exponent for the remainder term $Q_R (l)$ should
also be $\frac{1}{2}$ for arithmetical systems. Thus the
inapplicability of the inverse quantum chaology reasoning
for arithmetic groups rather seems to be of a technical nature
than of a fundamental one. Regarding convergence properties of
the Dirichlet series for the Selberg zeta function, however,
the difference between the arithmetic and the non-arithmetic case
seems to be not void of consequences. Using Hejhal's
{\it Theorem 10.10} \cite{Hejhal} of inverse quantum chaology
and Berry's result on $\De_\infty (E)$ for generic systems
revealed a convergence of the Dirichlet series at least until
directly before the critical line, which is the physically
interesting domain. Again, the strong spectral fluctuations
present for arithmetical systems prevented an application of this
method to the latter. The statistical model for the convergence
properties that was introduced in \cite{entropy} now hints at
the reason for this obstruction. The exponentially increasing
multiplicities of geodesic lengths yield an exponential growth
of the multiplicities of pseudo-lengths. The exponent $\al$
describing the latter increase is a measure for the distance
of the domain of convergence to the critical line.

In conclusion, the arithmetical systems that were studied in
the present text excel by properties of important classical
and quantum mechanical quantities that distinguish them
from those of strongly chaotic systems that are commonly considered
as generic. Rather convincing heuristic reasons for the
exceptional spectral statistics could be derived from the
classical properties of arithmetical chaos. It was mainly a
combination of heuristic reasoning with the intuition gained
from numerical observations that could be used in conjunction
with rigorous results. This amalgam of different methods
proved to be particularly fruitful. The arithmetical systems
thus turned out to provide a convenient test-ground for
the ideas and methods developed in the framework of
periodic-orbit theory.

\vfill
\subsection*{Acknowledgements}
I want to thank Professor Frank Steiner for uncountable
fruitful discussions and for his constant support throughout
the time I have been his student.

I enjoyed many useful discussions with Ralf Aurich and Holger
Ninnemann. Furthermore, Ralf Aurich, Claudia Matthies, Holger
Ninnemann and Gunther Steil generously supplied me with
their numerical data. Many thanks are due to them.

I acknowledge financial support by the University of Hamburg
through a Doktorandenstipendium.

\setcounter{section}{0}
\renewcommand{\thesection}{\Alph{section}}
\xsection{The Riemann Zeta Function}
In the main body of the present work the Selberg trace
formula and the Selberg zeta function have been extensively used.
Historically, Selberg introduced his formalism in close analogy
to the theory of the Riemann zeta function and the
distribution of prime numbers. Since many techniques appearing
in the latter theory can be carried over to the case of the
Selberg zeta function, some important tools that were developed
to study the Riemann zeta function will be briefly introduced
in this appendix. Those details that will be omitted can be
found in e.g.\ \cite{Titchmarsh,Ingham}.

The Riemann zeta function $\ze (s)$ is a meromorphic function
for all $s\in\C$ that has a simple pole at $s=1$ with residue
one. For $Re\,s>1$ it is defined by
\beq
\label{Riemannzeta}
\ze (s)=\sum_{n=1}^\infty \frac{1}{n^s}=\prod_p (1-p^{-s})^{-1}\ .
\eeq
The {\it Dirichlet series} for $\ze (s)$ extends over all positive
integers $n$, whereas its {\it Euler product} runs over all primes $p$.
The analogy to Selberg's zeta function $Z(s)$ leads to an
identification of the primes $p$ with $e^{l(\ga)}$, where
$l(\ga)$ denotes the length of the primitive closed geodesic
related to $\ga\in\Ga$, and an identification of the integers
$n$ with the pseudo-orbits $\ro$. These identifications describe
what for the Selberg zeta function is the ``classical'' side
of its theory. The ``quantum'' side is missing in the theory
of $\ze (s)$ in that there is no self-adjoint operator known,
whose eigenvalues are related to the non-trivial zeros of
the Riemann zeta function. The trivial zeros of $\ze (s)$
are, however, explicitly known to be located at $s_k =-2k$,
$k\in\nz$. If a self-adjoint operator related to the
non-trivial zeros were known, the Riemann hypothesis (RH)
would be true, since then (depending on how the operator is
defined) either $\ga_n$ or $\ga_n^2$ is a real eigenvalue
of it. This means that the non-trivial zeros $s_n =\frac{1}{2}
\pm i\ga_n$ lie on the critical line $Re\,s=\frac{1}{2}$.
But, the RH still being unproven, the $\ga_n$ can be complex.
It is only known that $0<Re\,s_n <1$. There is, however,
tremendous evidence in favour of the RH from extensive
numerical computations of non-trivial zeros \cite{Odlyzko}.
Therefore it seems to be justified to assume the validity
of the RH throughout, if not stated otherwise.

The importance of the Riemann zeta function derives not only
from its connection to the RH, being one of the most
famous unsolved problems in mathematics, but also from its
decisive role played in the proof of the {\it prime number theorem}
(PNT) and for estimating the remainder term appearing in the PNT.
Thus the analytic properties of $\ze (s)$ are essential for
describing the distribution of prime numbers, which is a
central issue of number theory. It will now be explained
how the validity of the RH influences the magnitude of the
remainder in the PNT. Let therefore
\beq
\label{primecount}
\pi (x):= \# \{ p;\ p\leq x\}
\eeq
be the counting function of prime numbers. The PNT now states
that
\beq
\label{PNT}
\pi (x)=li(x)+Q(x)\ ,\ \ \ \ \ Q(x)=o(x/\log x)\ ,\ \ \ x\rto\infty\ .
\eeq
$\pi (x)$ may also be expressed by two other functions that have
been introduced by Chebyshev. Using the {\it von Mangoldt function}
\beq
\label{Mangoldt}
\La (n):= \left\{ \begin{array}{ccl} \log p & , & n=p^k \\
0 & , & \mbox{otherwise} \end{array} \right.
\eeq
these are defined as
\beqa
\label{Chebyshev}
\th (x)  &:=& \sum_{p\leq x}\log p \ , \nonumber \\
\psi (x) &:=& \sum_{n\leq x}\La (n) =\sum_{k\geq 1}\sum_{p^k \leq x}
              \log p \\
         & =& \sum_{k\geq 1} \th (x^{\frac{1}{k}}) \nonumber \ .
\eeqa
When $x^{\frac{1}{k}}<2$ is reached the last series breakes off, i.e.\
when $k>\frac{\log x}{\log 2}$. Writing $\psi (x)=\th (x)
+\cR (x)$ one can easily estimate that $\cR (x)=O( \sqrt{x}
(\log x)^2 )$, $x\rto\infty$.
{}From the definition of $\th (x)$ one obtains
\beq
d\th (x)=\sum_p \log p\ \de (x-p)\,dx\ ,
\eeq
and thus
\beq
\int_2^x \frac{d\th (t)}{\log t}=\sum_p \log p \int_2^x
\frac{dt}{\log t}\,\de (t-p)=\sum_{p\leq x}1=\pi (x)\ .
\eeq
An integration by parts yields
\beq
\pi (x)=\int_2^x \frac{d\th (t)}{\log t}=\frac{\th (x)}{\log x}+
\int_2^x dt\,\frac{\th (t)}{t(\log t)^2}+O(1)\ .
\eeq
Using $\th (x)=\psi (x)+O(\sqrt{x}(\log x)^2 )$ leads to
\beq
\label{PsiPNT}
\pi (x)=\frac{\psi (x)}{\log x}+\int_2^x dt\,\frac{\psi (t)}
{t(\log t)^2}+O(\sqrt{x}\log x)\ .
\eeq
Proving the PNT is thus equivalent to determining the leading
asymptotic behaviour of the Chebyshev function $\psi (x)$
for $x\rto\infty$, and estimating the remainder term $Q(x)$
can be achieved by knowing the remainder to the asymptotics
of $\psi (x)$.

At this stage now the Riemann zeta function enters the game.
Using its Euler product for $Re\,s>1$ one observes that
\beq
-\frac{\ze'(s)}{\ze (s)}=\sum_{n=1}^\infty \frac{\La (n)}{n^s}\ .
\eeq
Employing the Cauchy integral theorem one can derive that
(for $b>1$, $a>0$, $a\not= 1$)
\beq
\label{Perron}
\frac{1}{2\pi i}\int_{b-i\infty}^{b+i\infty}ds\,\frac{a^s}{s}=
\left\{ \begin{array}{ccl} 1 & , & a>1 \\ 0 & , & a<1 \end{array}
\right. \ \ .
\eeq
This result may be used to show that $(x\not\in \nz$)
\beq
\label{psirz}
\psi (x)=\sum_{n=1}^\infty \La (n)\,\frac{1}{2\pi i}\int_{b-i\infty}^{
b+i\infty}\frac{ds}{s}\left( \frac{x}{n}\right)^s =-\frac{1}{2\pi i}
\int_{b-i\infty}^{b+i\infty}ds\, \frac{x^s}{s}\frac{\ze'(s)}{\ze (s)}
\ ,
\eeq
a relation that together with (\ref{PsiPNT}) clearly shows
how the PNT is related to the analytic properties of $\ze (s)$.
One can now use the Weierstra\3 representation of $\ze (s)$
as a product over its zeros to obtain from (\ref{psirz})
the {\it explicit formula} of Riemann and von Mangoldt,
\beq
\label{explicitformula}
\psi (x)=x-\sum_{s_n}\frac{x^{s_n}}{s_n}-\frac{1}{2}\log (1-x^{-2})
-\frac{\ze'(0)}{\ze (0)}\ ,
\eeq
where the (conditionally convergent) sum runs over all non-trivial
zeros $s_n =\be_n +i\ga_n $ of $\ze (s)$. Denoting $\si_0 :=\sup\
\{ \be_n ;\ s_n =\be_n+i\ga_n \}$, the RH is equivalent to $\si_0 =
\frac{1}{2}$. An estimate for the sum over the non-trivial
zeros in (\ref{explicitformula}) can be found in \cite{Ingham},
$\sum_{s_n}\frac{x^{s_n}}{s_n}=O(x^{\si_0}(\log x)^2 )$,
leading to
\beq
\label{psiasym}
\psi (x)=x+P(x)\ ,\ \ \ \ \ P(x)=O(x^{\si_0}(\log x)^2 )\ .
\eeq
Since $\int_2^x \frac{dt}{(\log t)^2}=li (x) -\frac{x}{\log x}
+O(1)$, one observes from (\ref{PsiPNT}), using (\ref{psiasym}),
that
\beq
\label{PNTemainder}
\pi (x)=li (x)+O( x^{\si_0}\log x)\ ,
\eeq
which is the PNT with an estimate for the remainder term $Q(x)$.
This relation shows the influence of the validity of the RH
on the PNT. Notice that in order that $Q(x)=o(x/\log x)$ one has to
show that $\si_0 <1$, i.e.\ $\ze (s)$ must not have a zero on
$Re\,s=1$, nor a subsequence of zeros with real parts accumulating
at $1$. To show this was the main achievement of
Hadamard and de la Vall\'ee Poussin in proving the PNT in 1896.

The point of view that will be taken now introduces a
connection of the RH (equivalently the PNT) to the convergence
properties of a certain Dirichlet series. The general theory
of Dirichlet series is presented in \cite{Hardy}, where one can
find the following results. A {\it generalized Dirichlet series}
is a series of the form $F(s)=\sum_{n=1}^\infty a_n e^{-\la_n s}$,
$a_n \in\kz$, $\la_1 <\la_2 <\la_3 <\dots$. If $\la_n =\log n$,
it is called an {\it ordinary Dirichlet series}. In any case
there exists a number $\si_c \in\rz\cup\{\pm\infty\}$ such that the
Dirichlet series converges for $Re\,s>\si_c$ and diverges for
$Re\,s<\si_c$. Since the series in addition
converges uniformly on compact
sets, $F(s)$ is a holomorphic function in the domain of convergence.
There also exists a number $\si_a$, $\si_a \geq\si_c$, such that the
Dirichlet series converges absolutely for $Re\,s>\si_a$.
The {\it abscissae of convergence} $\si_c$ and $\si_a$ are
determined by
\beqa
\label{absconv}
\si_c &=& \lsup \frac{1}{\la_N}\log \left| \sum_{n=1}^N a_n \right|
          \ ,\nonumber \\
\si_a &=& \lsup \frac{1}{\la_N}\log \sum_{n=1}^N |a_n | \ .
\eeqa
Applying (\ref{absconv}) to the (ordinary) Dirichlet series
for $\ze (s)$ one obtains $(a_n =1,\ n\in\nz)$ that $\si_c =1=
\si_a$. The location of the abscissae of convergence follows from
the existence of the pole of $\ze (s)$ at $s=1$, which prevents
the series to converge for $Re\,s\leq 1$.

Our goal now is to define a Dirichlet series that yields a
meromorphic function for $s\in\C$ and whose abscissa of convergence
$\si_c$ is given by $\si_0$, i.e.\ by the non-trivial zero of
$\ze (s)$ with largest real part. In the vicinity of $s=1$ the
Riemann zeta function behaves like $\ze (s)=\frac{1}{s-1}+\ga
+O((s-1))$, $\ga$ being the Euler constant. Thus $\frac{\ze'(s)}
{\ze (s)}=-\frac{1}{s-1}+\mbox{regular terms}$, $s\rto 1$,
is a meromorphic function
with poles at $s=1$ and at the zeros of $\ze (s)$. Then
\beq
f(s):=\ze (s)+\frac{\ze'(s)}{\ze (s)}
\eeq
defines a meromorphic function that is holomorphic for $Re\,s>\si_0$.
In the {\it critical strip} $0<Re\,s<1$ its poles are located at
the non-trivial zeros $s_n$ of $\ze (s)$. Inserting for $Re\,s>1$
the Dirichlet series' for $\ze (s)$ and $\frac{\ze'(s)}{\ze (s)}$
yields
\beq
f(s)=\sum_{n=1}^\infty \frac{a_n}{n^s}\ ,\ \ \ \ \ a_n =1-\La(n)\ .
\eeq
By (\ref{absconv}) the abscissae of convergence for this
Dirichlet series are
\beqa
\label{fvonsconv}
\si_a &=& \lsup\ \frac{1}{\log N}\ \log \sum_{n=1}^N |1-\La (n)|
          \ , \nonumber \\
\si_c &=& \lsup\ \frac{1}{\log N}\ \log \left| \sum_{n=1}^N (
          1-\La (n))\right| \nonumber \\
      &=& \lsup\ \frac{1}{\log N}\ \log |N-\psi (N)| \\
      &=& \lsup\ \frac{\log |P(N)|}{\log N}\ . \nonumber
\eeqa
Since $\La (n)\not= 0$ only for $n=p^k$, and $\hat\pi (x)=\#
\{p^k ;\ p^k \leq x\}\sim \pi(x)\sim \frac{x}{\log x}$, $x\rto\infty$,
one concludes that $\sum_{n\leq N} |1-\La (n)| \sim 2N$
for $N\rto\infty$.
Therefore the Dirichlet series for $f(s)$ converges absolutely
for $Re\,s>\si_a =1$. The interesting observation one makes with
(\ref{fvonsconv}) is that the abscissa of conditional convergence
is determined by the first pole that is encountered when moving with
the axis $Re\,s=const.$ to the left. Namely, since $\si_c$ has
obviously to fulfill $\si_c \geq \si_0$, but the asymptotics
$P(x)=O(x^{\si_0}(\log x)^2 )$ gives $\si_c \leq\si_0$, one concludes
$\si_c =\si_0$. The Dirichlet series for $f(s)$ therefore
converges in the maximal possible domain. In sloppy terms one
could call this Dirichlet series a ``detector'' for the
non-trivial zero of $\ze (s)$ with largest real part,
indicating through its convergence properties.

The condition $\si_c \geq\si_0$ now also gives a lower bound
for $P(x)$, namely
\beq
P(x)=\Om (x^{\si_0 -\ve})\ \ \ \ \forall \ve >0\ .
\eeq
Thus the true magnitude of the remainder term $Q(x)$ in the PNT
is asymptotically bounded from below by
$x^{\si_0 -\ve}$ for all $\ve >0$ and from above
by $x^{\si_0}\log x$. Therefore $Q(x)=x^{\si_0}\cdot\om (x)$,
where $\om (x)$ is some combination of logarithmic functions.
It is thus the ``leading'' term $x^{\si_0}$ that determines the
fine structure in the PNT. If now the RH were true, the remainder
term in the PNT would have the ``minimal'' asymptotic behaviour
$|Q(x)|\propto \sqrt{x}\cdot\om (x)$. In this case the best lower
bound available is $\om (x)=\Om_\pm (\frac{\log \log \log x}{\log x})$
\cite{Ingham}.

\xsection{Desymmetrizing the Hyperelliptic Involution}
The general procedure of desymmetrizing the quantum problem of a
particle on a hyperbolic surface possessing symmetries
has been reviewed in section 4.1. This appendix now contains
an explicit application of the general formulation derived in
\cite{Venkov,VenkovZograf} to a rather simple case, namely
the so-called hyperelliptic involution, emphasizing the point
of view employing the Selberg trace formula. A discussion
of the example used here can be found in \cite{AurichBolte}.
The following presentation, however, differs a little from
the one given in \cite{AurichBolte} by being closer to
\cite{Venkov,VenkovZograf} in order to serve more explicitly as
an example for the general situation.

The symmetry under consideration is the {\it hyperelliptic involution}
(see e.g.\ \cite{Farkas}) present on all hyperelliptic (compact) Riemann
surfaces. The latter ones can be realized as two-sheeted
coverings of the sphere. If the surface $\GaH$ is hyperelliptic
and of genus $g$, this covering is branched at the $2g+2$
Weierstra\3 points. The operation that interchanges the two sheets
of the covering is an involution (i.e.\ the symmetry group is
isomorphic to $\gz_2$ if no other symmetries are present, as
will be assumed henceforth), called the hyperelliptic
involution. The hyperelliptic surfaces form a $(2g-1)$--dimensional
subvariety of the $(3g-3)$--dimensional moduli space of compact
Riemann surfaces of genus $g$; for $g=2$ all compact surfaces
are hyperelliptic.

For the following it proves useful to change the model of
hyperbolic geometry and to pass to the Poincar\'e unit-disc
$\cD =\{w\in\C;\ |w|<1\}$ by mapping the upper half-plane $\cH$
via $z\mapsto w=Cz$, $C=\frac{1}{\sqrt{2}}\left({1\atop i}
{i\atop 1}\right)$, for $z\in\cH$. $SL(2,\rz)$ is then mapped to
$SU(1,1)=\{ \left({\al\atop\bar\be}{\be\atop\bar\al}\right);
\ |\al|^2 -|\be|^2 =1\}$ by conjugation with $C$. A $g\in SU(1,1)$
operates on $\cD$ via fractional linear transformations of the
form
\beq
\label{fraclinSU}
g\,w=\frac{\al w+\be}{\bar\be w+\bar\al}\ ,\ \ \ \ \ w\in\cD\ .
\eeq
The conjugation of a Fuchsian group $\Ga\subset SL(2,\rz)$ by
$C$ yields a discrete subgroup of $SU(1,1)$ that will also be
denoted as $\Ga$ by abuse of notation. The geodesics of the
appropriately transformed hyperbolic metric are the half-circles
and straight lines perpendicular to $\partial\cD =\{ w\in\C ;\
|w|=1\}$. \cite{BalazsVoros} may serve as a reference to find
further details concerning the geometry of this model.

Any Fuchsian group $\Ga$ leading to a hyperelliptic surface $\Ga
\backslash\cD$ may be obtained in the following manner (see also
\cite{Aurich90}). One can choose a fundamental domain $\cF
\subset\cD$ whose boundary $\partial\cF$ consists of $4g$
geodesic segments. The corner points $w_1 ,\dots,w_{4g}$ are
enumerated in ascending order when going counterclockwise
along $\partial\cF$ and starting with $w_1$ on the positive
real axis. $w_2 ,\dots,w_{2g}$ are placed in the upper half
of $\cD$ ($Im\ w_i >0$). The remaining corner points are
obtained as $w_{2g+1}=-w_1 ,\dots, w_{4g}=-w_{2g}$. $\cF$ is
therefore symmetric under the operation $w\mapsto -w$. Having
fixed $w_2 ,\dots,w_{2g}$ one has to vary $w_1$ (hence also
$w_{2g+1}=-w_1$) on the real axis until the constraint
$\mbox{area}(\cF)=4\pi (g-1)$ is fulfilled.
The fractional linear transformations of the form (\ref{fraclinSU})
that identify opposite geodesic segments comprising $\partial\cF$
then serve as generators for a strictly hyperbolic Fuchsian
group $\Ga$ possessing $\cF$ as a fundamental domain.
This identification of pairs of edges of $\cF$ is compatible
with the symmetry $w\mapsto -w$, so that
$\Ga\backslash\cD$ is a compact hyperelliptic surface of genus $g$.
The above construction yields every such surface by choosing
the corner points $w_2 ,\dots,w_{2g}$ appropriately in the
upper half of $\cD$, therefore clearly showing that the subvariety
of hyperelliptic surfaces in the moduli space of compact
surfaces of genus $g$ is of (complex) dimension $2g-1$.

The generators $b_1 ,\dots,b_{4g}$ will be enumerated such
that $b_i$ identifies the geodesic segment connecting $w_i$
and $w_{i+1}$ with the one connecting $-w_i$ and $-w_{i+1}$,
$i=1,\dots,2g$.
By their very construction then $b_{i+2g}=b_i^{-1}$
for $i=1,\dots,2g$.
The generators obey the constraint
\beq
\label{constraint}
b_1 b_2^{-1}b_3 b_4^{-1}\dots b_{2g-1}b_{2g}^{-1}\ b_1^{-1}b_2 \dots
b_{2g-3}^{-1}b_{2g-2}b_{2g-1}^{-1}b_{2g}=\unmat \ .
\eeq
The hyperelliptic involution is realized as
the mapping $w\mapsto -w$, which can be represented by the matrix
$S=\left({-i\atop 0}{0\atop i}\right)\in SU(1,1)$, $S\not\in\Ga$.
Since the $b_i$'s identify opposite edges of $\cF$, one notices that
$Sb_i S=b_i^{-1}$, where $S^2 =\unmat$ as an identity in $PSU(1,1)=
SU(1,1)/\{\pm\unmat\}$ has been used. The $2g+2$ fixed points
of $S$ on $\Ga\backslash\cD$, i.e.\ the Weierstra\3 points,
can now be identified. Obviously, $u_1 =0$ is fixed by $S$. The
second Weierstra\3 point $u_2$ is represented by the corner
points of $\cF$ that are all being identified to one another
by $\Ga$. The remaining fixed points $u_3 ,\dots,u_{2g+2}$
are given by the mid-points of the $2g$ pairs of opposite
edges of $\cF$ as determined by the hyperbolic metric.

A new Fuchsian group $\Ga'$ is now introduced by adjoining
$S$ to $\Ga$, i.e.\ $\Ga'$ consists of all words in the generators
$b_1 ,\dots,b_{4g},S$, subject to the constraint (\ref{constraint})
and fulfilling $b_i^{-1}=b_{i+2g}$, $S^2 =\unmat$, $Sb_i S=b_i^{-1}$.
It is thus possible to rewrite any word in those generators as
being of the form $S^\ve b_{i_1}\dots b_{i_n}$, $\ve\in\{0,1\}$.
One therefore obtains that $\Ga$ is a normal subgroup of index
two in $\Ga'$. The latter decomposes disjointly according to
\beq
\label{decompga'}
\Ga' =\Ga \dcup S\Ga\ ,
\eeq
thus reproducing for the hyperelliptic involution $S$ the algebraic
setting of symmetries reviewed in section 3.5. The symmetry group
$\Si$ is yielded as $\Si =\{\unmat,S\}$, $S^2 =\unmat$. Thus
$\Si \cong \gz_2$, and also for the unitary dual $\Si^* \cong \gz_2$.
Explicitly, $\Si^*$ is given by the two representations $\chi_+$
and $\chi_-$ of $\Ga'$; $\chi_+$ denotes the trivial representation
$\chi_+ (\ga')=1$ for all $\ga'\in\Ga'$, whereas $\chi_-$
is defined as
\beq
\chi_- (\ga'):=\left\{ \begin{array}{rcl} +1 &,& \ga'\in\Ga \\
-1 &,& \ga'\in S\Ga \end{array} \right. \ .
\eeq

Since $S\in SU(1,1)$ is elliptic ($|tr\,S|<2$), $\Ga'$ is not
strictly hyperbolic. Fortunately, it is possible to identify
all elliptic conjugacy classes of $\Ga'$ explicitly by their
fixed points on the surface $\Ga'\backslash\cD$. An
elliptic $R\in\Ga'$ has one fixed point $z_R$ in the interior of $\cD$,
and its conjugacy class $\{R\}_{\Ga'}$ fixes the set of points
$\Ga' z_R$ that are identified under $\Ga'$. To each elliptic
conjugacy class there hence corresponds the point $\Ga' z_R$ on
the surface $\Ga'\backslash\cD$; but this must be one of the
Weierstra\3 points $u_1 ,\dots,u_{2g+2}$. The Fuchsian group
$\Ga'$ therefore contains $2g+2$ elliptic conjugacy classes
$\{R\}_{\Ga'}$, all of them of order $m(R)=2$ (meaning the minimal
positive integer with $R^{m(R)}=\pm\unmat$). One can easily
determine representatives for the elliptic classes,
\begin{itemize}
\item $u_1$ is fixed by $S$,
\item $u_2$ is fixed by $Sb_1 b_2^{-1}b_3 b_4^{-1}\dots b_{2g-1}
b_{2g}^{-1}$,
\item $u_i$ is fixed by $Sb_{i-2}^{-1}$, for $i=3,\dots,2g+2$.
\end{itemize}
All other conjugacy classes in $\Ga'$ are hyperbolic ones.

The general receipt of \cite{VenkovZograf} now proceeds in
constructing the unitary representation of $\Ga'$ that is received
as being induced from the trivial representation of its subgroup
$\Ga$. To this end define the one dimensional representation
(on $\C$)
\beq
\overline{\chi}(\ga'):=\left\{ \begin{array}{rcl} 1 &,& \ga'\in\Ga \\
0 &,& \ga'\in S\Ga \end{array} \right. \ .
\eeq
The {\it induced representation} $\ro :\ \Ga'\rto\mbox{End}(\C\oplus
\C)$ is then obtained as
\beq
\ro (\ga')(v\oplus w):=\left[ \overline{\chi}(\ga')v+\overline{\chi}
(\ga'S)w\right]\oplus\left[ \overline{\chi}(S\ga')v+\overline{\chi}
(S\ga'S)w\right] \ ,
\eeq
for $\ga'\in\Ga'$ and $v,w\in\C$. On $\Ga$ this representation
operates trivially, $\rho(\ga)(v\oplus w)=v\oplus w$, $\ga\in\Ga$;
on $S\Ga$, however, it acts according to $\rho(S\ga)(v\oplus w)=
w\oplus v$, $\ga\in\Ga$. As a matrix representation on $\C^2$
the induced representation is hence given by
\beq
\ro (\ga')=\left\{ \begin{array}{rcl} {1\ 0\choose 0\ 1}&,&\ga'
\in\Ga \\ {0\ 1\choose 1\ 0}&,&\ga'\in S\Ga \end{array}\right. \ .
\eeq
Later, in the Selberg trace formula, $tr\,\ro (\ga')^k$, $k\in\nz$,
is needed. For $\ga\in\Ga$, clearly $tr\,\ro (\ga)^k =2$, but for
$\ga'\in S\Ga$ one obtains that $tr\,\ro (\ga')^k =1+(-1)^k$.
Altogether, one can reformulate this as $tr\,\ro (\ga')^k =\chi_+
(\ga')^k +\chi_- (\ga')^k$ for all $\ga' \in\Ga'$.

Venkov and Zograf now demonstrate \cite{VenkovZograf} that the
hyperbolic and the elliptic contributions to the Selberg trace
formulae for $\Ga$ endowed with the trivial representation and
for $\Ga'$ endowed with the induced representation $\ro$
coincide, respectively. The hyperbolic terms thus yield
\beqa
\sum_{\{\ga\}_{\Ga}}\sum_{k=1}^\infty \frac{l(\ga)\ g(kl(\ga))}{2
\sinh(\frac{k}{2}l(\ga))} &=& \sum_{\{\ga'\}_{\Ga'}}\sum_{k=1}^\infty
\frac{tr\,\ro (\ga')^k \ l(\ga')\ g(kl(\ga'))}{2\sinh(\frac{k}{2}
l(\ga'))} \\
  &=& \sum_{\{\ga'\}_{\Ga'}}\sum_{k=1}^\infty \frac{l(\ga')\ g(kl
  (\ga'))}{2\sinh(\frac{k}{2}l(\ga'))}+
  \sum_{\{\ga'\}_{\Ga'}}\sum_{k=1}^\infty \frac{\chi_- (\ga')^k \
  l(\ga')\ g(kl(\ga'))}{2\sinh(\frac{k}{2}l(\ga'))} \nonumber \ ,
\eeqa
where the outer sums extend over the respective primitive hyperbolic
conjugacy classes. This result yields a factorization formula
for the Selberg zeta function,
\beq
Z_\Ga (s)=Z_{\Ga'}^+ (s)\cdot Z_{\Ga'}^- (s)\ ,
\eeq
where for $Re\,s>1$
\beqa
\label{factorpm}
Z_{\Ga'}^+ (s) &:=& \prod_{\{\ga'\}_{\Ga'}}\prod_{n=0}^\infty
                    \left( 1-e^{-(s+n)l(\ga')}\right) \ ,\nonumber \\
Z_{\Ga'}^- (s) &:=& \prod_{\{\ga'\}_{\Ga'}}\prod_{n=0}^\infty
                    \left( 1-\chi_- (\ga')\,e^{-(s+n)l(\ga')}\right)\ ,
\eeqa
which should be compared with (\ref{zetafactor}) and
(\ref{zetachi}).
The factorization (\ref{factorpm}) comprises the desymmetrization
with respect to the hyperelliptic involution on the level of the
Selberg zeta function because the non-trivial zeros of $Z_{\Ga'}^\pm
(s)$ correspond to the eigenvalues of $-\De$ referring to the
two symmetry classes of $S$. The reason for this is that the
respective wavefunctions transform under $\chi_\pm$ according to
$\psi_\pm (\ga' z)=\chi_\pm (\ga')\psi_\pm (z)$, $\ga'\in\Ga'$. Hence
the $\psi_\pm (z)$ are invariant under $\Ga$ ($\chi_\pm (\ga)=1$
for $\ga\in\Ga$), and transform under $S$ as $\psi_\pm (Sz)=
\chi_\pm (S)\psi_\pm (z)=\pm \psi_\pm (z)$.

Finally it should be remarked that the explicit knowledge of the
elliptic conjugacy classes of $\Ga'$ allows to determine the
elliptic contributions to the Selberg trace formulae for $\Ga'$
endowed with the representations $\chi_\pm$ as
\beq
\pm\frac{g+1}{4}\int_{-\infty}^{+\infty}dp\ \frac{h(p)}{\cosh(\pi p)}\ ,
\eeq
see \cite{AurichBolte}. There one can also find a numerical evaluation
of the Selberg zeta functions $Z_{\Ga'}^\pm (s)$ on the critical line
for a specific surface of genus $g=2$.

\xsection{Estimates of Remainder Terms}
In analytic number theory several estimates are used to describe
remainder terms to the leading asymptotic behaviour of
functions of interest, the most prominent example being the
remainder $Q(x)$ in the PNT, see appendix A. Since in other fields,
like in physics, some of these estimates are not so commonly
used, their definitions will be supplied in this appendix.

Let therefore $f(x)$ be a function of the real variable $x\geq 0$
that shall be estimated for $x\rto\infty$ and compared to the
positive and monotonic function $g(x)$.

The first estimate is
\beq
f(x)=O(g(x))\ \ \ :\Leftrightarrow \ \ \ \limsup_{x\rto\infty}
\frac{|f(x)|}{g(x)}<\infty \ .
\eeq
Asymptotically $|f(x)|$ is thus bounded by $g(x)$, which might
therefore also be referred to as an {\it upper bound}.

Another, stronger, upper bound is
\beq
f(x)=o(g(x))\ \ \ :\Leftrightarrow \ \ \ \limsup_{x\rto\infty}
\frac{|f(x)|}{g(x)}=0\ .
\eeq

A sort of asymptotic {\it lower bounds} is provided by $\Om$--estimates,
which are given as
\beqa
f(x)=\Om (g(x)) &:\Leftrightarrow& \limsup_{x\rto\infty}\frac{|f(x)|}
                                   {g(x)}>0 \nonumber \ , \\
f(x)=\Om_+ (g(x)) &:\Leftrightarrow& \limsup_{x\rto\infty}\frac{f(x)}
                                     {g(x)}>0  \ , \\
f(x)=\Om_- (g(x)) &:\Leftrightarrow& \liminf_{x\rto\infty}
                                     \frac{f(x)}{g(x)}<0 \nonumber \ .
\eeqa

The $O$--estimate can also be formulated slightly differently:
$f(x)=O(g(x))$, if there exist some $x_0 \geq 0$ and $M>0$ such
that $|f(x)|\leq M\,g(x)$ for $x\geq x_0$. Analogously, $f(x)=
\Om (g(x))$, if $|f(x)|\geq M\,g(x)$ for $x\geq x_0$, etc.

\clearpage

\end{document}